\documentclass[preprint]{aastex6}

\def\lapp{\ifmmode\stackrel{<}{_{\sim}}\else$\stackrel{<}{_{\sim}}$\fi}
\def\gapp{\ifmmode\stackrel{>}{_{\sim}}\else$\stackrel{>}{_{\sim}}$\fi}

\usepackage{amsmath}

\begin{document}

\title{A New Electron Density Model for Estimation of Pulsar and FRB
  Distances}

\author{J. M. Yao\altaffilmark{1,3}}
\author{R. N. Manchester\altaffilmark{2}}
\and
\author{N. Wang\altaffilmark{1,4}}
\altaffiltext{1}{Xinjiang Astronomical Observatory, Chinese Academy of
  Sciences, 150, Science 1-Street, Urumqi, Xinjiang 830011,
  China;}
\altaffiltext{2}{CSIRO Astronomy and Space Science, Australia Telescope
  National Facility, P.O.~Box~76, Epping NSW~1710, Australia} \email{}
\altaffiltext{3}{University of Chinese Academy of Sciences, 19A Yuquan
  Road, Beijing 100049, China}
\altaffiltext{4}{Key Laboratory of Radio Astronomy, Chinese Academy of
  Science, Nanjing 210008, China}

\begin{abstract}
We present a new model for the distribution of free electrons in the
Galaxy, the Magellanic Clouds and the intergalactic medium (IGM) that
can be used to estimate distances to real or simulated pulsars and
fast radio bursts (FRBs) based on their dispersion measure (DM). The
Galactic model has an extended thick disk representing the so-called
warm interstellar medium, a thin disk representing the Galactic
molecular ring, spiral arms based on a recent fit to Galactic HII
regions, a Galactic Center disk and seven local features including the
Gum Nebula, Galactic Loop I and the Local Bubble. An offset of the Sun
from the Galactic plane and a warp of the outer Galactic disk are
included in the model.   Parameters of the Galactic model are
determined by fitting to 189 pulsars with independently determined
distances and DMs. Simple models are used for the Magellanic Clouds
and the IGM. Galactic model distances are within the uncertainty
range for 86 of the 189 independently determined distances and within
20\% of the nearest limit for a further 38 pulsars. We estimate that
95\% of predicted Galactic pulsar distances will have a relative error
of less than a factor of 0.9. The predictions of YMW16 are compared to
those of the TC93 and NE2001 models showing that YMW16 performs
significantly better on all measures. Timescales for pulse broadening
due to interstellar scattering are estimated for (real or simulated)
Galactic and Magellanic Cloud pulsars and FRBs.
\end{abstract}
 

\section{Introduction}\label{sec:intro}
Distances to astronomical objects are often difficult to estimate, but
are of fundamental significance. They determine, for example, the
source luminosity, the location in the Galaxy or the Universe and the
space velocity, all important for studies of the origin, evolution and
emission properties of the object in question. Pulsars come with a
built-in distance indicator, interstellar
dispersion, which results in a radio-frequency-dependent delay $\Delta t$
in the pulse arrival times:
\begin{equation}\label{eq:tdm}
  \Delta t = \frac{e^2}{2\pi mc} \nu^{-2} \int^D_0 n_e dl
\end{equation}
where $e$ and $m$ are the charge and mass of the electron, $c$ is the
velocity of light, $\nu$ is the radio frequency, $n_e$ is the local
free electron density and the integral is along the path to the pulsar
at distance $D$. The electron column density along the path is known
as the dispersion measure (DM) and can be measured using:
\begin{equation}\label{eq:dm}
  {\rm DM}= \int^D_0 n_e dl = 2.410\times 10^{-16} (t_2 -
  t_1)/(\nu_2^{-2} - \nu_1^{-2}) \;\;{\rm cm^{-3} pc}
\end{equation}
where $t_1$ and $t_2$ are observed pulse arrival times at frequencies
$\nu_1$ and $\nu_2$, respectively.

\begin{figure}[ht]
\includegraphics[angle=270,width=85mm,]{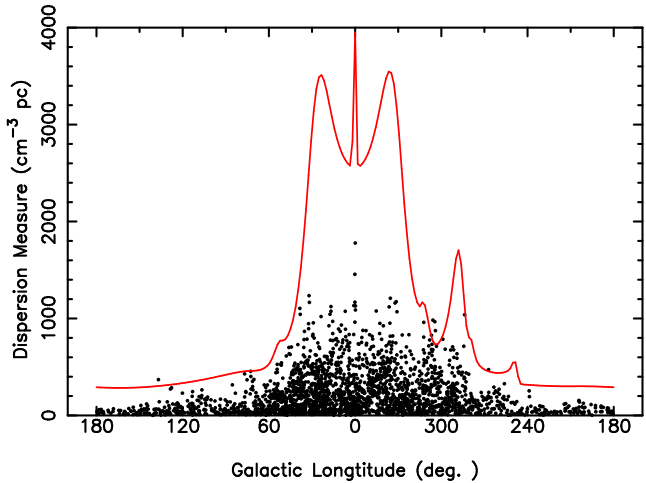}
\caption{Dispersion measure plotted against Galactic longtitude for
  the 2430 Galactic pulsars with known DM.  As will be discussed
    further below, the line is the total DM
  obtained by integrating the YMW16 electron density model through the
  Galaxy at $b=0$. \label{fg:dmgl}}
\end{figure}

Most of the $\sim$2540 currently known pulsars\footnote{See the ATNF
  Pulsar Catalogue:
  \url{http://www.atnf.csiro.au/research/pulsar/psrcat}, V1.54. Note
  that this total includes rotating radio transients (RRATs) even when
  no pulsational period has been identified, anomalous X-ray pulsars
  (AXPs), soft-gamma-ray repeaters (SGRs) with detected periodic
  pulsations, and X-ray isolated neutron stars (XINS).} are located
within our Galaxy -- the 29 known extra-galactic pulsars are all in
the Magellanic Clouds. Of the known pulsars, $\sim$2430 in the Galaxy
and 27 in the Magellanic Clouds have known DMs (the remainder were
discovered at high energies and currently have no radio
counterpart). Figure~\ref{fg:dmgl} shows the distribution of DMs as a
function of Galactic longitude $l$. 

 Fast radio bursts (FRBs) are isolated millisecond-duration radio
 bursts that have DMs that are much higher than expected from the
 Galaxy, suggesting an extra-galactic origin. The first FRB,
 discovered by \citet{lbm+07}, remained enigmatically alone for
 several years, but in the past few years many more have been
 detected\footnote{For a catalog of currently known FRBs, see
   \url{http://www.astronomy.swin.edu.au/pulsar/frbcat/} and
   \citet{pbj+16}}, firmly establishing them as an extra-galactic
 phenomenon. One FRB has been found to emit repeated pulses
 \citep{ssh+16}, suggesting a relationship with pulsars, especially
 RRATS \citep[cf.,][]{kp15}, but their origin remains a topic of great
 debate. However, this uncertainty about the nature of the source does
 not affect use of their DM as a distance indicator.

To make use of DMs for pulsar or FRB distance estimation, a model for
the distribution of free electrons in the intervening interstellar
medium (ISM) or intergalactic medium (IGM) is required. In principle,
at least for our Galaxy and nearby galaxies, such a model could be
derived directly from observations of recombination lines of ionised
interstellar gas or radio thermal continuum radiation. In practice
though, this is difficult. The intensity of both these types of
radiation is related to the emission measure (EM), which is
proportional to the integral of $n_e^2$ along the path. Since the
filling factor $\langle n_e^2\rangle/\langle n_e\rangle^2$ is largely
unknown and variable along typical interstellar paths, conversion from
EM to DM is problematic. Observed recombination lines and radio
thermal continuum are dominated by regions of high density such as HII
regions,  whereas, except at low Galactic latitudes, pulsar dispersion
largely originates in the more widely distributed ``warm ionized
medium'' (WIM) which can be investigated using wide-field H$\alpha$
observations \citep[e.g.,][]{hrt+03}. Even these observations are not
useful for inner regions of the Galaxy since they are affected by
interstellar dust extinction which is difficult to quantify,
especially when optical depths become high.

The main components of a reasonable Galactic $n_e$ model can be
identified from our general knowledge of the distribution of molecular
gas and star-formation regions in the Galaxy, as well as H$\alpha$
surveys of more local gas. For example, these can be used to define
the spiral-arm structure in the Galaxy \citep[e.g.,][]{hh14} and
suggest the presence of a ``thick'' electron disk with scale height
$\sim$1.5~kpc in addition to the ``thin'' disk defined by HII regions
which has a scale height $\sim$70~pc
\citep[e.g.,][]{hbk+08}. However, the parameters describing the $n_e$
distribution generally cannot be independently determined with
sufficient precision. Hence we are forced to use the observed DMs of
pulsars with independently known distances to calibrate the $n_e$
model. As is discussed further in \S\ref{sec:inddist} below, such
independent distances can be obtained from measurements of pulsar
parallax, pulsar associations with globular clusters or supernova
remnants, optical observations of binary companions and kinematic
distances based on observations of 21-cm HI absorption in pulsar
spectra combined with a model for Galactic rotation.

An accurate model for the $n_e$ distribution in the Galaxy has other
important uses besides the direct estimation of pulsar distances. For
example, it can help to identify pulsars that are likely to be
in other galaxies such as the Magellanic Clouds by their excess DM
over that expected from the Galaxy \citep[e.g.,][]{mfl+06}, and to
firmly establish that ``fast radio bursts'' (FRBs) are distant
extra-galactic sources \citep[e.g.,][]{tsb+13}. A knowledge of the
$n_e$ distribution along the path is important for the interpretation
of Faraday rotation in pulsars and extra-galactic sources when
investigating the large-scale structure of the Galactic magnetic field
\citep[e.g.,][]{hml+06,vbs+11} and for the interpretation of
measurements of interstellar scattering
\citep[e.g.,][]{bdd+14,lrk+15}. It is also important in investigations
of the intergalactic medium, especially intergalactic magnetic
fields \citep[e.g.,][]{ptu15}.

One of the first attempts to construct a Galactic $n_e$ model was by
\citet{mt81} and \citet{lmt85}. Their model (LMT85) consisted of two
main components: a thin disk of scale height 70~pc and a
$z$-independent component (where $z$ is the perpendicular distance
from the Galactic plane), both of which had a mid-plane electron
density that tapered off with increasing
Galactocentric radius $R$. In addition, the Gum Nebula \citep{gum52}
was recognised as significantly modifying the DMs of pulsars lying
within or behind it and hence was included in the model as a separate
term. The model was calibrated using the 36 pulsars with independent
distances known at the time, mostly kinematic distances from HI
absorption. At about the same time \citet{vn82} used data from the
Second Molonglo pulsar survey to investigate the form of the $n_e$
distribution at $R>$~5~kpc, concluding that, apart from an HII-region
layer, the disk thickness was large. They presented a simple model
that is very similar to the LMT85 model.

The next major step forward was by \citet{tc93}, building on the work
of \citet{cwf+91}, who included the effect of spiral structure in the
model and also made use of the increased number of independent
distances (74) and the overall dependence of pulsar DMs on Galactic
longitude. For the first time, they also took into account
observations of interstellar scattering. Their ``TC93'' model included
both thin and thick Galactic disks and made use of the sech$^2(x)$
function\footnote{$\mathrm{sech}^2(x) = 1/\mathrm{cosh}^2(x)=
4/(e^x + e^{-x})^2$} for the $z$-dependencies and some
$r$-dependencies of $n_e$ \citep[see also][]{gbc01}. The sech$^2(x)$
function has a physical basis \citep[see][]{spi42} and (unlike the
exponential function) has no cusp at $x=0$, but is asymptotically
exponential at large $x$.

Since 2002, the ``NE2001'' model \citep{cl02,cl03} has been the {\it
  de facto} standard for estimation of pulsar distances. This model
makes use of 112 independent pulsar distances and scattering measures
(SMs) for 269 pulsars to define a model which includes both a
quasi-smooth $n_e$ distribution and large-scale variations in the
strength of the fluctuations in $n_e$ that result in interstellar
scattering. The model includes multiple components: thin and thick
axisymmetric disks, spiral arms, and local components including a
local arm, a local hot bubble surrounding the Sun and relatively large
super-bubbles in the first and third Galactic quadrants, together with
over-dense components representing the Gum Nebula, the Vela supernova
remnant, Galactic Loop I and a small region around the Galactic
Center. Finally, NE2001 adds clumps toward pulsars with excess DM or
scattering and toward active galactic nuclei (AGNs) that have excess
scattering, and voids toward pulsars that have DMs below those
predicted by the quasi-smooth component. The model also uses sech$^2(z)$
dependencies and includes 67 parameters describing the main
components, including parameters that describe the $n_e$ fluctuation
amplitudes in each component. Another 493 parameters were used to
describe 82 clumps and 137 parameters to describe 17 voids. An
iterative approach was used to fit the parameters to the available
data, first solving for the parameters of the large-scale quasi-smooth
components, then adding the clumps and voids and other local features
and refitting as necessary \citep{cl02,cl03}. 

Compared to the TC93 model, distances obtained from NE2001 are
generally smaller, especially for high-latitude pulsars which in many
cases had unbounded distances in the TC93 model. This mainly results
from NE2001 having a denser thick disk ($n_{1_0}=0.035$~cm$^{-3}$,
$H_1 = 950$~pc) compared to TC93 ($n_{1_0}=0.019$~cm$^{-3}$, $H_1 =
880$~pc), where $n_{1_0}$ is the mid-plane electron density for the
thick disk and $H_1$ is its scale height.

In the next few years, the population of known pulsars increased
significantly, mainly as a result of the Parkes Multibeam Pulsar
Survey, and several groups \citep[e.g.,][]{lfl+06} began to notice
that NE2001 systematically under-estimates the $z$-distance of
high-latitude pulsars. To address this, \citet{gmcm08} analysed just
high-latitude pulsars ($|b|>40\degr$) that are not contaminated by HII
regions or spiral arms along the path. They obtained a smaller
mid-plane density, $n_{1_0}=0.014\pm 0.001$~cm$^{-3}$, and a much
larger (exponential) scale height $H_1 = 1830^{+120}_{-250}$~pc.
However, \citet{sw09} pointed out a bias in the fitting method used by
\citet{gmcm08} and obtained a revised scale height of
$1410^{+260}_{-210}$~pc. For both the exponential and sech$^2(z)$
distributions, the $n_{1_0} H_1$ product is equal to the
``perpendicular DM'' (DM$_\perp$), i.e., the DM integrated to infinity
perpendicular to the Galactic disk. For the TC93, N2001,
\citet{gmcm08} and \citet{sw09} fits, DM$_\perp$ is, respectively,
16.5~cm$^{-3}$~pc, 33.0~cm$^{-3}$~pc, 25.6~cm$^{-3}$~pc and
21.9~cm$^{-3}$~pc. \citet{gmcm08} also estimated a mid-plane volume
filling factor ($f_v$) for the thick-disk electrons of 0.014, rising
to around 0.3 at $z\sim$1000~pc.\footnote{The volume filling factor
  ($f_v$) is related to the line-of-sight filling factor $\langle
  n_e^2\rangle/\langle n_e\rangle^2$ by a form-dependent factor which
  is of order unity \citep{bmm06}.}

Based on accurate parallax measurements using the VLBA, \citet{cbv+09}
also showed that the NE2001 model under-predicts distances for several
high-latitude pulsars by a factor of two, but also over-predicts
distances for several relatively local pulsars. These results
illustrate the clumpy nature of the ISM on scales of 100 --
1000~pc. An analysis by \citet{sch12} considered a range of modified
TC93 and NE2001 models with different parameters for the thick
disk. These were tested using a sample of 41 pulsars with
$|b|>5\degr$, having well-determined independent distances and
unaffected by HII regions in the path. A modified TC93 model with a
scale height of $1590\pm 300$~pc and the same $n_{1_0} H_1$ product as
TC93 gave the most consistently accurate predictions.

In this work we present a new ``YMW16'' model for the large-scale
distribution of free electrons in the Galaxy, the Magellanic Clouds
and the intergalactic medium. The Galactic part of the model has the
same basic structure as TC93 and NE2001 but also some important
differences compared to these earlier models.  For example, we adopt a
four-armed spiral pattern (plus a ``local arm'') with the location and
form of the arms as given by \citet{hh14} based on observations of
more than 1800 HII regions across the Galaxy. The YMW16 model is
fitted to a compilation of 189 independent pulsar distance estimates,
using global and local optimization algorithms. There are seven local
features in the YMW16 model: the Local Bubble (LB), two regions of
enhanced $n_e$ on the periphery of the Local Bubble (LB1, LB2), the
Gum Nebula (GN), Loop I (LI), a region of enhanced $n_e$ in the Carina
arm and a region of reduced $n_e$ in the Sagittarius tangential
region. Most of these features are also included in the NE2001 model.

In contrast to both TC93 and NE2001, we do not make
use of observations of interstellar scattering in building the
model. Numerous observations \citep[e.g.,][]{sti06,tr07,bmg+10} have
shown that interstellar scattering is typically dominated by just a
few regions of very strong $n_e$ fluctuations along the path to a
pulsar. This makes it essentially impossible to satisfactorily model
the large-scale distribution of interstellar scattering and leads to
the very large scatter seen in plots of SM and scattering delay
$\tau_{\rm sc}$ versus DM \citep[e.g.,][]{lrk+15,kmn+15}. 

Another major difference between NE2001 and YMW16 is that we do not
attempt to correct discrepant model distances for individual pulsars
by invoking clumps or voids in their direction. We believe that this
procedure should be deprecated as it is likely to lead to poor
distance estimates for future discoveries (or even for currently known
pulsars) that have adjacent lines of sight. We have only invoked
additional features beyond the large-scale model where a number of
pulsars in a given region have discrepant distances and/or where there
is good independent evidence for such features. This policy of course
leads to more pulsars having over-estimated or under-estimated
distances, but this is an unavoidable consequence of our current
inability to adequately model the small-scale structure in the ISM.

With the advent of FRB astronomy and the likely increase in the number
and distribution of extra-galactic pulsars with increasingly sensitive
searches, there is a strong motivation to include extra-galactic
components in the electron-density model. We model the DM
contributions of the Large and Small Magellanic Clouds (LMC, SMC)
based on prior information on their shape, size and distance and
making use of the DMs of the 27 pulsars believed to be associated with
either the LMC or SMC.  We also include components for the IGM and the
host galaxy to allow distance estimates for FRBs.

The arrangement of our paper is as follows. Pulsars with
DM-independent distance estimates or limits and the methods by which
these are obtained are discussed in \S\ref{sec:inddist}. In
\S\ref{sec:model} we describe the functional form of the model
components and in \S\ref{sec:model-fit} we describe the model fitting
procedure and the algorithms used for global optimization and local
optimization. Results from the model fitting are presented and
compared with the predictions of the TC93 and NE2001 models in
\S\ref{sec:results}. A summary our results and concluding remarks are
given in \S\ref{sec:summary}. Tables of DM-independent distances and
limits, and corresponding model distances are given in Appendix
A. Details of the algorithms used to compute perpendicular distances
from spiral arms and the Gum Nebula are given in Appendix
B. Coordinate conversions for the LMC are given in Appendix C and the
{\sc ymw16} program code and outputs are described in Appendix D.

\section{Model-independent distances}\label{sec:inddist}
There are a variety of ways that pulsar distances can be estimated
independent of their DM and the Galactic $n_e$ model. As described
above, these independent distances are essential for the calibration
of any model for the $n_e$ distribution. We have obtained independent
distance data for 189 pulsars from the ATNF Pulsar Catalogue (V1.54),
only including pulsars with both upper and lower limits on the
estimated distance and known DMs. Only one pulsar from each globular
cluster association (and the Double Pulsar) is included. Parallax
measurements of low significance are also omitted. In this
section we discuss the different methods of obtaining independent
distances; the adopted distances are listed in Appendix A for each
category.

\subsection{Distances from annual parallax}
Measurements of annual parallax give a direct measure of the pulsar
distance. Such measurements can be made in two different ways: a) fitting
for annual parallax in precision timing solutions and b) direct
measurements of position shifts due to annual parallax using
very-long-baseline interferometry (VLBI). Of the 70 parallax
measurements listed in Table~\ref{tb:a1_px}, about 40 are based on
VLBI observations, one (PSR J0633+1746) on optical astrometry with the
Hubble Space Telescope, and the remainder on pulse timing analyses. Measured
parallaxes of low significance are biased by the ``Lutz-Kelker''
effect \citep{lk73}. This has been considered in the context of pulsar
parallax measurements by \citet{vwc+12} and we have adopted their
corrected distance estimates where available. Otherwise, we have
ignored parallax measurements with a value less than three times the
quoted uncertainty.

For three of the entries in Table~\ref{tb:a1_px} (PSRs J0437$-$4715, J1537+1155 and
J2129$-$5721), the distance estimate is based on a measure
of the time-derivative of the orbital period $\dot P_b$. If a
signficant excess $\dot P_b$ (over that expected from
general-relativistic decay of the orbit) is measured, then this excess can be
attributed to the Shklovskii effect \citep{shk70} and a distance
estimate obtained \citep{bb96}. For these three pulsars, the $\dot
P_b$ distance is the most precise distance estimate available. 

\subsection{Kinematic distances}
For low-latitude pulsars, 21-cm absorption spectra resulting from
dense HI clouds in the path can be used to obtain pulsar distance
limits by using a Galactic rotation model to convert radial velocities
of clouds to distances. Lower and upper distance limits correspond to
the presence and absence of absorption features, but upper distance
limits are always more difficult to estimate. Also for ``inner'' lines
of sight ($0\degr < l < 90\degr$ and $270\degr < l < 360\degr$)
difficulties arise due to ambiguities about whether a given spectral
feature corresponds to absorbing gas at the ``near'' distance (closer
than the tangent point) or the ``far''distance (beyond the tangential
point). Generally, emission features with $T_B \gapp 35$~K correspond
to significant absorption and this has been used a criterion to decide
on distance limits \citep{wbr79,fw90}. Table~\ref{tb:a1_kin} gives
kinematic distances derived from HI absorption and emission spectra
for 62 pulsars. Parallax distances are always underestimated because
of the Lutz-Kelker bias but kinematic distances tend to be
over-estimated because of a luminosity bias. Again, this bias was
considered by \citet{vwc+12} and we have adopted their limits when
available. Because of the difficulties in interpretation of observed
spectra, HI kinematic distances are generally less reliable than those
derived using other methods.

\subsection{Association with globular clusters}
Many (mostly) millisecond pulsars (MSPs) are associated with globular
clusters - it is clear that the dense cores of these clusters have
conditions that are favourable for the recycling of pulsars to form
MSPs. Distances to clusters are obtained by a variety of methods
including astrometry \citep[e.g.,][]{mam+06} and photometry
\citep[e.g.,][]{obb+07}. In our database we include just one pulsar
for each cluster, but assign to it a DM averaged over the pulsars in
that cluster. Table~\ref{tb:a1_gc} lists the 27 clusters and pulsars for which
distances have been obtained from observations of the associated
globular cluster. Globular clusters are all at large distances from
the Sun and these associations are of great importance in defining the
distance model, especially for larger $z$-distances.

\subsection{Association with nebulae}\label{sec:neb}
About 60 young pulsars are associated with supernova remnants (SNR)
but many of these were discovered at X-ray or $\gamma$-ray wavelengths
and have no radio counterpart. Consequently their DM is unknown and
they are not useful as independent distances for $n_e$
modelling. Table~\ref{tb:a1_neb} lists 14 pulsars that are associated with SNR or
pulsar wind nebulae (PWN), have a known DM and an estimated distance
based on the association. In this Table we also include a pulsar
(PSR J0248+6021) that is believed to lie in or near the HII region W5
\citep{tpc+11} and the magnetar PSR J1745$-$2900 that is believed to lie
close to the Galactic Center \citep{efk+13,rep+13}.

\subsection{Binary companion stars}
Although about 250 pulsars are members of a binary system, in orbit
with another star, in only a few cases is the companion star optically
identified. Table~\ref{tb:a1_stars} gives the nine published
independent distance estimates based on optical identifications.

\subsection{Extra-galactic pulsars}
Up to now, pulsars believed to lie outside our Galaxy have only been
found in the Magellanic Clouds, our nearest neighbor galaxies. A total
of 29 pulsars are known in the Clouds, six in the SMC and 23 in the
LMC. However, only 27 of these have known DM, with one pulsar in each
of the SMC and LMC being detected only at high energies. These 27
pulsars are listed in Table~\ref{tb:a1_mc} along with their nominal distance
estimates and limits which refer to the center of each Cloud.  The
Magellanic Clouds form an important step in the extra-galactic
distance ladder and so extensive studies of Cepheid and RR Lyrae
variables have been carried out in order to estimate their
distances. For the LMC we have adopted the distance modulus of
$18.48\pm0.05$ given by \citet{wal12}, which corresponds to a distance
of $49.7\pm1.1$~kpc. For the SMC, we adopt the value from
\citet{scg+04}: distance modulus $18.88\pm0.13$ corresponding to a
distance of $59.7^{+3.8}_{-3.5}$~kpc.

\subsection{Fast radio bursts}
Currently, there are 17 known FRBs as listed in Table~\ref{tb:a1_frb}
\citep{pbj+16}. Only one of these, FRB150418, has an identification of
the host galaxy, an elliptical galaxy at $z=0.492\pm 0.008$
\citep{kjb+16}, although this identification has been disputed
\citep{ak16,wb16}.

\section{Components of the model}\label{sec:model}
In this section we describe the various components of the YMW16 model
for the distribution of free electrons in the Galaxy, the Magellanic
Clouds and the intergalactic medium. We also describe the model for
interstellar scattering delays. For the Galaxy we use a right-handed
coordinate system with origin at the Galactic Center, $x$ axis
parallel to $l=90\degr$ and $y$ axis parallel to $l=180\degr$ where
($l,b$) are the usual Sun-centred Galactic coordinates.   The Sun
  is located at ($x=0$, $y=R_\odot$, $z=z_\odot$). We adopt a distance
  of the Sun from the Galactic Center, $R_\odot = 8300$~pc
  \citep{brm+11}\footnote{Both \citet{tc93} and \citet{cl02} used
    $R_\odot=8500$~pc. We ignore the small systematic offset
    introduced by this difference.}. Recent estimates of $z_\odot$
  range between +6~pc and +30~pc \citep{jdpj16}. Based on an analysis
  of nearby open clusters \citet{jdpj16} obtain $z_\odot = +6.2\pm
  1.1$~pc; we fix $z_\odot$ at +6.0~pc. The Galactic Cartesian
  coordinates of any object are ($x$, $y$, $z$) = ($D\sin l \cos b$,
  $R_\odot - D\cos l \cos b$, $z_\odot + D\sin b$), where $D$ is its
  distance from the Sun. We also use the Galactocentric cylindrical
  coordinate system ($R$, $\phi$, $z$), which is defined with
  $R=(x^2+y^2)^{1/2}$ and $\phi$ measured counter-clockwise from the
  $+x$ direction, i.e., toward the $+y$ direction.

  There is extensive evidence from HI surveys \citep[see][for a
  review]{kk09c} and stellar distributions \citep[e.g.,][]{ufm+14}
that the Galactic disk has a pronounced warp in its outer regions, most
probably induced by gravitational interactions with the Magellanic
Clouds. \citet{rrdp03} have modelled the warp as follows:
\begin{equation}
  z_w = z_c \cos(\phi - \phi_w),
\end{equation}
\begin{equation}
  z_c = \gamma_w(R - R_w),
\end{equation}
with the parameters $\phi_w = 0\degr$,\footnote{Note that
  \citet{rrdp03} measure $\phi$ counterclockwise from the $-y$
  direction, whereas we measure it counterclockwise from the $+x$
  direction.} where we take $R_w = 8400$~pc and fit for
$\gamma_w$. That is, the warp begins just outside the solar radius and
is toward positive $z$ in the $+x$ direction.

\subsection{Thick disk}\label{sec:thickdisk}
The presence of an extensive diffuse ionised medium in the disk of our
Galaxy was proposed more than five decades ago by \citet{he63} based
on evidence for free-free absorption in the spectrum of the Galactic
synchrotron background. Further evidence came from observations of
diffuse H$\alpha$ and other recombination lines
\citep{rsr73,hrt99,mrh06}, pulsar DMs
\citep{mt81,nct92b,cl02,gmcm08,sw09,sch12} and interstellar scattering
\citep{rd75,cwf+91,cl02}, all suggesting mid-plane electron densities
of 0.02 -- 0.03~cm$^{-3}$ and scale heights of $\sim$1000~pc with no
strong dependence of either mid-plane $n_e$ or scale height on
Galactocentric radius out to the edge of the Galactic
disk. \citet{hrt99} used observations of diffuse H$\alpha$ from the
Perseus arm region, about 2.5~kpc from the Sun at Galactic longitudes
$125\degr$ -- $150\degr$, corresponding to $R \sim$10.5~kpc, to
estimate a scale height $H_{n_e}$ for the WIM in this region of
$1000\pm 100$~pc. This estimate assumes a constant filling factor
$f_v$ for the ionised gas, so that $H_{n_e} =
2H_{n_e^2}$. \citet{hbh+14} used similar methods to investigate the
WIM in the vicinity of the Scutum-Centaurus arm at a distance of
$3.5\pm0.3$~kpc from the Sun at Galactic longitudes $320\degr$ --
$340\degr$, corresponding to $R\sim$5~kpc. They estimate an $n_e^2$
scale height for the WIM in this region of about 430~pc, corresponding
to $H_{n_e} \sim$860~pc given the same assumptions. These results
suggest that there is no strong dependence of $H_{n_e}$ on
Galactocentric radius out to the edge of the Galactic disk. They also
support the relative independence of WIM H$\alpha$ intensity and hence
of $n_e$ as a function of Galactocentric radius. Preliminary fits to
the independent distance data including $R$-dependencies of $n_e$
and/or $H_{n_e}$ were less successful than those omitting them.

  We therefore model the thick disk as a plane-parallel disk of scale
height $H_1$:
\begin{equation} 
  n_1=n_{1_0}\;g_d\;\mathrm{sech}^2\left(\frac{z-z_w}{H_1}\right)
\end{equation}
where $n_{1_0}$ is the mid-plane electron density and $H_1$ is the
scale height. The extent of the Galactic disk is defined by the
parameters $A_d$ and $B_d$ which are respectively the scale length of
the cutoff and the Galactocentric radius at which the cutoff
begins. The factor $g_d = 1$ for $R < B_d$, and
\begin{equation}\label{eq:disk}
  g_d = \mathrm{sech}^2 \left(\frac{R-B_d}{A_d}\right)
\end{equation}
for $R \ge B_d$.

 The radial extent of the thick disk (and other disk components)
 affects the model distance of distant pulsars and FRBs, especially
 those located toward the Galactic anticenter region. We have no
 independent distances for pulsars with $R > 15$~kpc and so we cannot
 calibrate the cutoff radius in this way. While there is ample
 evidence of Galactic HI extending to $R\sim 20$~kpc
 \citep[e.g.,][]{mdgg04}, the distribution of young stars that could
 ionize this gas is evidently more limited. A survey of distant HII
 regions by \citet{aaj+15} finds the last significant concentration
 peaking at $R\sim 12$~kpc, with only a few at $R>15$~kpc. We
 therefore fix $B_d$ at 15~kpc and $A_d$ at 2.5~kpc, which results in
 the disk density being reduced to $\sim 15$\% of its $R=15$~kpc value
 at $R\sim 19$~kpc. This allows for some ionisation beyond most of the
 known HII regions.

Despite the assumption of a plane-parallel thick disk, there are good
reasons to suggest that the density of the thick disk is reduced in
the vicinity of the Galactic Center. Firstly, there is a group of
globular cluster pulsars with $|l| \lapp 10\degr$, $|b|\gapp
10\degr$ at distances comparable to the Galactic Center, e.g., PSR
J1835$-$3259A in NGC6652 and PSR J1823$-$3021A in NGC6624, for which
the mean $n_e$ along the path is less than the nominal mid-plane
density of the thick disk. Secondly, observations of the so-called
``Fermi Bubbles'' \citep[see, e.g.,][and references therein]{cbtc15}
indicate that powerful winds from the Galactic Center region
have evacuated large bubbles extending to $z$-distances of $\sim$8~kpc
on both sides of the Galactic plane. In a slight modification of the
prescription given by \citet{ssf10}, we model the Fermi Bubbles as
ellipsoidal cavities in the thick disk that extend to $l=\pm 20\degr$
and $b=\pm 50\degr$ and touch at the Galactic centre.  The semi-major
and semi-minor axes of each ellipsoid are therefore
$a_{\rm FB}=0.5 R_{\odot}\tan 50\degr$ and $b_{\rm FB}=R_{\odot}\tan 20\degr$
respectively, and their centers are at
$(x_{\rm FB},y_{\rm FB},z_{\rm FB})=(0,0,\pm 0.5 R_{\odot}\tan 50\degr)$,
respectively. The boundary of each bubble is then defined by:
\begin{equation}\label{eq:fb}
  P_{\rm FB} = \left(\frac{x}{b_{\rm FB}}\right)^2 +
  \left(\frac{y}{b_{\rm FB}}\right)^2 + \left(\frac{z-z_{\rm FB}}{a_{\rm FB}}\right)^2 =1.
\end{equation}

If $P_{\rm FB}<1$, then $n_1$ is replaced by $n_1' = J_{\rm FB}\;n_1$, where $J_{\rm FB}$ is a constant scaling factor relating the electron
density inside the bubbles to the unperturbed density of the thick
disk at the same position.

\subsection{Thin disk}\label{sec:thindisk}
As for the earlier TC93 and NE2001 models, the thin disk in our model
represents the region of high gas density and massive star formation
often referred to as the ``molecular ring''. Studies of the
distribution of massive stars, molecular gas and HII regions in the
inner Galaxy \citep[e.g.,][]{ns06,ufm+14,aaj+15} show that the gas
density peaks at a Galactocentric radius of 4 -- 5 kpc.  Analyses of
the $z$-distribution of tracers of high-density gas in the Galactic
disk and spiral arms such as neutral hydrogen (HI)
\citep[e.g.,][]{kdkh07} and molecular gas \citep[e.g.,][]{ns06} give
scale heights in the range 50 -- 70~pc at Galactocentric radii of 4 --
6~kpc, whereas mid-infrared observations of massive stars in the
Galactic disk \citep[e.g.,][]{ufm+14} and open star clusters
\citep{jdpj16} give scale heights less than half as large, 20 --
25~pc.\footnote{The thickness of the Galactic disk is parameterised by
  different authors in different ways. An exponential distribution
  with scale height $h_e$, i.e., $n_e \sim \exp(-|z|/h_e)$, is often
  assumed, whereas studies of the distribution of atomic and molecular
  gas generally give either the full width at half-maximum or the
  half-width at half-maximum (HWHM). In this paper we use the square
  of the hyperbolic secant, $\mathrm{sech}^2(|z|/h_s)$ to represent
  $z$-distributions (as well as some other distributions). The
  different definitions of scale height are significantly different
  with HWHM~$=h_e \ln(2) \sim 0.693\;h_e$ and HWHM~$=h_s \ln(\sqrt
  2+1) \sim 0.881\;h_s$, although these differences are often
  ignored.}  All of these tracers show an increasing scale height with
increasing Galactocentric radius, with a pronounced flaring at radii
$\gapp 10$~kpc. Based on the CO observations of \citet{ns06}, we
parameterise this dependence of scale height (in pc) on $R$ by a
quadratic function:
\begin{equation}\label{eq:zRG}
   H = 32 + 1.6\times10^{-3}R + 4.0\times 10^{-7}R^2
\end{equation}
where $R$ is in pc. The scale height is then $K_i H$, where $K_i$
is a constant factor for the $i$th component.

 We model the thin disk with sech$^2(x)$ functions for both the radial and
$z$ variations as follows:
\begin{equation}
  n_2=n_{2_0}\;g_d\;\mathrm{sech}^2\left(\frac{R-B_2}{A_2}\right)
  \mathrm{sech}^2\left(\frac{z-z_w}{K_2 H}\right)
\end{equation}
where $n_{2_0}$ is mid-plane electron density at the ring central
radius $R=B_2$, $g_d$ is defined in \S\ref{sec:thickdisk}.
We fix $A_2=1200$~pc and $B_2=4000$~pc and fit for
$n_{2_0}$ and $K_2$. Since the Galactic tracers basically determine or
are determined by the distribution of $n_e^2$, we expect $K_2\sim 2$.

\subsection{Spiral arms}\label{sec:spiral}
Outside the molecular ring, the distribution of free electrons in the
Galactic disk is dominated by spiral structure. This structure was
first identified in our Galaxy with observations of HI \citep{okw58}
and was first clearly delineated by \citet{gg76} using kinematic
distances of HII regions supplemented by optical observations of
massive stars. Many other tracers such as CO emission, especially from
giant molecular clouds \citep[e.g.][]{gcbt88} and methanol masers
\citep[e.g.][]{gcm+11} have helped to define the spiral structure.
The importance of spiral structure for models of the Galactic $n_e$
distribution was first indicated \citep{jlm+92,tc93} by the asymmetry
in the distribution of pulsar DMs between the first (northern) and
fourth (southern) Galactic quadrants as illustrated in
Figure~\ref{fg:dmgl}.

In a recent study, \citet{hh14} updated the catalogs of spiral arm
tracers, including more than 2500 HII regions, 1300 giant molecular clouds,
and 900 6.7-GHz methanol masers. They used these data to
investigate the spiral structure of the Galaxy, assuming a logarithmic
spiral form:
\begin{equation}\label{eq:spiral}
\ln\left(\frac{R}{R_{a_i}}\right)=(\phi-\phi_{a_i})\tan\psi_{a_i}
\end{equation}
where $R_{a_i}$, $\phi_{a_i}$ and $\psi_{a_i}$ are the initial radius,
the start azimuth angle and the pitch angle for the $i$th arm.  As the
HII-region data have the most complete and reliable distance estimates
and four-armed spirals are preferred, we have adopted the
corresponding \citet{hh14} fit with solar galactocentric radius
$R_\odot = 8.3$~kpc and circular velocity for the Local Standard of
Rest of 239~km~s$^{-1}$ to define the spiral structure in our model
(apart from a minor modification to the start position of the Perseus
arm).  The adopted arm parameters are given in
Table~\ref{tb:spiral}.
\begin{deluxetable}{lcrrr}
\tabletypesize{\small}
\tablecaption{Adopted spiral-arm parameters\label{tb:spiral}}
\tablehead{
\colhead{Arm name}&\colhead{Index}&\colhead{$R_{a_i}$}&\colhead{$\phi_{a_i}$}&\colhead{$\psi_{a_i}$}\\
& & \colhead{(kpc)} & \colhead{($\degr$)} & \colhead{($\degr$)}}
\startdata
Norma -- Outer & 1 & 3.35 & 44.4 & 11.43 \\
Perseus & 2 & 3.71 & 120.0 & 9.84 \\
Carina -- Sagittarius & 3 & 3.56 & 218.6 & 10.38 \\
Crux -- Scutum & 4 & 3.67 & 330.3 & 10.54 \\
Local & 5 & 8.21 & 55.1 & 2.77 \\
\enddata
\end{deluxetable}
We adopt a sech$^2(s_a)$ cross-section for the electron density in the
arms, where $s_a$ is the perpendicular distance to the arm axis in the
$x-y$ plane.  We also
adopt a sech$^2(x)$ dependence for the radial variation in $n_e$,
joining with the thin disk at $R = B_2$. The radial dependence of
scale height defined by Equation~\ref{eq:zRG} is assumed for all
arms. Although we solve for the spiral-arm scale factor $K_a$, an
indication of its size can be obtained from the $z$-dependence of
H$\alpha$ intensity shown by \citet{hbh+14}. This suggests a FWHM for
the Scutum arm HII-region component at $R \sim$5~kpc of about
170~pc, corresponding to a sech$^2(x)$ scale height for $n_e^2 \sim
95$~pc. Hence, the $n_e$ scale height $H_{a_4} \sim$190~pc, corresponding
to $K_a \sim 3.8$. Although the spiral-arm $n_e$ is dominated by
individual HII regions, except for a few relatively local features, we
assume that these average to a uniform arm density over the
typically long paths to pulsars.

The electron density contributed by the spiral
arms is therefore defined as follows:
\begin{equation}\label{eq:arm_ne}
 n_a=\sum_{i=1}^n\;g_d\;n_{a_i}\;\mathrm{sech}^2
 \left(\frac{s_{a_i}}{w_{a_i}}\right)
 \mathrm{sech}^2\left(\frac{R-B_2}{A_a}\right)\mathrm{sech}^2
 \left(\frac{z-z_w}{K_aH}\right)
 \end{equation}
where, for arm $i$, $n_{a_i}$ is the mid-plane density at $R=B_2$.
The electron densities $n_{a_i}$ refer to
the point were the arms join the thin disk, at $B_2 = 4.0$~kpc. The
arms are assumed to have the same radial termination $g_d$ as the
thick disk (Equation~\ref{eq:disk}). Where the thin disk and an arm
overlap, the larger of the two densities $n_2$ and $n_{a_i}$ is
taken. Because of covariance with the arm electron densities
$n_{a_i}$, the widths of the arms, $w_{a_i}$ were held fixed in the
final fit at values determined from preliminary global fits.

Preliminary model fits showed consistently over-estimated model
distances for pulsars in and beyond the Carina tangential zone ($l
\sim 285\degr - 300\degr$) and, conversely, consistently
under-estimated model distances for pulsars in and beyond the
Sagittarius tangential region ($l \sim 45\degr - 55\degr$).  This
asymmetry is also visible in Figure~\ref{fg:dmgl} which shows that DMs
in the Sagittarius region are lower than those in the Carina
region. To overcome these problems, we have defined an over-dense
region in the Carina arm and an underdense region in Sagittarius, modifying
the density of spiral arm 3, as follows:

\begin{equation}\label{eq:arm3}
  n_{a_3}'=\left\{\begin{array}{ll}
  n_{a_3}\left\{1+n_{\rm CN}\exp\left[-\left(\frac{\phi-\phi_{\rm CN}}
     {\Delta\phi_{\rm CN}}\right)^2\right]\right\}
  \left\{1-n_{\rm SG}\exp\left[-\left(\frac{\phi-\phi_{\rm SG}}
     {\Delta\phi_{\rm SG}}\right)^2\right]\right\}, & \phi < \phi_{\rm
    CN} \vspace{2pt} \\
  n_{a_3}\left\{1+n_{\rm CN}\right\}
  \left\{1-n_{\rm SG}\exp\left[-\left(\frac{\phi-\phi_{\rm SG}}
     {\Delta\phi_{\rm SG}}\right)^2\right]\right\}, & \phi > \phi_{\rm
    CN}
  \end{array}\right.
\end{equation}
where $n_{\rm CN}$ and $n_{\rm SG}$ represent the over-density and
under-density of the Carina arm and Sagittarius region, respectively,
and $\phi_{\rm CN}$, $\Delta\phi_{\rm CN}$ and $\phi_{\rm SG}$,
$\Delta\phi_{\rm SG}$ are the central azimuth and half-width in
azimuth of the low-$\phi$ side of the Carina arm and the Sagittarius
under-density, respectively.

\subsection{Galactic Center}\label{sec:galctr}
The Galactic Center region is unique in the Galaxy because of the
$4\times 10^6$~M$_\odot$ black hole, seen as the point source Sgr
A$^*$, at its center, its high gas density, high star-formation rate,
high magnetic fields and numerous other energetic phenomena occuring
in the region \citep[see][for a review]{geg10}. These phenomena lead
to a high density of ionised gas which will affect signals from any
pulsar in or behind the region. In the ATNF Pulsar Catalogue (V1.54),
there are eight known pulsars that have $|l|<1\degr$ and
$|b|<0.25\degr$. All eight have DMs $\gapp 1000$~cm${-3}$~pc and could
plausibly lie close to the Galactic Center. However, only for the
closest one, PSR J1745$-$2900 which is only 2\farcs4 from Sgr
  A$^*$ \citep{bdd+15}, has a case been made for a physical
  association \citep{rep+13,bdd+15}. As well as having the highest DM
of the eight, it also has a very large rotation measure
\citep{sj13,efk+13}, strengthening the case for a physical
association. With only one pulsar, we cannot constrain the form or
extent of the ionised gas from the pulsar observations. However, we
can use observations of molecular gas \citep[e.g.][]{oon+12}, radio
thermal continuum emission \citep[e.g.][]{lycm08} and radio
recombination lines \citep{lbym09,acd+15} to establish a sufficiently
reliable definition of the region as follows:
\begin{equation}
R_{\rm GC}=[(x-x_{\rm GC})^2+(y-y_{\rm GC})^2]^{1/2}
\end{equation}
\begin{equation}
n_{\rm GC}=n_{\rm GC_0}\;
\exp\left[-\left(\frac{R_{\rm GC}}{A_{\rm GC}}\right)^2\right]
\mathrm{sech}^2\left(\frac{z-z_{\rm GC}}{H_{\rm GC}}\right)
\end{equation}
where $(x_{\rm GC},y_{\rm GC},z_{\rm GC})$ = (+50~pc, 0~pc, $-$7~pc) is the center
of the disk, $R_{\rm GC}$ is the radial distance from the center of the
disk in the $x-y$ plane, $A_{\rm GC}$ is its radial scale length and the
$H_{\rm GC}$ is its scale height. We fix $A_{\rm GC}$ =
160~pc and $H_{\rm GC}$ = 35~pc, based largely on the CO observations of
\citet{oon+12}.

\subsection{Gum Nebula}\label{sec:gn}
The Gum Nebula is the largest known optical emission nebula in the
southern sky and contributes significantly to the DM of pulsars within
and behind it. All pulsar-based $n_e$ models since \citet{mt81} have
included the Gum Nebula as a component. On the basis of an image from
the Southern H$\alpha$ Sky Survey Atlas, \citep{fin03}, \citet{pgs+15}
modelled the Gum Nebula with a spherical shell of outer angular radius
$\sim 23\degr$ as seen from the Sun (corresponding to a physical radius
of about 170~pc) and thickness about 20~pc. Based
on the DMs of pulsars in the region and the H$\alpha$ image
\citep{pgs+15}, we adopt a modified version of the \citet{pgs+15}
model with an ellipsoidal shell, extended in the $z$-direction,
and centered at $l=264\degr$, $b=-4\degr$ at distance 450~pc:

\begin{equation}\label{eq:gn}
  \left(\frac{x-x_{\rm GN}}{A_{\rm GN}}\right)^2
  +\left(\frac{y-y_{\rm GN}}{A_{\rm GN}}\right)^2
  +\left(\frac{z-z_{\rm GN}}{K_{\rm GN}A_{\rm GN}}\right)^2=1
\end{equation}
where ($x_{\rm GN}$, $y_{\rm GN}$, $z_{\rm GN}$) = ($-446$~pc,
$R_\odot+47$~pc, $-31$~pc) is the center of the shell, $K_{\rm GN}$ is the
ratio of the $z$-axis dimension to that in the $x-y$ plane and
$A_{\rm GN}$ is the mid-line radius of the shell in the $x-y$
plane. The shell electron density is assumed to have a gaussian profile with 1/e
half-width $W_{\rm GN}$:
 \begin{equation}\label{eq:gn_ne}
n_{\rm GN}=n_{\rm GN_0}\;\exp\left[-\left(\frac{s_{\rm GN}}{W_{\rm GN}}\right)^2\right]
\end{equation}
where $s_{\rm GN}$ is the perpendicular distance to the
mid-point of the ellipsoidal shell.   Eight pulsars with
model-independent distances and known DMs are affected by the Gum
Nebula. The value of $K_{\rm GN}$ was fixed at 1.4 in the global
parameter fit.

\subsection{Local Bubble region}\label{sec:lb}
Several tracers of the local ISM indicate that the Sun resides in a
relatively low-density cavity often called the Local Bubble. Low
densities within 100~pc or so of the Sun are indicated by low
interstellar reddening of nearby stars \citep{rcds11,lvv+14}, HI 21-cm
spectral-line observations \citep[see][]{skk+15}, NaI and CaII
absorption lines in the spectra of nearby early-type stars
\citep[e.g.,][]{wlvr10} and observations of the ``diffuse interstellar
bands'' \citep{bvf+16}. A background of diffuse soft X-ray emission
\citep{sfp+95} is generally attributed to hot ($10^6$~K) gas in the
Local Bubble \citep{plvs14,skk+15} although this remains somewhat
controversial with X-ray emission from solar-wind charge-exchange
interactions possibly contributing to the background
\citep[see][]{rcds11}. The stellar absorption-line observations of
\citet{wlvr10} also indicate the presence of partially ionised
``cloudlets'' of lower temperature gas within the Local Bubble as well
as clouds of colder and denser gas around its boundary that would be
expected to have ionised outer layers. Even if much of the volume of
the Local Bubble is occupied by high-temperature low-density gas,
these ionised lower-temperature regions will contribute to pulsar
dispersion within the bubble.

Observations of NaI absorption in nearby stars with accurate
distances, either from EUV observations \citep{wssl99} or {\it Hipparcos}
parallaxes \citep{lwv+03}, led to a model for the Local Bubble
consisting of a tilted chimney-like cavity extending out of the
Galactic disk on both sides and bounded by regions of high-density gas
within the disk. We have implemented this model with a cylinder of
radius $R_{\rm LB} = 110$~pc, tilted at $20\degr$ toward the Galactic anticentre
(i.e., in the $+y$ direction) above the plane and centered 40~pc from the Sun in the same
direction, together with two regions of enhanced electron density on
its boundary. The radial distance from the cylinder axis is given by:
\begin{equation}\label{eq:r_lb}
r_{\rm LB}=\{[0.94(y-R_\odot-40)-0.34z]^2+x^2\}^{1/2} 
\end{equation}
and the electron density within the cylinder (i.e., $r_{\rm LB}<R_{\rm
  LB}$ and [$n_1+\rm max(n_2, n_a)]> [n_{\rm LB1}+n_{\rm LB2}$]) is:
\begin{equation}\label{eq:n_lb}
n_{\rm LB} = J_{\rm LB}\;n_1 + \max(n_2, n_a)
\end{equation}
where $n_1$ is the electron density of the thick disk, $n_a$ is the
summed density of spiral arms (Equation~\ref{eq:arm_ne}) and
  $J_{\rm LB}$ is a scale factor similar to $J_{\rm FB}$
  (\S\ref{sec:thickdisk}). The two regions of enhanced density on the
boundary of the Local Bubble, LB1 and LB2, are defined by
\begin{equation}
n_{\rm LB1}=n_{\rm LB1_0}\;\mathrm{sech^2}\left(\frac{l-l_{\rm
    LB1}}{\Delta l_{\rm LB1}}\right)
\mathrm{sech^2}\left(\frac{r_{\rm LB}-R_{\rm LB}}{W_{\rm LB1}}\right) 
\mathrm{sech^2}\left(\frac{z}{H_{\rm LB1}}\right)\end{equation}
\begin{equation}
n_{\rm LB2}=n_{\rm LB2_0}\;\mathrm{sech^2}\left(\frac{l-l_{\rm
    LB2}}{\Delta l_{\rm LB2}}\right)
\mathrm{sech^2}\left(\frac{r_{\rm LB}-R_{\rm LB}}{W_{\rm LB2}}\right)
\mathrm{sech^2}\left(\frac{z}{H_{\rm LB2}}\right).
\end{equation}
Although guided by the stellar absorption-line results, the
central longitudes and angular widths of LB1 and LB2 are fitted for in
the global model (along with the respective densities) and hence are
primarily determined by the DMs of relatively local pulsars.

\subsection{Loop I}\label{sec:loop1}
After fitting for the model components described above, we noticed
that a number of relatively nearby pulsars in the region $0\degr \lapp
l \lapp 30\degr$ and $b\gapp 0\degr$ had over-estimated
distances. This region roughly defines the location of the North Polar
Spur, apparently the brightest feature of the bubble known as Loop I
\citep{bhs71}. Although not universally accepted \citep[see,
  e.g.,][]{hc03,plvs14}, we adopt the view that the North Polar Spur
is associated with Loop I and that Loop I is relatively local,
probably associated with outflow from the Scorpio -- Centaurus OB
association \citep[see, e.g.,][and references
  therein]{wol07}. Following \citet{wol07}, we model Loop I as a
spherical shell centered 200~pc from the Sun in the direction
$l=346\degr$, $b=3\degr$ and the ionisation as a spherical cap centred
in the north-eastern part of the shell. The defining equations for the
electron density at a point $(r_{\rm LI},\theta_{\rm LI})$, where $r_{\rm LI}$ is the radial
distance from the center of the shell and $\theta_{\rm LI}$ is the angle to the
$+x$ axis, are therefore:
\begin{equation}\label{eq:loop1}
n_{\rm LI}=n_{\rm LI_0}\;\exp\left[-\left(\frac{r_{\rm LI}-R_{\rm LI}}{W_{\rm LI}}\right)^2\right]
\exp\left[-\left(\frac{\theta_{\rm LI}}{\Delta\theta_{\rm LI}}\right)^2\right]
\end{equation}
\begin{equation}
r_{\rm LI}=[(x-x_{\rm LI})^2+(y-y_{\rm LI})^2+(z-z_{\rm LI})^2]^{1/2}
\end{equation}
\begin{equation}
\theta=\arccos\left[\frac{(x-x_{\rm LI})\cos\theta_{\rm LI}+(z-z_{\rm LI})
    \sin\theta_{\rm LI}}{r_{\rm LI}}\right]
\end{equation}
where $R_{\rm LI}$ is the shell mid-line radius, $W_{\rm LI}$ is the shell
half-thickness, $\theta_{\rm LI}$ is the angle between the direction of
the center of the spherical cap and the $+x$ direction,
$\Delta\theta_{\rm LI}$ is the extent of the cap in $\theta_{\rm LI}$,
($x_{\rm LI}, y_{\rm LI}, z_{\rm LI}$) =
($-48{\rm\,pc},8106{\rm\,pc},10{\rm\,pc}$) is the center of the shell.

\subsection{Total Galactic electron density}
Given the above equations defining each component of the Galactic
electron density, the $n_e$ at any ($x,y,z$) within the Galaxy is defined to be:
\begin{equation}\label{eq:ne}
n_{\rm Gal} = \{1-w_{\rm LB}\}\;\{[1-w_{\rm GN}]\;[(1-w_{\rm
    LI})\;(n_0+n_{\rm GC})+ w_{\rm LI}\;n_{\rm LI}]+ w_{\rm GN}\;n_{\rm GN}\}+ w_{\rm LB}\;(n_{\rm LB1}+n_{\rm LB2})
\end{equation}
where
\begin{equation}\label{eq:n_0}
n_0 = n_1 + \max(n_2 , n_a),
\end{equation}
except that within the Fermi Bubbles (Equation~\ref{eq:fb})
\begin{equation}
n_0 = J_{\rm FB}n_1 + \max(n_2 , n_a).
\end{equation}
Within the Local Bubble, i.e., $r_{\rm LB}<R_{\rm
  LB}=110$~pc (Equation~\ref{eq:r_lb}),
\begin{equation}
n_0 = J_{\rm LB}n_1 + \max(n_2 , n_a),
\end{equation}
and, if $(n_{\rm LB1}+n_{\rm LB2})>n_0$, $w_{\rm LB}=1$, otherwise $w_{\rm LB}=0$.
Outside the Local Bubble, if $(n_{\rm LB1}+n_{\rm LB2})>n_0$ and
$(n_{\rm LB1}+n_{\rm LB2})>n_{\rm GN}$, $w_{\rm LB}=1$, otherwise
$w_{\rm LB}=0$. For Loop I, if $n_{LI}>n_0$, $w_{\rm LI}=1$, otherwise $w_{\rm
  LI}=0$, and similarly for the Gum Nebula. 

In total, the Galactic model has 96 parameters, of which 64 are fixed
and 32 are fitted.

\subsection{The Magellanic Clouds}\label{sec:mc}
Since the known extra-galactic pulsars are all associated with the
Magellanic Clouds, we include these Clouds in our electron-density
model. Following \citet{vc01} and \citet{hgd12a}, we model the LMC as
a thick inclined disk with inclination angle $i = 32\degr$ and line of
nodes at astronomical position angle $\Theta = 116\degr$. It is
convenient to define coordinate systems ($x_c,y_c,z_c$) with $x_c$,
$y_c$ in the plane of the sky, $x_c$ in the direction of decreasing
right ascension (toward west), $y_c$ toward north and $z_c$ toward the
observer, and ($x',y',z'$), with $x'$ along the line of nodes at an
angle $\theta = \Theta + 90\degr$ to the $x_c$ axis, $y'$ in the plane
of the galaxy on the far side and $z'$ normal to the galaxy plane
toward the observer. Both coordinate systems have their origin at the
center of the LMC, which we define to lie in the direction
($\alpha_{\rm LMC},\delta_{\rm LMC}$) = (05$^{\rm h}$~$24^{\rm m}$,
$-69\degr$ $45\arcmin$) at distance $D_{\rm LMC} =49.7$~kpc
\citep{wal12}. Relations for converting between Galactic ($l,b,D$),
Galactic ($x,y,z$), the celestial ($x_c,y_c,z_c$) system and the
($x',y',z'$) system are given in Appendix C.

We model the electron density of the LMC, $n_{\rm LMC}$, as a thick disk
centered at the origin of the ($x',y',z'$)
frame, together with an additional component, $n_{\rm 30D}$,
representing the giant HII complex 30 Doradus:
\begin{equation}
R_{\rm LMC}=(x^{\prime 2} + y^{\prime 2})^{1/2}
\end{equation}
\begin{equation}
n_{\rm LMC}=n_{\rm LMC_0}\;
\exp\left[-\left(\frac{R_{\rm LMC}}{A_{\rm LMC}}\right)^2\right]
\mathrm{sech}^2\left(\frac{z'}{H_{\rm LMC}}\right)
\end{equation}
where $A_{\rm LMC}$ is the radial scale of the disk, $H_{\rm LMC}$ is its
scale height.

 The giant HII region 30 Doradus is a major feature of the LMC and
 warrants separate treatment. It is almost certainly responsible for
 the anomalously high DM of PSR J0537$-$69 (Table~\ref{tb:a1_mc}). As
 seen in radio continuum \citep{hsk+07} and H$\alpha$ \citep{gmrv01},
 30 Doradus is not far from the nominal center of the LMC at
 ($\alpha_{\rm 30D}$, $\delta_{\rm 30D})$ = (05$^{\rm h}$~40$^{\rm
   m}$, $-69\degr$~$00\arcmin$) and it is believed to be at a similar
 distance. We model its electron density as a spherical Gaussian
 distribution:

\begin{equation}
R_{\rm 30D}=[(x' - x'_{\rm 30D})^2 + (y' -
  y'_{\rm 30D})^2 + z'^2]^{1/2}
\end{equation}
\begin{equation}
n_{\rm 30D}=n_{\rm 30D_0}\;g_{\rm 30D}\;
\exp\left[-\left(\frac{R_{\rm 30D}}{A_{\rm 30D}}\right)^2\right]
\end{equation}
where $A_{\rm 30D}$ is the radial scale length. Note that we
have placed the center of 30 Doradus in the plane of the LMC, i.e.,
$z'_{\rm 30D} = 0$.

We fix $A_{\rm LMC} = 3000$~pc and $H_{\rm LMC}$ = 800~pc, based largely on
the distribution of the Cepheid variables in the LMC, which represent the younger
stellar population \citep{hgd12a}, and fit for $n_{\rm LMC_0}$ based on
the LMC pulsars with known DM. Similarly we fix $A_{\rm 30D} = 450$~pc
based on the radio continuum size and fit for $n_{\rm 30D_0}$.

Compared to the LMC, the SMC has a much greater depth and, while there
is some evidence for a flattened distribution of young stars
\citep[e.g.,][]{hgd12b}, the parameters are relatively uncertain. Because of
this and the relatively small number of associated radio pulsars, we
have chosen a simple model for the SMC, a spherical Gaussian nebula
centered at ($\alpha_{\rm SMC}$, $\delta_{\rm SMC}$) = (00$^{\rm h}$~$51^{\rm m}$,
$-73\degr$ $06\arcmin$) at distance $D_{\rm SMC}$ = 59.7~kpc:

\begin{equation}
R_{\rm SMC}=[(x - x_{\rm SMC})^2 + (y - y_{\rm SMC})^2 + (z - z_{\rm SMC})^2]^{1/2}
\end{equation}
\begin{equation}
n_{\rm SMC}=n_{\rm SMC_0}\;
\exp\left[-\left(\frac{R_{\rm SMC}}{A_{\rm SMC}}\right)^2\right]
\end{equation}
where $A_{\rm SMC}$ is the radial scale length. We fix $A_{\rm SMC} =
3000$~pc based on the distribution of Cepheid variables \citep{hgd12b}
and fit for $n_{\rm SMC_0}$.

In total, the Magellanic Cloud model has 18 parameters of which 15 are
fixed and three are fitted. 

\subsection{Intergalactic medium}\label{sec:IGM}
In order to provide a convenient method for estimating the redshifts
and distances of current, future and simulated FRBs from their DM (and
vice versa), YMW16 includes a model for the free electron density in
the intergalactic medium.  Following \citet{kat16}, in the approximation
of zero curvature, the comoving distance $D$ to an FRB is given by:
\begin{equation}\label{eq:IGD}
  D = \frac{c}{H_0} \ln(1+z)
\end{equation}
where $H_0$ is the Hubble Constant and the redshift $z$ is given by:
\begin{equation}\label{eq:IGz}
  z = \frac{\rm DM_{IGM}\;H_0}{c\;n_{\rm IGM}}
  =\frac{[{\rm DM-(DM_{Gal}+DM_{MC}+DM_{Host}})]\;H_0}{c\;n_{\rm IGM}}
\end{equation}
where DM is the observed FRB dispersion measure, $\rm DM_{Gal}$ is the
total Galactic DM along the path to the FRB, $\rm DM_{MC}$ is any
Magellanic Cloud contribution, $\rm DM_{IGM}$ is the contribution from
the IGM and $\rm DM_{Host}$ is the contribution of the FRB host galaxy
to the observed DM. We adopt a value for $H_0$ of
67.3~km~s$^{-1}$~Mpc$^{-1}$ \citep{aaa+14a} and adopt the
\citet{kat16} value for $n_{\rm IGM}$, $0.16$~m$^{-3}$.

We know little about the host galaxies of FRBs and so it is difficult
to estimate $\rm DM_{Host}$. Furthermore, the observed dispersive
delay is a factor ($1+z$) larger than the delay at the host galaxy,
where $z$ is the host galaxy redshift, and the radio frequency at the
host galaxy is a factor of ($1+z$) larger than the observed radio
frequency. The net effect of these two factors is that, in the host
galaxy frame, DM$_{\rm Host,z} = (1+z){\rm DM}_{\rm
  Host}$. \citet{lbm+07} assumed a $\rm DM_{Host}$ of 200~cm$^{-3}$~pc
in their analysis of the original FRB010724, but DM contributions from
elliptical galaxies or randomly oriented spirals are likely to be less
than that, especially given the time dilation factor. Following
\citet{tsb+13} and \citet{cfb+16} we adopt $\rm DM_{Host} =
100$~cm$^{-3}$~pc as the default value, but allow the optional entry
of a different value.

To avoid pulsars with DMs that are larger than the maximum YMW16 model
prediction in that direction being placed at (obviously incorrect)
cosmological distances, Equations~\ref{eq:IGD} and \ref{eq:IGz} are
invoked only if an ``IGM'' input flag is set in the DM -- $D$
routine. Otherwise, reasonable upper bounds are placed on the model
distance.

\subsection{Interstellar and intergalactic scattering}
As discussed in \S\ref{sec:intro}, we do not include interstellar
scattering results as an input to the YMW16 model. However, it is
useful to give an estimate of scattering timescales ($\tau_{\rm sc}$) as
part of the output of the model.  In the absence of reliable data about
the distribution and strength of scattering regions in the Galaxy, we adopt a
simple approach to this, using the relation obtained by \citet{kmn+15}
for the DM dependence of observed $\tau_{\rm sc}$ values (in units of
seconds and scaled to 1 GHz assuming $\tau_{\rm sc}\propto \nu^{-4.0}$):
\begin{equation}\label{eq:tscat}
  \tau_{\rm sc}=4.1\times 10^{-11}\;{\rm DM}^{2.2}\;(1.0 + 0.00194\;{\rm
    DM}^{2.0}).
\end{equation}
Observed values of $\tau_{\rm sc}$ have an rms
scatter about the fitted line of close to an order of magnitude, but the
variation of $\tau_{\rm sc}$ over the range of observed DMs is about
eight orders of magnitude, so it is a useful predictor.

For the Magellanic Clouds, scattering will occur in both the Galaxy
and the Clouds. (Dispersion and scattering in the intervening IGM is
negligible.) Equation~\ref{eq:tscat} estimates the scatter-broadening
for a pulsar having a given DM. On average, the scattering screen will
be about half-way between the Sun and the pulsar. For a pulsar in the
Magellanic Clouds, the wave incident on the Galaxy is essentially
plane. Therefore, the Galactic component of the scattering delay is
just half of the $\tau_{\rm sc}$ computed from Equation~\ref{eq:tscat}
with ${\rm DM} = {\rm DM_{Gal}}$, again assuming that the scattering
screen is halfway along the Galactic path \citep{ric90}, that is,
$\tau_{\rm Gal}=0.5\tau_{\rm sc}({\rm DM_{Gal}}$). If we assume that
the scattering properties of the Magellanic Clouds are roughly the
same as those of our Galaxy, we can apply these considerations to the
Clouds as well, just reversing the direction of propagation. That is,
$\tau_{\rm MC}=0.5\tau_{\rm sc}({\rm DM_{MC}}$). Since the two
scattering distributions are statistically independent, the net effect
is given by the convolution of the two scattering functions. This
convolved distribution is given by
\begin{equation}
  f(t) = \frac{1}{\tau_{\rm Gal}} \exp(-t/\tau_{\rm Gal}) *
  \frac{1}{\tau_{\rm MC}} \exp(-t/\tau_{\rm MC})
\end{equation}
where $\tau_{\rm Gal} = 0.5\tau_{\rm sc}$ from Equation~\ref{eq:tscat}
with DM = DM$_{\rm Gal}$, and similarly for $\tau_{\rm MC}$ with DM$_{\rm MC}$. The
convolution integral for $t\ge0$ is
\begin{equation}
  f(t) = \frac{1}{\tau_{\rm Gal}\tau_{\rm MC}} \int^t_0 dt^\prime
  \exp[-(t-t^\prime)/\tau_{\rm Gal}]\; \exp[-t^\prime/\tau_{\rm MC}]
\end{equation}
and hence
\begin{equation}\label{eq:scat_fn}
   f(t) = \frac{1}{(\tau_{\rm MC}-\tau_{\rm Gal})}[\exp(-t/\tau_{\rm MC})-\exp(-t/\tau_{\rm Gal})].
\end{equation}
In the limit of $\tau_{\rm Gal} = \tau_{\rm MC} = \tau$, this reduces
to
\begin{equation}
  f(t) = \frac{t}{\tau^2}\;\exp(-t/\tau)
\end{equation}
\citep[cf.,][]{wil72}.  Equation~\ref{eq:scat_fn} shows that, for
short times, the rise of $f(t)$ is determined by the inverted decay of
the term with the shorter scattering time whereas, at longer times,
the decay of $f(t)$ is dominated by the longer scattering time (or by
the scattering time of each screen if they are equal). This
longer timescale will determine measured values of FRB scattering
times. We therefore model the scattering of Magellanic Cloud pulsars
using
\begin{equation}\label{eq:MCscat}
  \tau_{\rm sc}({\rm MC}) = \max(\tau_{\rm Gal},\tau_{\rm MC}).
\end{equation}

Scattering timescales have been measured for ten of the 17 known FRBs,
with upper limits for the remaining seven (Table~\ref{tb:a1_frb}). The
origin of this broadening is a matter of considerable debate. With the
possible exceptions of the three known FRBs at low Galactic latitude
(Table~\ref{tb:a1_frb}), scattering in our Galaxy (or the SMC for
FRB010724) is too small to account for the observed $\tau_{\rm sc}$.
This leaves the IGM and the host galaxy as possible sources of the
observed scattering.

For the host galaxy, scattering could arise in the ISM along the path
\citep[see, e.g.,][]{xz16,cws+16} or in the immediate environs of the
FRB source \citep[see, e.g.,][]{mls+15}. Because of the extremely
small lever-arm factor $a(1-a)$, where the screen is at $aD$ and $D$
is the distance to the source \citep[cf.,][]{wil72}, origin of the
observed scattering in the FRB environs would require extremely large
scattering angles. While this is not impossible, it requires most (but
not all) FRB sources to lie in very dense and turbulent nebulae. This
is exacerbated by time dilation of the observed scatter
broadening. Similar to the effect on DM as discussed above, the
observed scatter broadening is a factor $(1+z)$ larger than the
broadening in the host frame and the radio frequency at host galaxy is
a factor of ($1+z$) larger than the observed radio frequency. Since
scatter broadening scales approximately as $\nu^{-4.0}$ the observed
scattering time is reduced by a factor $\sim (1+z)^3$
\citep[cf.,][]{mk13a}. In view of these factors, we believe that FRB
scattering is unlikely to arise in the immediate environs of the FRB
source.

Scattering in the ISM of the host galaxy can be modelled in a similar
way to scattering in our Galaxy, that is, based on the $\tau_{\rm sc}$
-- DM relation (Equation~\ref{eq:tscat}). This scatter broadening is
also reduced by the $\sim (1+z)^3$ factor, but there is a significant
additional factor. As discussed above in \S\ref{sec:IGM}, the DM in
the host galaxy frame is a factor ($1+z$) larger than DM$_{\rm Host}$,
the observed DM contribution attributed to the host galaxy. Following
the same logic as for scattering in the Magellanic Clouds and using
Equation~\ref{eq:tscat} in the frame of the host galaxy, we adopt
$\tau_{{\rm Host},z} = 0.5 \tau_{\rm sc}[(1+z){\rm DM_{Host}}]$. Since
$\tau_{\rm sc}$ is roughly proportional to DM$^{2.2}$
(Equation~\ref{eq:tscat}), the net effect of time dilation is that the
observed scatter broadening due to the host galaxy $\tau_{\rm Host}
\sim \tau_{{\rm Host},z}/(1+z)$. Consequently, even for large
star-forming galaxies similar to our own Galaxy, the host-galaxy ISM
is unlikely to contribute significantly to the observed FRB
scattering.

The remaining location for FRB scattering is the IGM between us and
the host galaxy. \citet{lg14} have argued that the level of turbulence
in the IGM is too small to account for the observed scattering
delays. On the other hand, \citet{mk13a} argue that relatively dense
intracluster gas or individual galaxies along the path to the FRB
could account for the observed scattering. For known FRBs with
estimated $z\lapp 2$, the expected number of such intervening screens
is low, $\lapp 2$. This small number of effective screens provides a
natural explanation for the large variation in observed scattering
times. FRBs with only upper limits on scatter broadening simply have
no strong scattering screen along the path.

\begin{figure}[ht]
\includegraphics[angle=270,width=85mm]{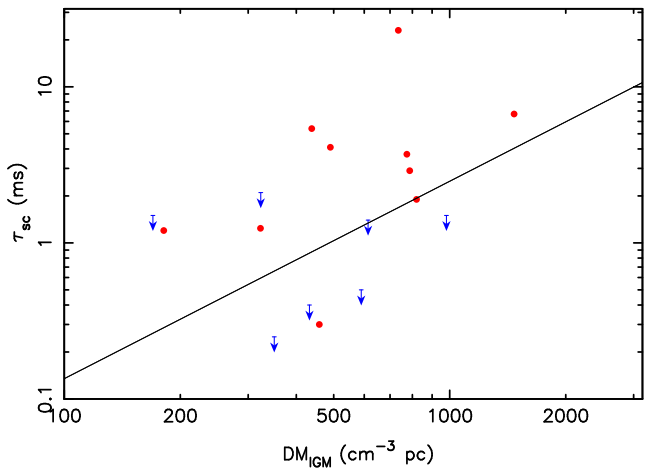}
\caption{Observed FRB scattering times, normalised to 1~GHz assuming
  $\tau_{\rm sc} \sim \nu^{-4.0}$, versus $\rm DM_{IGM}=\rm DM - \rm
  DM_{Gal}-\rm DM_{MC}-\rm DM_{Host}$, where we have assumed $\rm
  DM_{Host} = 100$~cm$^{-3}$~pc. The fitted line is derived using a
  ``survival analysis'' technique to properly account for the upper
  limits on $\tau_{\rm sc}$.\label{fg:frb_sc}}
\end{figure}

Observationally, as Figure~\ref{fg:frb_sc} shows, there is a clear
tendency for scattering times to increase with ${\rm DM_{IGM}}$,
supporting the idea that the IGM is a significant, if not the major
contributor to the observed DM. In view of the relatively large number
of upper limits, we have used a ``survival analysis'' technique,
implemented in the routine {\sc
  asurv} \citep{lif92}, to fit a straight line to this
  trend, giving the result
\begin{equation}\label{eq:tsc_igm}
  \log(\tau_{\rm IGM}) = (1.27\pm 0.72)  \log({\rm DM_{IGM}}) - (3.4\pm 2.0),
\end{equation}
with a 90\% probability that the correlation is real.
Compared to the \citet{kmn+15} relation for Galactic scattering which
is basically proportional to DM$^{2.2}$, FRB scattering evidently has
a somewhat flatter DM-dependence, $\sim {\rm DM}^{1.3}$, although the
index has large uncertainty. 
The observed $\tau_{\rm sc}$ values implicitly include the lever-arm
effect, and given the small number and large scatter of the observed
$\tau_{\rm sc}$ values, we do not attempt to take the redshift
dependencies \citep{mk13a} into account at this stage.

Therefore, following the same
reasoning as that used for the Magellanic Clouds, we take
\begin{equation}
 \tau_{\rm FRB} = \max(\tau_{\rm Gal},\tau_{\rm MC},\tau_{\rm
   IGM},\tau_{\rm Host}).
\end{equation}
as a predictor of FRB scattering times. Finally, we note that the
ability to vary the assumed $DM_{\rm Host}$, goes some way toward
accommodating possible different assumptions about the origin of FRB
scattering.

\section{Data fitting}\label{sec:model-fit}
We wish to fit the electron density model described in the previous
section to the data set comprising $N_D = 189$ Galactic pulsars with
independently determined distances and the 27 Magellanic Cloud pulsars
with known DMs as described in \S\ref{sec:inddist} and listed in
Tables~\ref{tb:a1_px} -- \ref{tb:a1_mc}.  Seeking a global solution to
this problem requires an algorithm that is optimised for non-linear,
derivative-free and multi-dimensional parameter optimisation. We have
investigated three such algorithms. The first is the modified
evolutionary algorithm {\sc esch} from the {\sc NLopt} library of
non-linear optimisation routines
\citep{sgh10}\footnote{\url{http://ab-initio.mit.edu/nlopt}} which
analyses the generational evolution of candidate solutions. The
second, {\sc PSwarm}
\citep{vv07}\footnote{\url{http://www.norg.uminho.pt/aivaz/pswarm}},
is based on a ``particle swarm'' algorithm and the third is {\sc
  PolyChord}
\citep{hhl15}\footnote{\url{https://ccpforge.cse.rl.ac.uk/gf/project/polychord}},
a Bayesian routine using Markov-Chain Monte Carlo nested sampling
methods. Both {\sc PSwarm} and {\sc PolyChord} are enabled for use
with the Message Passing Interface {\sc MPI}\footnote{For a
  description of {\sc MPI}, see, e.g.,
  \url{https://software.intel.com/en-us/articles/mpi-parallelizes-work-among-multiple-processors-or-hostsand}}
enabling faster execution on multi-processor computing systems. In
each case, we have followed the global optimisation with a local
parameter optimisation using the Nelder-Mead simplex algorithm {\sc
  mead} routine from the {\sc NLopt} library. 

\begin{figure}[ht]
\includegraphics[angle=270,width=85mm]{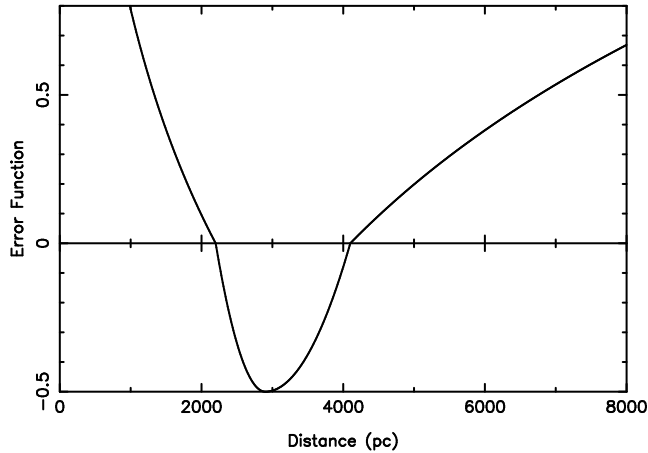}
\caption{Error function $F$ for PSR J1048$-$5832. For this pulsar the
  DM-independent distance estimate is $D_i=2900$~pc, with lower and
  upper limits of $D_l=2200$~pc and $D_u=4100$~pc, respectively
  \citep{vwc+12}.\label{fg:F}}
\end{figure}

As a goodness-of-fit statistic we seek to minimise the following function:
\begin{equation}
F=\sum_{n=1}^{N_D}{F_n}
\end{equation}
where $F$ is summed over all (Galactic) pulsars with DM-independent distances and
\begin{equation}
F_n=\left\{\begin{array}{ll}
\ln\left(\frac{D_l}{D_m}\right), & D_m < D_l \vspace{2pt}\\
0.5\left[\left(\frac{D_i-D_m}{D_i-D_l}\right)^2-1\right], & D_l \le
   D_m< D_i \vspace{2pt} \\
0.5\left[\left(\frac{D_m-D_i}{D_u-D_i}\right)^2-1\right], & D_i \le
   D_m\le D_u  \vspace{2pt} \\
\ln\left(\frac{D_m}{D_u}\right), & D_m > D_u \\
\end{array}\right.
\end{equation}
with $D_m$ is the model distance for a given pulsar, $D_i$ is the best
estimate of its DM-independent distance and $D_l$, $D_u$ are the lower
and upper limits to the independent distance,
respectively. Figure~\ref{fg:F} shows the goodness-of-fit function for
PSR J1048$-$5832, chosen to illustrate the function for a pulsar with
asymmetric and widely spaced limits.

Although all three optimisation routines we tested gave consistent
global solutions, we found that {\sc PSwarm} + {\sc mead} gave the
best results in terms of the final value of the goodness-of-fit
statistic $F$ and speed of execution, and so the results presented
here are based on these routines.

As described in \S\ref{sec:model}, we fix many parameters of the model
based on independent information, e.g., the form and location of the
spiral arms, the size and shape of the Galactic Center disk, and the
central position, distance and form of features such as the Gum Nebula
and the Magellanic Clouds. There are no free parameters in the model
for the intergalactic medium (see \S\ref{sec:IGM}). Even after fixing
these parameters, the number of model parameters to be fitted for is a
significant fraction of the number of data values, which is a recipe
for large covariances between parameters and poor convergence of the
fitting process.  For some parameters that are poorly constrained
because of the small number of pulsars whose line-of-sight path passes
through the particular model component, we fix additional parameters
based on the results of preliminary fits to the data set.

The Fermi Bubbles are a special case. As described in
\S\ref{sec:thickdisk}, we modeled these as large cavities in the thick
disk above and below the Galactic Center with electron densities given
by $J_{\rm FB} n_1$, where $n_1$ is the local thick-disk electron
density. Only two pulsars (PSRs J1748$-$2021A and J1750$-$3703A,
associated with globular clusters NGC~6440 and NGC~6441 at distances
$8.2\pm 0.6$~kpc and $13.5\pm 1.0$~kpc, respectively) are within or
behind the Fermi Bubbles. Preliminary fits for $J_{\rm FB}$ had relatively large
uncertanties, but were consistent with a value of 1.0. In view of the
large uncertainties, we decided to fix $J_{\rm FB}$ at 1.0. This
obviously means that the Fermi Bubbles currently have no
effect. However, we did not remove them from the model, as future
pulsar discoveries and distance determinations will undoubtedly be
made, allowing a more realistic determination of $J_{\rm FB}$ in a
future version of the model.

 The three adjustable parameters of the model for the Magellanic
  Clouds are separately fitted using just the pulsars known or
  believed to lie in the Magellanic Clouds (Table~\ref{tb:a1_mc})
  while holding the Galactic model fixed. As is discussed further in
  \S\ref{sec:mc_results}, J0131$-$7301 was omitted from the fit
  because of its relatively large DM compared to the DMs of the other
  SMC pulsars.

Overall, the YMW16 model has 117 parameters, of which 35 are fitted by
the parameter optimisation and 82 are fixed. Of the fitted
parameters, 32 are for Galactic components of the model and three are
for the Magellanic Clouds. Of the fixed parameters, 64 define the
location, size and shape of Galactic components, including 15 defining
the location of the spiral arms (Table~\ref{tb:spiral}). In addition
to the three fitted parameters, the Magellanic Cloud components are
defined by 15 fixed parameters. The model for the IGM has three fixed
parameters, one of which, DM$_{\rm Host}$, the contribution of the FRB
host galaxy to the observed DM, is adjustable but not fitted for.

As a means of assessing the performance of the model (and earlier
models) we define a ``distance error'', $D_{err}$ for each pulsar
having an independent distance estimate as follows:
\begin{equation}\label{eq:derr}
D_{err}=\left\{\begin{array}{lll}
0, & D_{l} \le D_{m} \le D_{u} \\
\frac{D_{l}}{D_{m}}-1, & D_{m}<D_{l} \vspace{2pt}\\
\frac{D_{m}}{D_{u}}-1, & D_{m}>D_{u}
\end{array}\right.
\end{equation}
In general, if $D_{err}$ is less than 20 per cent, we consider the
model distance satisfactory.

As a check of the convergence of the model fit, we repeated the
fitting 100 times randomly choosing initial values for the parameters
within their allowed range. There was no significant evidence for
secondary error-function minima in any of the parameters, although a
few of the more poorly determined ones had a relatively wide and
non-Gaussian distribution. Except for these cases, the peak of the
distribution was used as the starting value for each parameter in the
final global fit. For the non-gaussian cases a central starting value
was chosen.

Parameter uncertainties are not easy to estimate because of the
relatively small number of data points compared to the number of
fitted parameters. The ``bootstrap with replacement''
method\footnote{See, e.g., \citet{efr81} for a description of bootstrap
  methods.} gave a good indication of uncertainties for some
parameters, but not in cases where the parameter was determined by
only a few data points, e.g., the central density of Galactic Center
disk, which is determined by just one pulsar, PSR J1745-2900.  As an
alternative, we chose to estimate parameter uncertainties by randomly
varying the values of $D_i$ within the range $D_l$ to $D_u$ for all
pulsars in the data set and then refitting for the model
parameters. This was repeated 200 times to give a distribution of
values for each model parameter. These distributions were generally
approximately Gaussian and hence gave reasonable values for the
1-$\sigma$ parameter uncertainties. The principal exceptions were
$n_{a_1}$, the electron density of the Norma-Outer arm, and $n_{\rm
  GC_0}$, the central density of Galactic Center disk. As discussed
above, $n_{\rm GC_0}$ was determined by just one pulsar and so this
procedure would not be expected to lead to reliable results. For
$n_{a_1}$, wide scatter may result from the fact that, within the zone
of the thin disk, the larger of $n_{a_i}$ and $n_2$, the density of
the thin disk, was taken as the model density (see
\S\ref{sec:spiral}). Despite these issues, we took the rms scatter of
the 200 trial values to represent the 1-$\sigma$ uncertainties for all
fitted parameters.

\section{Results}\label{sec:results}
\subsection{The Galaxy}
Table~\ref{tb:params} gives the allowed ranges, final-fit values and
uncertainties for the 35 fitted parameters of the model and the
assumed values for the 19 of the fixed parameters that are based on
preliminary fits to the data set or independent evidence about the
structure of model components. Other fixed parameters are described in
\S\ref{sec:model}.

\begin{deluxetable}{llcccc}
  \tabletypesize{\footnotesize}
  \tablecaption{Parameters for the YMW16 electron-density model\label{tb:params}}
  \tablehead{
  \colhead{Component} & \colhead{Parameter} & \colhead{Lower
    limit} & \colhead{Upper limit} & \colhead{Value} & \colhead{Uncertainty} }
\startdata 
Thick disk & $A_d$~(pc) & 2500 & 2500 & 2500 & -- \\ 
           & $B_d$~(pc) & 15000 & 15000 & 15000 & -- \\        
           & $n_{1_0}$~(cm$^{-3}$) & 0.008 & 0.016 & 0.01132 & 0.00043  \\ 
           & $H_1$~(pc) & 1200 & 2000 & 1673 & 53\\
Thin disk  & $A_2$~(pc) & 1200 & 1200 & 1200 & -- \\
           & $B_2$~(pc) & 4000 & 4000 & 4000 & -- \\
           & $n_{2_0}$~(cm$^{-3}$) & 0.3 & 0.5 & 0.404 & 0.051 \\ 
           & $K_2$ & 1.0 & 3.0 & 1.54 & 0.16 \\
Spiral arms & $n_{a_1}$~(cm$^{-3}$) & 0.03 & 0.15 & 0.135 &0.024 \\ 
          & $n_{a_2}$~(cm$^{-3}$) & 0.03 & 0.15 & 0.129 & 0.011\\
          & $n_{a_3}$~(cm$^{-3}$) & 0.03 & 0.15 & 0.103 & 0.006\\
          & $n_{a_4}$~(cm$^{-3}$) & 0.03 & 0.15 & 0.116 & 0.008\\
          & $n_{a_5}$~(cm$^{-3}$) & 0.003 & 0.02 & 0.0057 & 0.0013 \\
          & $w_{a_1}$~(pc)  & 300 & 300 & 300 &  -- \\ 
          & $w_{a_2}$~(pc)  & 500 & 500 & 500 &  -- \\ 
          & $w_{a_3}$~(pc)  & 300 & 300 & 300 &  -- \\ 
          & $w_{a_4}$~(pc)  & 500 & 500 & 500 &  -- \\ 
          & $w_{a_5}$~(pc)  & 300 & 300 & 300 & -- \\ 
          & $A_{a}$~(pc) & 7000 & 15000 & 11680 & 1490 \\
          & $K_a$ & 3.0 & 6.0 & 5.01 &  0.15\\
          & $n_{\rm CN}$ & 1.0 & 3.0 & 2.40 & 0.26 \\           
          & $\phi_{\rm CN}$~($\degr$) & 109 & 109 & 109 &  -- \\
          & $\Delta\phi_{\rm CN}$ & 2.0 & 15.0 & 8.2 & 1.1\\ 
          & $n_{\rm SG}$ & 0.1 & 1.0 & 0.626 & 0.068\\ 
          & $\phi_{\rm SG}$~($\degr$) & 60 & 100 & 75.8 & 2.2 \\
          & $\Delta\phi_{\rm SG}$~($\degr$) & 20 & 20 & 20 & -- \\ 
  Galactic warp & $\gamma_w$ & 0 & 0.3 & 0.140 & 0.066\\
   Galactic Ctr & $n_{\rm GC_0}$~(cm$^{-3}$) & 1.0 & 10.0 & 6.2 & 2.6\\
          &  $A_{\rm GC}$~(pc) & 160 & 160 & 160 & -- \\ 
          & $H_{\rm GC}$~(pc) & 35 & 35 & 35 &  -- \\ 
   Gum Nebula  & $n_{\rm GN_0}$~(cm$^{-3}$) & 1.0 & 3.0 & 1.84 & 0.12 \\
        & $W_{\rm GN}$~(pc) & 10 & 20 & 15.1 & 0.8 \\  
        & $A_{\rm GN}$~(pc) & 120 & 130 & 125.8 & 0.8 \\ 
        & $K_{\rm GN}$ & 1.4 & 1.4 & 1.4 & -- \\ 
  Local Bubble  & $J_{\rm LB}$ & 0.4 & 1.2 & 0.480 & 0.063 \\
                & $n_{\rm LB1_0}$~(cm$^{-3}$) & 0.5 & 1.5 & 1.094 & 0.073 \\ 
                & $\theta_{\rm LB1}$~($\degr$) & 190 & 200 & 195.4& 1.2 \\ 
                & $\Delta\theta_{\rm LB1}$~($\degr$) & 20 & 40 & 28.4 & 1.1 \\ 
                & $W_{\rm LB1}$~(pc) & 10 & 20 & 14.2 & 0.9 \\ 
                & $H_{\rm LB1}$~(pc) & 80 & 130 & 112.9 & 3.9 \\ 
                & $n_{\rm LB2_0}$~(cm$^{-3}$) & 1.0 & 3.0 & 2.33 & 0.15 \\
                 & $\theta_{\rm LB2}$~($\degr$) & 260 & 300 & 278.2 & 1.1\\ 
                 & $\Delta\theta_{\rm LB2}$~($\degr$) & 10 & 60 & 14.7& 0.7\\ 
                & $W_{\rm LB2}$~(pc) & 10 & 20 & 15.6 &  1.1 \\ 
                 & $H_{\rm LB2}$~(pc) & 10 & 60 & 43.6 & 2.6 \\ 
Loop I  & $n_{\rm LI_0}$~(cm$^{-3}$) & 0.0 & 3.0 & 1.907 & 0.096\\
        & $R_{\rm LI}$~($pc$) & 80 & 80 & 80 & -- \\ 
        & $W_{\rm LI}$~($pc$) & 15 & 15 & 15 & -- \\ 
        & $\Delta\theta_{\rm LI}$~($\degr$) & 30 & 30 & 30 & -- \\ 
        & $\theta_{\rm LI}$~($\degr$) & 40 & 40 & 40 & -- \\ 
Fermi Bubbles & $J_{\rm FB}$ & 1.0 & 1.0 & 1.0 & -- \\
LMC & $n_{\rm LMC_0}$~(cm$^{-3}$) & 0.05 & 0.3 & 0.066 &0.007 \\
    & $n_{\rm 30D_0}$~(cm$^{-3}$) & 0.05 & 0.5 & 0.32 & 0.17 \\
SMC & $n_{\rm SMC_0}$~(cm$^{-3}$) & 0 & 0.3 & 0.045 & 0.017\\
\enddata
\end{deluxetable}

\begin{figure}[ht]
\includegraphics[angle=270,width=170mm]{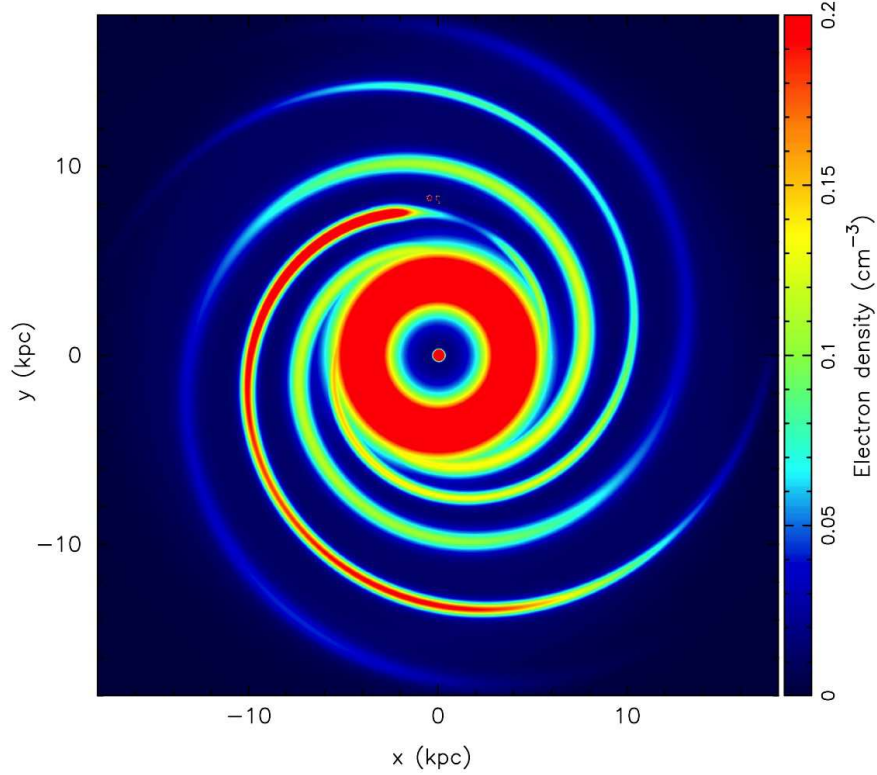}
\caption{Electron density in the Galactic plane ($z=0$) for the YMW16
  model. The Galactic Center is at the origin and the Sun is at $x=0$,
  $y=+8.5$~kpc. The dense annulus of central radius 4~kpc is the
  ``thin disk'' which represents the Galactic molecular ring. Spiral
  arms have a pre-determined logarithmic spiral form and generally
  decay exponentially at large Galactocentric radii. Exceptions are in
  the Carina -- Sagittarius arm where there are over-dense and
  under-dense regions in Carina and Sagittarius respectively. The Gum
  Nebula and Local Bubble features are faintly visible near the
  position of the Sun.\label{fg:ne_model}}
\end{figure}

The model electron density in the Galactic plane ($z=0$) is
illustrated in Figure~\ref{fg:ne_model}. This figure emphasizes the
high degree of symmetry of our model. As discussed in
\S\ref{sec:intro}, we have tried to minimise the number of special
features, only defining one when a localised group of pulsars showed
consistently over-estimated or under-estimated model
distances. Figure~\ref{fg:ne_model} also illustrates the way in which
the spiral arms emerge from the thin disk annulus. 

\begin{figure}[ht]
\includegraphics[angle=270,width=85mm]{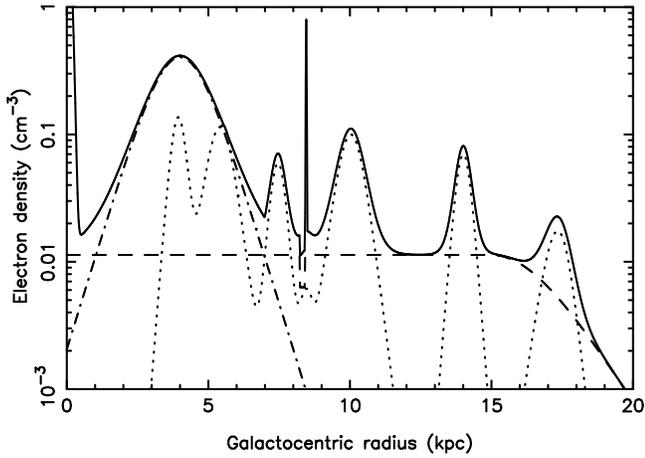}
\caption{Electron density versus Galactocentric radius along the $+y$
  axis (toward the Sun) for the YMW16 model. The major components are
  shown separately: thick disk (dashed), thin disk (dot-dashed) and
  spiral arms (dotted). The Galactic Center component peaks at
  5.8~cm$^{-3}$~pc and the Local Bubble depression and associated
  over-density LB1 can be seen at 8.5~kpc. \label{fg:ne_r}}
\end{figure}

The radial dependence of the electron density on the Galactic plane
from the Galactic Center outwards in the direction of the Sun is shown
in Figure~\ref{fg:ne_r}. In the inner Galaxy the Galactic Center disk
and the thin disk dominate, whereas in the outer Galaxy, the spiral
arms dominate. At higher Galactic latitudes the thick disk is
dominant.  We have also plotted the total model Galactic DM at $b=0$
as a function of Galactic longitude $l$ in Figure~\ref{fg:dmgl}. This
clearly shows the asymmetry in the spiral structure with much greater
integrated electron densities in the fourth Galactic quadrant,
especially for the Carina arm. Since most known pulsars are relatively
local and at non-zero Galactic latitude, most points in
Figure~\ref{fg:dmgl} are well below the maximum DM line. Exceptions
are PSR J0248+6021 (at $l=137\degr$), believed to lie close to the HII
region W5 (\S\ref{sec:neb}), and four pulsars near $l=305\degr$ in the
Carina region. Our model already has an increased electron density in
the Carina arm (\S\ref{sec:spiral}) but these four pulsars clearly
have a greater-than-average contribution from HII regions in the path.

For the thick disk, the YMW16 model gives $n_{1_0} = 0.01132\pm
0.00043$~cm$^{-3}$ and $H_1 = 1673\pm 53$~pc (Table~\ref{tb:params}),
corresponding to a $n_{1_0} H_1$ product of $18.9\pm
0.9$~cm$^{-3}$~pc. This is just over half of the perpendicular DM for
the NE2001 model, but comparable to those for the TC93, \citet{gmcm08}
and \citet{sw09} models. Correspondingly, the YMW16 mid-plane density
is about one third the NE2001 value and the scale height is about 70\%
larger than the NE2001 value. However, the YMW16 scale-height estimate is
consistent within the uncertainties with the estimates of
\citet{gmcm08}, \citet{sw09} and \citet{sch12}. 

\begin{figure}[ht]
\includegraphics[angle=270,width=85mm]{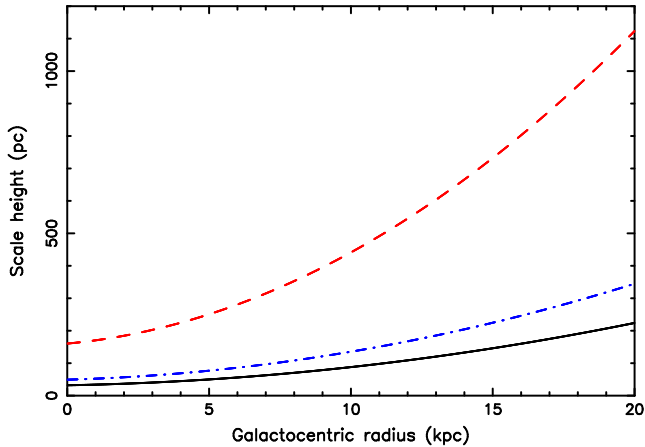}
\caption{Dependence of thin-disk and spiral-arm scale height on
  Galactocentric radius. The full line corresponds to the function
  $H$ (Equation~\ref{eq:zRG}), the dot-dashed line to the thin disk
  with $K_2 \approx 1.5$ and the dashed line is for the spiral arms
  with $K_a \approx 5.0$.\label{fg:shsp}}
\end{figure}

As Figures~\ref{fg:ne_model} and \ref{fg:ne_r} show, the thin disk is
relatively dense with a peak $n_e \sim 0.4$~cm$^{-3}$ at $R =
4$~kpc. Although the thin disk doesn't extend very far in
Galactocentric radius, based on the independent measurements discussed
in \S\ref{sec:thindisk}, we use Equation~\ref{eq:zRG} to define the
scale height dependence on $R$ and fit for the scale factor
$K_2$. This scale height dependence is illustrated in
Figure~\ref{fg:shsp}.

Except for the Local arm, the reference spiral arm electron densities
$n_{a_i}$ at $R = 4$~kpc are all about 0.1~cm$^{-3}$
(Table~\ref{tb:params}). The Local arm is much less dense, with
$n_{a_5}$ just 0.0057~cm$^{-3}$, only about half of the density
of the thick disk. The radial scale length of the spiral arms $A_a$,
common to all arms, is relatively large, $11.7\pm 1.5$~kpc, and the
scale-height factor $K_a$ is $5.01\pm 0.15$ (Figure~\ref{fg:ne_r}).
The over-dense Carina region and under-dense Sagittarius region
reflect the asymmetry seen in the latitude distribution of DMs
(Figure~\ref{fg:dmgl}). Parameters of these regions, defined by
Equation~\ref{eq:arm3}, are given in Table~\ref{tb:params}. Angular
widths were estimated from preliminary fits and were held fixed for
the global fit.

The Galactic warp only affects only a few pulsars with our present data
set. We never-the-less have fitted for its amplitude and obtain a value for
$\gamma_w = 0.140\pm 0.066$. This is a little less than, but consistent
with, the value of 0.18 adopted by \citet{rrdp03}. The largest change
resulting from introduction of the warp is for PSR~J2229+6114, where
the model distance is reduced by $\sim870$~pc to $\sim 5040$~pc,
moving it closer to the independent distance of 3000~pc
(Table~\ref{tb:a1_neb}). Eleven pulsars with independent distances
have their model distance changed by 100~pc or more.  

The Galactic Center disk and the Gum Nebula ellipsoid are relatively
straight-forward. As mentioned in \S\ref{sec:model-fit}, the Galactic
Center electron density is currently determined by just one pulsar,
but eight pulsars (including the Vela pulsar and the Double Pulsar)
lie within or beyond the Gum Nebula and have a DM contribution from
it. The derived radius of the ellipsoidal shell in the $x-y$ plane is
about 125~pc, the shell half-thickness is $W_{\rm GN}\sim 15$~pc and
the electron density at the shell mid-line is $n_{\rm GN_0}\sim
1.8$~cm$^{-3}$. Consequently, the DM contribution of a perpendicular
traverse of one side of the shell is $\sqrt{\pi}\;n_{\rm GN_0} W_{\rm
  GN}\sim 50$~cm$^{-3}$~pc.

As discussed in \S\ref{sec:lb}, there is good evidence for a
relatively low-density region surrounding the Sun. The size and shape
of this region in our model is determined by the stellar absorption
results and we just fit for the internal electron density as a
multiple of the thick-disk density as part of the global fit. The
derived scale factor $J_{\rm LB} \sim 0.48$ indicating that, on
average, $n_e$ within the Local Bubble is about half the density of
the surrounding thick disk plus local spiral arm contributions. 
The stellar absorption results also indicate the presence of
relatively dense swept-up regions surrounding the Local Bubble. We
modelled these as described in \S\ref{sec:lb}, obtaining the
parameters listed in Table~\ref{tb:params} as part of the global
fit. The two modelled over-dense regions cover about half of the
circumference of the local bubble with central electron densities of
1.0~cm$^{-3}$ and 2.3~cm$^{-3}$ respectively.

Most of the parameters of Loop I (and the roughly hemispherical
ionised shell which appears to be associated with the North Polar
Spur) were held fixed as discussed in \S\ref{sec:loop1}. Only the
reference electron density, defined by Equation~\ref{eq:loop1}, was
solved for as part of the global fit; the derived value is quite high, about
1.9~cm$^{-3}$, suggesting the presence of a relatively dense ionised
shell associated with the North Polar Spur.

\begin{figure}[ht]
\includegraphics[angle=270,width=170mm]{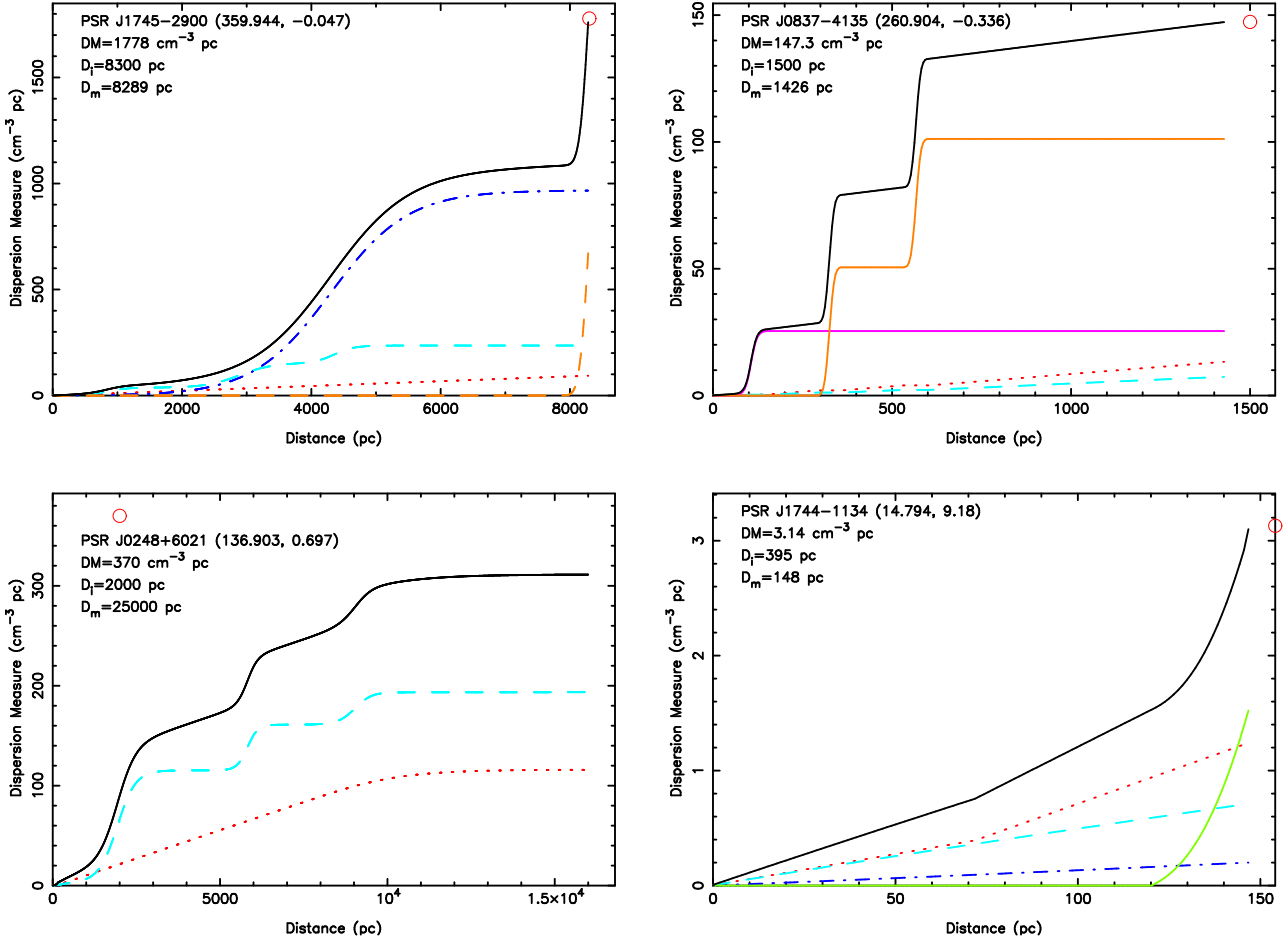}
\caption{Plots showing the contributions from different components of
  the model to the pulsar DM as a function of distance along the path
  for four selected pulsars. For each subplot, the pulsar name, the
  Galactic coordinates, the DM, the independent distance estimate and
  the model distance estimate are given. The upper
  two plots are for pulsars with $D_{err}\le20$\%, the lower-left plot
  is for a pulsar with an over-estimated distance and the lower-right
  plot is for a pulsar with an under-estimated distance. Model
  components are as follows: thick disk (red dotted line), thin disk
  (blue dot-dashed line), spiral arms (cyan dashed line), Local Bubble
  shell (magenta line), Gum Nebula (orange line), Loop I (green line),
  Galactic Center disk (orange dashed line) and the total (full black
  line).  The red circle on each subplot marks the independent
  distance $D_i$ and the observed DM. For PSR J1744$-$1134, the actual
  $D_i$ is beyond the right edge of the subplot. \label{fg:4dm-d}}
\end{figure}

Four representative plots showing the contribution of most model
components to the DM as a function of distance along the path to the
pulsar are shown in Figure~\ref{fg:4dm-d}. The top-left plot is for
PSR J1745$-$2900 which is located very close to the Galactic
Center. The dominant contributors to the DM are the thin disk, spiral
arms and, near the pulsar, the Galactic Center disk. This pulsar is
the only one determining the density of the Galactic Center disk which
is adjusted by the fit to correctly model the pulsar distance. The
top-right plot is for a pulsar behind the Gum Nebula with the plot
showing the increments of about 50~cm$^{-3}$~pc with the traversal of
the near side and far side of the ellipsoidal shell. At $\sim 100$~pc
distance, the Local Bubble feature LB2 contributes about
30~cm$^{-3}$~pc.

A pulsar, PSR J0248+6021, whose distance is greatly over-estimated is
shown in the bottom-left plot. Only the thick disk and spiral arms
contribute to the DM of this pulsar and they are insufficient to
account for the DM of 370~cm$^{-3}$~pc. In the discovery paper for
this pulsar, \citet{tpc+11} argue that this pulsar is associated with
the giant HII region W5 at a distance of approximately 2~kpc and this
distance is adopted as the independent distance.  The excess DM (i.e.,
DM - DM$_{\rm Gal}$), about 160~cm$^{-3}$~pc with the NE2001 model and
about 60~cm$^{-3}$~pc with our model can be attributed to W5. In
contrast, the bottom-right plot shows the DM components for PSR
J1744$-$1134, a relatively nearby millisecond pulsar whose distance is
significantly under-estimated by the model. This plot shows that the
under-estimate is largely due to a contribution from Loop I, building
up from about 120~pc distance. This contribution is clearly not needed
for this pulsar. However, of the 13 pulsars within or behind Loop I,
ten have satisfactory model distances ($D_{err}<20$\%), one is
over-estimated and two (including PSR J1744$-$1134) are
under-estimated. This shows that, on average, Loop I makes a useful
contibution to the model but that, not surprisingly, reality is more
complex than our relatively simple hemispherical shell model. The PSR
J1744$-$1134 plot is also interesting in that it shows the effect of
the Local Bubble on the contribution from the thick disk, with the DM
gradient increasing sharply at approximately 70~pc, the edge of the
Local Bubble in this direction.

\begin{figure}[ht]
\includegraphics[angle=270,width=170mm]{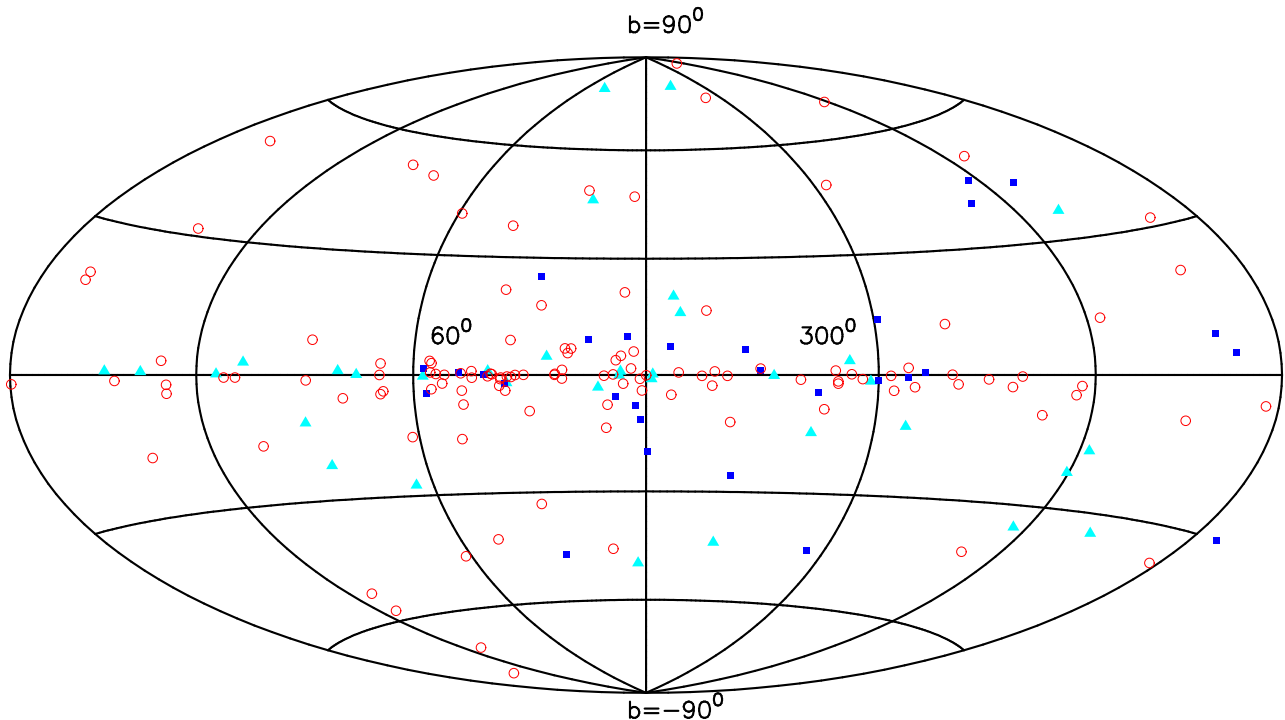}
\caption{Positions of the 189 pulsars with model-independent distances
  plotted in Galactic coordinates. Pulsars with $D_{err} < 20$\% are
  plotted as red open circles, those with under-estimated distances as
  blue squares and those with over-estimated distances as green
  triangles. \label{fg:glgb_pd}}
\end{figure}

\begin{figure}[ht]
\includegraphics[angle=270,width=170mm]{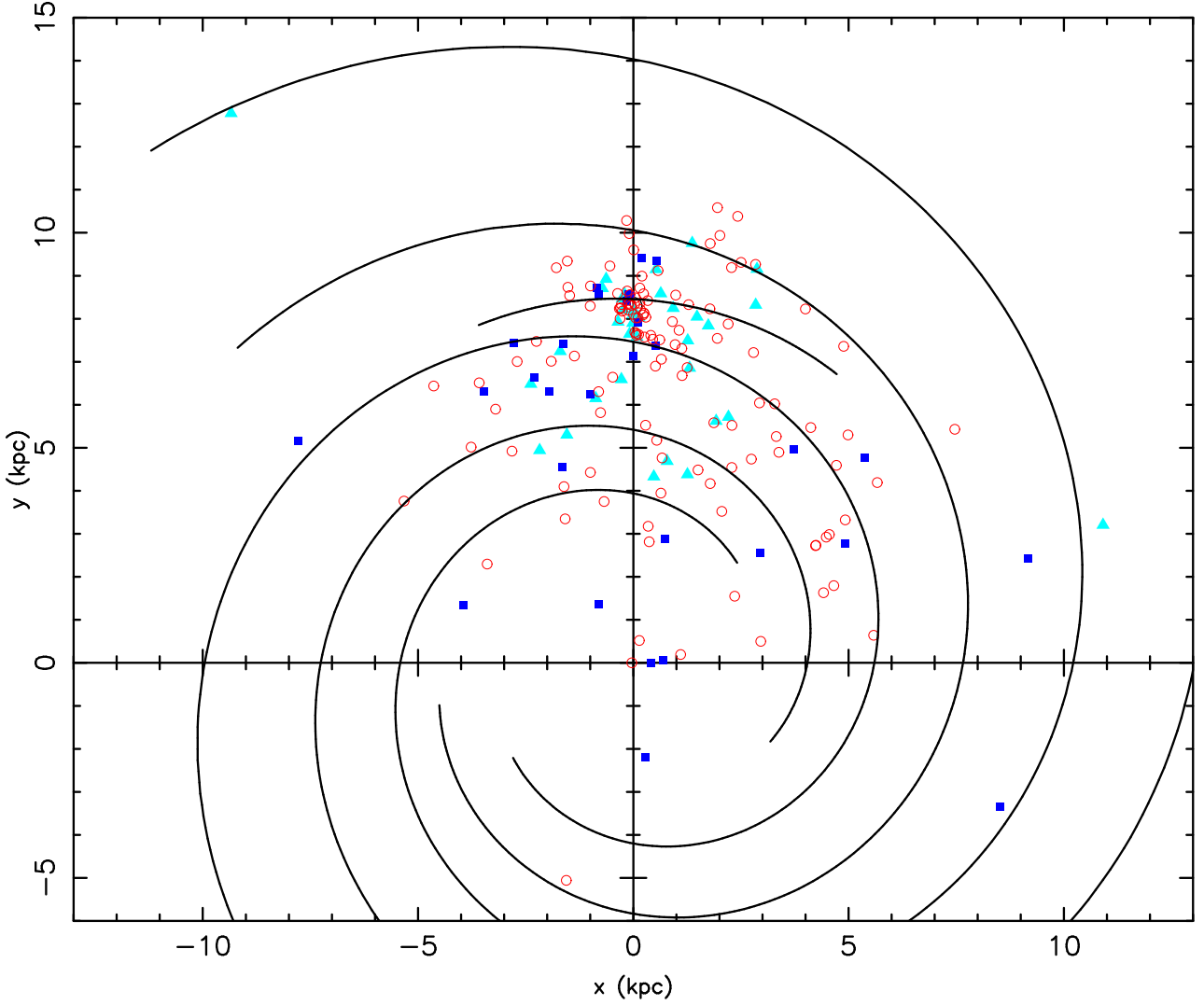}
\caption{Positions of the 189 pulsars with model-independent distances
  projected onto the Galactic plane at the position of their
  independently estimated distance. Pulsars are plotted with the same symbols as
  in Figure~\ref{fg:glgb_pd}.\label{fg:gd_pd}}
\end{figure}

Model distances and associated $D_{err}$ values are listed for all 189
Galactic pulsars with independent distances in Appendix
Tables\ref{tb:a1_px} -- \ref{tb:a1_stars}. Of these 189
Galactic pulsars, 86 have model distances within the uncertainties of
the independent distances and hence have $D_{err}=0$ and a further 38
have $D_{err}\le20$\%, giving 124 independent distances, or 65\% of the
total, that are satisfactorily represented by the model. Of the 65
pulsars with $D_{err}>20$\%, 35 have over-estimated distances, that is
insufficient $n_e$ along the path to account for the DM, and 30 have
under-estimated distances. These pulsars are listed in
Tables~\ref{tb:oe} and \ref{tb:ue}
respectively. Figures~\ref{fg:glgb_pd} and \ref{fg:gd_pd} show the
distribution of the 189 pulsars in Galactic coordinates and projected
onto the Galactic plane, respectively, with over-estimated and
under-estimated distances marked by different symbols. Within
statistical fluctuations, under-estimated, over-estimated and
correctly estimated distances are reasonably evenly distributed in
both projections. 

\begin{deluxetable}{lDDDrrrrr}
\tabletypesize{\scriptsize}
\tablecaption{Parameters for 35 pulsars with overestimated distances\label{tb:oe}}
\tablehead{\colhead{J2000} & \multicolumn2c{$l$} & \multicolumn2c{$b$} &
  \multicolumn2c{DM} & \colhead{$D_{\rm l}$}
    & \colhead{$D_{i}$} & \colhead{$D_{u}$} & \colhead{$D_{m}$} & \colhead{$D_{err}$} \\
  \colhead{Name} & \multicolumn2c{($\degr$)} & \multicolumn2c{($\degr$)}
  & \multicolumn2c{(cm$^{-3}$~pc)} & \colhead{(pc)}
  & \colhead{(pc)} & \colhead{(pc)} & \colhead{(pc)} & \colhead{(\%)}
}
\decimals
\startdata
J2144$-$3933  & 2.794 & $-$49.466 & 3.35  & 150 & 160 & 180 & 289 & 61  \\
J1756$-$2251  & 6.499 & 0.948 & 121.18  & 490 & 730 & 1330  & 2806  & 111 \\
J1801$-$2304  & 6.837 & $-$0.066  & 1073.9  & 3000  & 4000  & 5000  & 6522  & 30  \\
J1824$-$1945  & 12.279  & $-$3.106  & 224.65  & 2800  & 3700  & 4300  & 5612  & 30  \\
J1543+0929    & 17.811  & 45.775  & 35.24 & 5400  & 5900  & 6500  & 25000 & 285 \\
J1820$-$0427  & 25.456  & 4.733 & 84.44 & 100 & 300 & 900 & 2918  & 224 \\
J1903+0135    & 35.727  & $-$1.955  & 245.17  & 2800  & 3300  & 3900  & 6000  & 54  \\
J1901+0716    & 40.569  & 1.056 & 252.81  & 2700  & 3400  & 4300  & 7237  & 68  \\
J1342+2822A     & 42.209  & 78.709  & 26.5  & 9600  & 9900  & 10200 & 25000 & 145 \\
J1939+2134    & 57.509  & $-$0.29 & 71.04 & 1200  & 1500  & 2000  & 2897  & 45  \\
J2129+1210A     & 65.012  & $-$27.312 & 67  & 12900 & 13550 & 14200 & 25000 & 76  \\
J2021+3651  & 75.222  & 0.111 & 367.5 & 400 & 1800  & 3500  & 10512 & 200 \\
J2032+4127    & 80.224  & 1.028 & 114.65  & 1400  & 1500  & 1700  & 4623  & 172 \\
J2214+3000    & 86.855  & $-$21.665 & 22.56 & 909 & 1000  & 1111  & 1674  & 51  \\
J2157+4017    & 90.488  & $-$11.341 & 70.86 & 2500  & 2900  & 3400  & 4750  & 40  \\
J2229+6114    & 106.647 & 2.949 & 204.97  & 2400  & 3000  & 3600  & 5037  & 40  \\
J2337+6151    & 114.284 & 0.233 & 58.41 & 600 & 700 & 800 & 2079  & 160 \\
J0248+6021    & 136.903 & 0.697 & 370 & 1800  & 2000  & 2200  & 25000 & 1036  \\
J0358+5413    & 148.19  & 0.811 & 57.14 & 900 & 1000  & 1200  & 1594  & 33  \\
J0452$-$1759  & 217.078 & $-$34.087 & 39.9  & 300 & 400 & 600 & 2710  & 352 \\
J0922+0638    & 225.42  & 36.392  & 27.27 & 1000  & 1100  & 1300  & 1908  & 47  \\
J0630$-$2834  & 236.952 & $-$16.758 & 34.47 & 280 & 320 & 370 & 2072  & 460 \\
J0614$-$3329  & 240.501 & $-$21.827 & 37.05 & 760 & 890 & 1020  & 2691  & 164 \\
J0514$-$4002A   & 244.514 & $-$35.036 & 52.15 & 12100 & 12650 & 13200 & 25000 & 89  \\
J1017$-$7156  & 291.558 & $-$12.553 & 94.22 & 196 & 256 & 370 & 1807  & 388 \\
J1243$-$6423  & 302.051 & $-$1.532  & 297.25  & 0 & 2000  & 4000  & 9411  & 135 \\
J1326$-$5859  & 307.504 & 3.565 & 287.3 & 2000  & 3000  & 5000  & 10688 & 113 \\
J1603$-$7202  & 316.63  & $-$14.496 & 38.05 & 370 & 530 & 570 & 1129  & 98  \\
J1550$-$5418  & 327.237 & $-$0.132  & 830 & 3500  & 4000  & 4500  & 6291  & 40  \\
J1312+1810      & 332.954 & 79.763  & 24  & 17200 & 18900 & 20600 & 25000 & 21  \\
J2129$-$5721  & 338.005 & $-$43.57  & 31.85 & 1700  & 3200  & 4700  & 6170  & 31  \\
J1623$-$2631    & 350.976 & 15.96 & 62.86 & 1600  & 1800  & 2000  & 3651  & 82  \\
J1614$-$2230  & 352.636 & 20.192  & 34.5  & 400 & 700 & 1000  & 1395  & 39  \\
J1740$-$3015  & 358.294 & 0.238 & 152.15  & 100 & 400 & 2100  & 2945  & 40  \\
J1745$-$3040  & 358.553 & $-$0.963  & 88.37 & 0 & 200 & 1300  & 2343  & 80  \\
\enddata
\end{deluxetable}

\begin{deluxetable}{lDDDrrrrr}
\tabletypesize{\scriptsize}
\tablecaption{Parameters for 30 pulsars with underestimated distances\label{tb:ue}}
\tablehead{\colhead{J2000} & \multicolumn2c{$l$} & \multicolumn2c{$b$} &
  \multicolumn2c{DM} & \colhead{$D_{\rm l}$}
    & \colhead{$D_{i}$} & \colhead{$D_{u}$} & \colhead{$D_{m}$} & \colhead{$D_{err}$} \\
  \colhead{Name} & \multicolumn2c{($\degr$)} & \multicolumn2c{($\degr$)}
  & \multicolumn2c{(cm$^{-3}$~pc)} & \colhead{(pc)}
  & \colhead{(pc)} & \colhead{(pc)} & \colhead{(pc)} & \colhead{(\%)}
}
\decimals
\startdata
J1835$-$3259A   & 1.532 & $-$11.371 & 63.35 & 10200 & 10700 & 11200 & 2711  & 276 \\
J1823$-$3021A   & 2.788 & $-$7.913  & 86.88 & 7800  & 8400  & 9000  & 3145  & 148 \\
J1721$-$1936    & 4.857 & 9.738 & 75.7  & 7800  & 8400  & 9000  & 3070  & 154 \\
J1824$-$2452A   & 7.797 & $-$5.578  & 120.5 & 5200  & 5500  & 5800  & 3737  & 39  \\
J1744$-$1134  & 14.794  & 9.18  & 3.14  & 384 & 395 & 406 & 148 & 159 \\
J2140$-$2310A   & 27.179  & $-$46.837 & 25.06 & 9200  & 9450  & 9700  & 3112  & 196 \\
J1713+0747    & 28.751  & 25.223  & 15.99 & 1136  & 1176  & 1220  & 919 & 24  \\
J1905+0154A     & 36.208  & $-$2.201  & 193.69  & 13950 & 14450 & 14950 & 5509  & 153 \\
J1906+0746    & 41.598  & 0.147 & 217.75  & 6000  & 7400  & 9900  & 4814  & 25  \\
J1917+1353    & 48.26 & 0.624 & 94.54 & 4000  & 5000  & 6000  & 2940  & 36  \\
J1953+1846A     & 56.744  & $-$4.563  & 117 & 6000  & 6450  & 6900  & 4505  & 33  \\
J1932+2220    & 57.356  & 1.554 & 219.2 & 10100 & 10900 & 12200 & 7999  & 26  \\
J0454+5543    & 152.617 & 7.547 & 14.49 & 1130  & 1180  & 1250  & 631 & 79  \\
J0337+1715    & 169.99  & $-$30.039 & 21.32 & 1220  & 1300  & 1380  & 817 & 49  \\
J0633+1746    & 195.134 & 4.266 & 2.89  & 170 & 250 & 480 & 138 & 23  \\
J0659+1414    & 201.108 & 8.258 & 13.98 & 250 & 280 & 310 & 159 & 57  \\
J0953+0755    & 228.908 & 43.697  & 2.96  & 256 & 261 & 266 & 186 & 37  \\
J1023+0038    & 243.49  & 45.782  & 14.32 & 1328  & 1367  & 1410  & 1057  & 26  \\
J1024$-$0719  & 251.702 & 40.516  & 6.49  & 800 & 1100  & 1500  & 381 & 110 \\
J1048$-$5832  & 287.425 & 0.577 & 129.1 & 2200  & 2900  & 4100  & 1793  & 23  \\
J1119$-$6127  & 292.151 & $-$0.537  & 707.4 & 8000  & 8400  & 8800  & 6414  & 25  \\
J1227$-$4853  & 298.965 & 13.796  & 43.42 & 1800  & 1900  & 2000  & 1244  & 45  \\
J1224$-$6407  & 299.984 & $-$1.415  & 97.47 & 2000  & 4000  & 6000  & 1534  & 31  \\
J0024$-$7204C   & 305.923 & $-$44.892 & 24.4  & 3650  & 4000  & 4350  & 2547  & 43  \\
J1453$-$6413  & 315.733 & $-$4.427  & 71.07 & 2000  & 2800  & 4100  & 1432  & 40  \\
J1602$-$5100  & 330.688 & 1.286 & 170.93  & 7300  & 8000  & 8900  & 3407  & 114 \\
J1559$-$4438  & 334.54  & 6.367 & 56.1  & 2000  & 2300  & 2800  & 1480  & 36  \\
J1910$-$5959A   & 336.525 & $-$25.73  & 33.28 & 4490  & 4550  & 4610  & 1642  & 174 \\
J1701$-$3006A   & 353.578 & 7.322 & 114.97  & 6470  & 7050  & 7630  & 4881  & 33  \\
J1909$-$3744  & 359.731 & $-$19.596 & 10.39 & 1230  & 1234  & 1239  & 564 & 118 \\
\enddata
\end{deluxetable}

\begin{figure}[ht]
\includegraphics[angle=270,width=85mm]{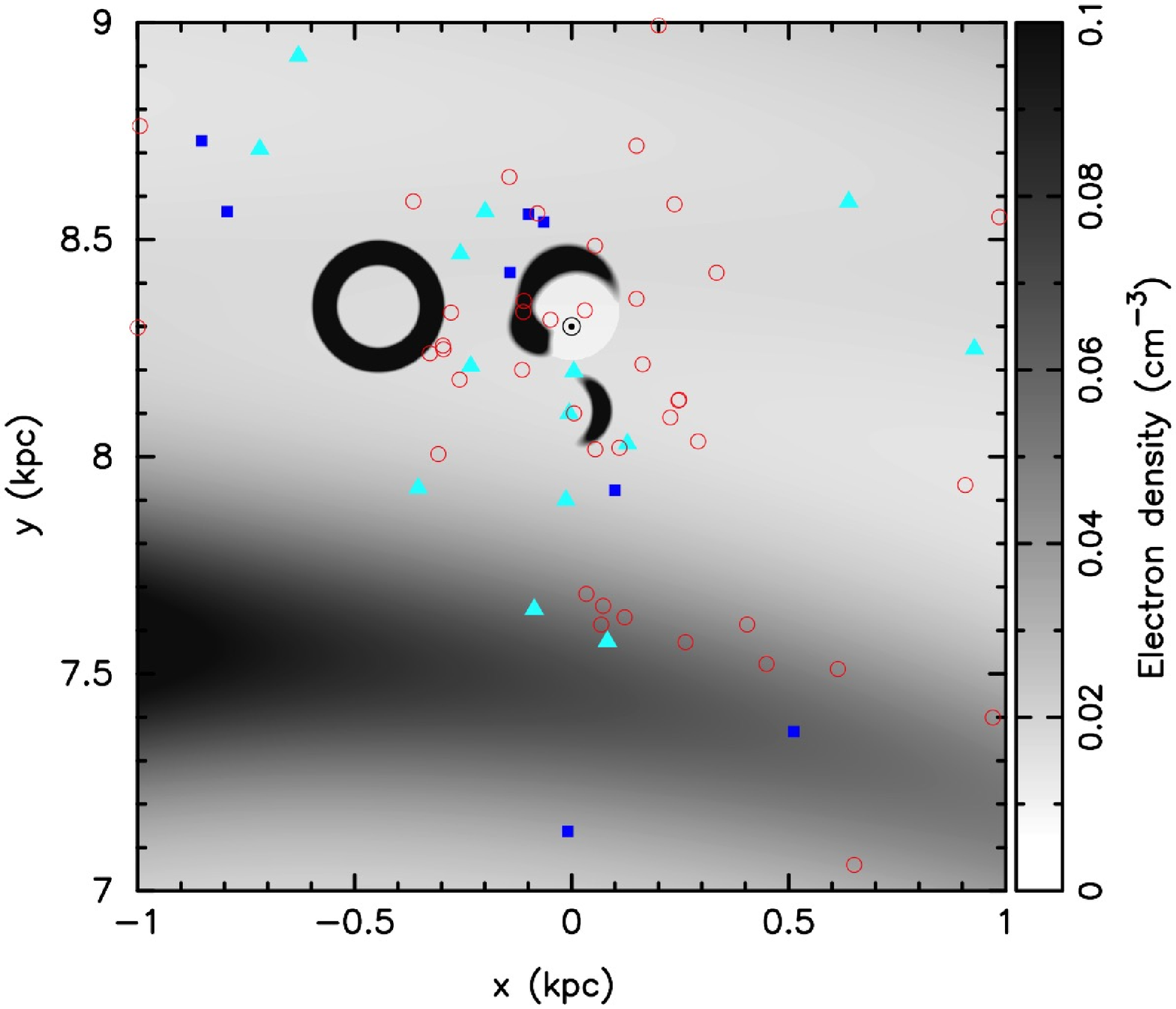}
\caption{Pulsar positions projected onto the Galactic plane and $n_e$
  distribution in the plane for the local region of the Galaxy. The
  position of the Sun is marked with $\odot$. Identifiable model
  features are the Carina enhancement and Carina -- Sagittarius spiral
  arm, the Gum Nebula, the Local Bubble and surrounding density
  enhancements and Loop I (at $x,y$ = 0.1, 8.1~kpc). Pulsars are
  plotted with the same symbols as in
  Figure~\ref{fg:glgb_pd}.\label{fg:ne_local}}
\end{figure}

\begin{figure}[ht]
\includegraphics[angle=270,width=85mm]{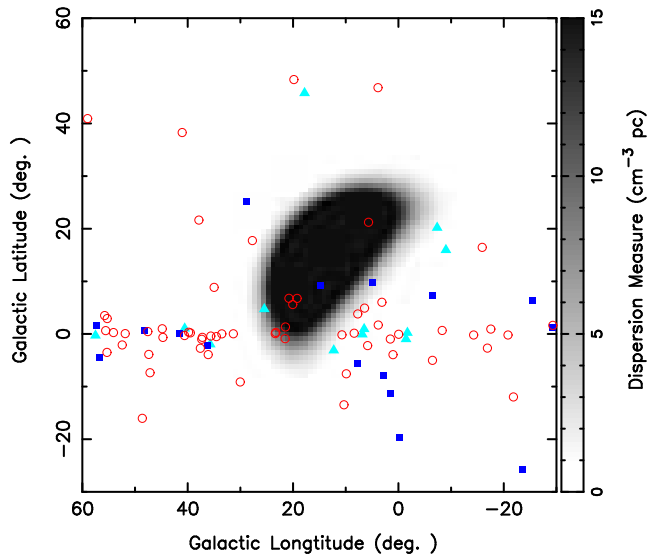}
\caption{Pulsar positions in Galactic coordinates in the vicinity of
  Loop I, with the grey-scale showing the DM contribution of Loop
  I. Pulsars are plotted with the same symbols as in
  Figure~\ref{fg:glgb_pd}.\label{fg:ne_li}}
\end{figure}

Figure~\ref{fg:ne_local} shows an expanded view of
Figure~\ref{fg:gd_pd} for the local region. The Gum Nebula and Local
Bubble are circular in the $x-y$ plane and, as Figure~\ref{fg:ne_li}
illustrates, only the tip of the Loop I shell crosses the $x-y$
plane. With a peak $n_e$ of just 0.0057~cm$^{-3}$, the Local spiral
arm is not visible in Figure~\ref{fg:ne_local}. Because of
their extent and proximity to the Sun, a relatively large number of
pulsars are affected by these local features.

The Local Bubble of course surrounds the Sun and so affects model
distances for all pulsars. However, it only has a significant effect
for those which are close to the Sun, say within 1~kpc. For PSR
J0437$-$4715, not only does the entire path lie within the Local
Bubble but also it has a very precisely measured distance. It clearly
has a strong influence on the value of $J_{\rm LB}$ since the model
distance is equal to the independent distance, both 156~pc. Of the
seven pulsars with $D_i \le 250$~pc, five have distances based on
parallax measurements and three of these are accurately modelled with
$D_{err}=0$. The other two, PSRs J2144$-$3933 and J0633+1746, have
model distances that are off by about a factor of two, PSR
J2144$-$3933 over-estimated and PSR J0633+1746 under-estimated. The
remaining two (PSRs J1745$-$3040 and J1752$-$2806) have kinematic
distance estimates which are less precise; for the first, $D_m$ is
about twice the upper limit of $D_i$, and for the second $D_{err}<20\%$
but $D_m$ is at the upper limit of $D_i$ (Table~\ref{tb:oe}). The Local
Bubble is a significant feature of the model, with most of the local
pulsars strongly affected by it having accurate model distances.

The associated density enhancements, LB1 and LB2, are relatively dense
and, because of their size and proximity to the Sun, they
significantly affect the model distances to a large number of pulsars
lying at Galactic longitudes between about $150\degr$ and
$315\degr$. Of the 55 pulsars with a significant DM contribution from
LB1/2, 35 have satisfactorily modelled distances with $D_{err}<20\%$, eight
have over-estimated distances and 12 have under-estimated
distances. Most of the over-estimated distances are for pulsars at
negative latitudes, whereas most of the under-estimated distances are
for pulsars at positive latitudes. This suggests that LB1/2 should be
centered at $z<0$ rather than at $z=0$ as in the present model. However,
at present we have insufficient data to quantify this offset. 
Most of the 35 well-modelled distances would be over-estimated in the
absence of LB1/2 and so these features are critical to the model.

As discussed above, the Gum Nebula shell has a relatively high density
and makes a signficant contribution to the DM of pulsars within or
behind it. Of the eight pulsars affected by the Gum Nebula, six have
model distances within the independent-distance limits. The two
exceptions, PSRs J0742$-$2822 and J0835-4510 (the Vela pulsar), have
slightly over-estimated distances (Tables~\ref{tb:a1_kin} and
\ref{tb:a1_px} respectively). Similarly, of
the 13 pulsars having a DM contribution from Loop I
(Figure~\ref{fg:ne_li}), two (PSRs J1721$-$1936 and J1744$-$1134) have
under-estimated distances (Table~\ref{tb:ue}), one (PSR J1820$-$0427)
has an over-estimated distance and the remaining ten have accurate
model distances. Clearly, both the Gum Nebula and Loop I make
important contributions to the model.

Table~\ref{tb:oe} shows that there are six pulsars for which the YMW16
model cannot account for the observed DM and therefore have nominal model
distances of 25000~pc. Only one of these, PSR J0248+6021, discussed
above in connection with Figure~\ref{fg:4dm-d}, is at low Galactic
latitude; the remaining five are all have $|b|\gapp 30\degr$. A
further five pulsars in Table~\ref{tb:oe} also have $|b|\gapp
30\degr$, but do have DMs within the model range. On the other hand,
six of the 30 pulsars with under-estimated distances
(Table~\ref{tb:ue}) are also at high Galactic
latitudes. These results show that the thick disk is well
modelled on average, but that structure within it results in some
over-estimated distances and some under-estimated distances.

Since nearly half of the 189 pulsars with independent distances have
$D_{err}=0$, that is, a model distance within the uncertainty range of
the independent distance, the distribution of $D_{err}$ is very
asymmetric.  The mean $D_{err}$ is
42\% but the median is just 5\%. It is difficult to estimate the
reliability of the model distances just using the known independent
distances, since the model has been fitted to these distances. To
overcome this problem, we adopt the following strategy:
\begin{enumerate}
  \item Randomly select five pulsars from the list of 189 having
    independently determined distances
  \item Do a full global plus local fit for model parameters as
    described in \S\ref{sec:model-fit} on the remaining 184 pulsars
  \item Use this model to compute distances $D_{mp}$ to the five
      omitted pulsars and store the results
  \item Repeat steps 1--3 100 times, generating 500
    ``predicted'' distances
  \item Compute the relative distance error $\epsilon_{mp} = (D_{mp} -
      D_i)/D_i$ for each of these 500 predicted distances.
\end{enumerate}

\begin{figure}[ht]
\includegraphics[angle=270,width=170mm]{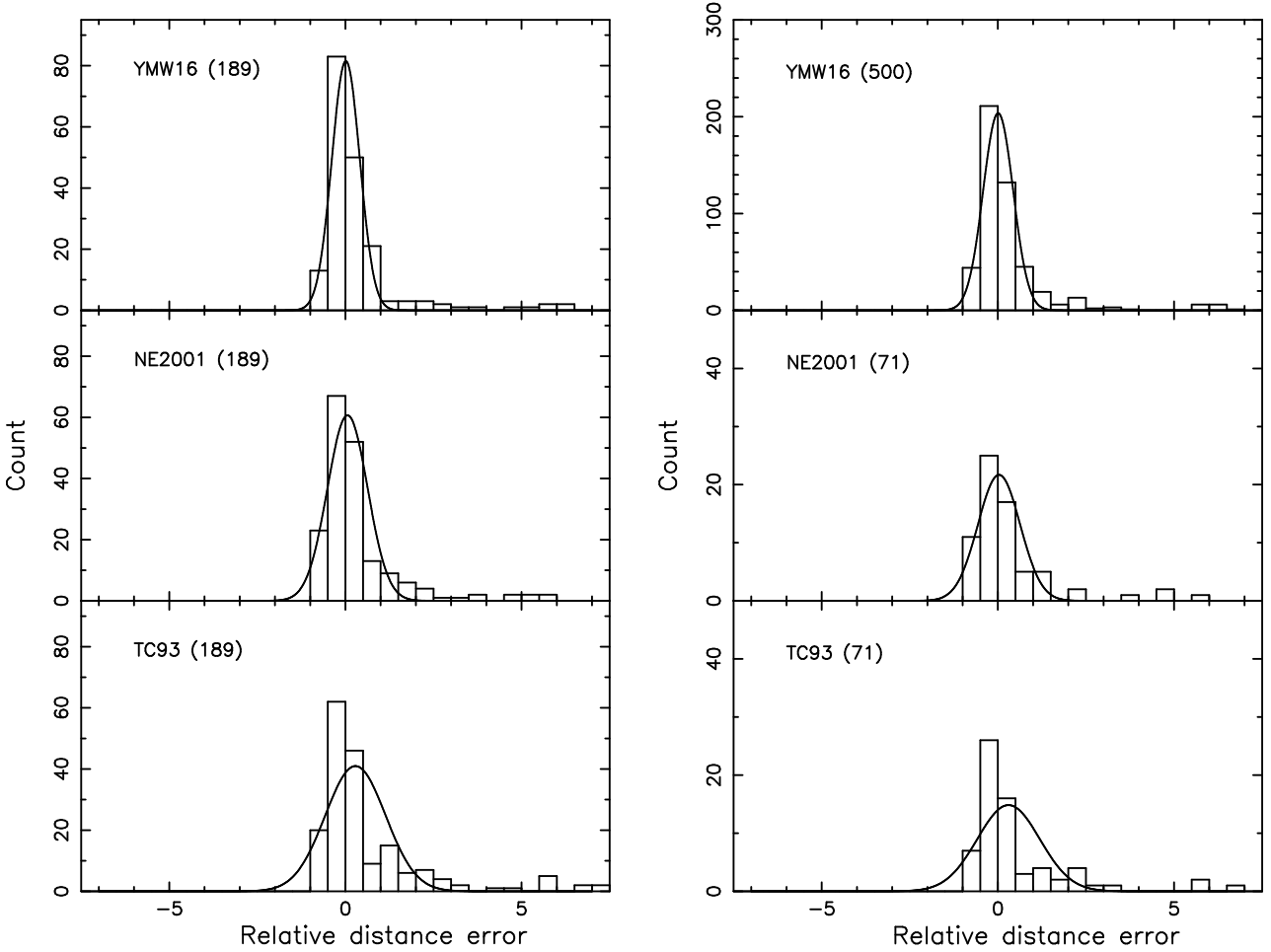}
\caption{Histograms of relative distance errors $(D_m - D_i)/D_i$,
  where $D_m$ is the model distance based on the observed DM and $D_i$
  is the independently determined distance, for three Galactic $n_e$
  models: YMW16, NE2001 and TC93. Plots in the left column show the
  distributions for all 189 pulsars with independent distances,
  whereas plots in the right column show distributions for the
  ``predictive'' data sets. For all cases, the results of a ``robust''
  fit of a Gaussian model to the distribution is shown. (See text for
  details.)  \label{fg:dhist}}
\end{figure}

The left-top panel of Figure~\ref{fg:dhist} shows the distribution of
relative distance errors $\epsilon_m =(D_m - D_i)/D_i$ for all 189
pulsars used to detemine the YMW16 model. Similarly, the right-top
panel shows the distribution of $\epsilon_{mp}$ for the
above-described predicted distances. In both cases, there is a
significant excess of over-estimated distances. For about half of this
excess, $D_m$ (or $D_{mp}$) is at the limit, 25000~pc, for pulsars
where the model cannot account for the observed DM. A Gaussian model
has been fitted to both distributions with the 10\% of pulsars having
the largest absolute relative errors omitted from the fit in order to
give a more ``robust'' estimate of the distribution
parameters. Table~\ref{tb:dfit} gives the results of this Gassian
fitting showing that the rms deviation of $\epsilon_m$ is a little
under 40\% and for $\epsilon_{mp}$ about 43\%. Based on the latter, we
conclude that the YMW16 model predicts pulsar distances based on their
DM with a 95\% confidence limit of approximately 90\%. That is, we
estimate that 95\% of all model predictions will have a relative error
of less than a factor of 0.9. 

\begin{deluxetable}{lDcc}
\tabletypesize{\small}
\tablecaption{Parameters of distance-error distributions\label{tb:dfit}}
\tablehead{
  \colhead{Model}&\multicolumn2c{Amp.}&\colhead{Mean}&\colhead{Rms
    devn}}
\decimals
\startdata
YMW16 (189 dist.)   & 81.5 & 0.012 & 0.398 \\
YMW16 (Pred. dist.) &203.7 & 0.008 & 0.426 \\
NE2001 (189 dist.)  & 60.7 & 0.060 & 0.571 \\
NE2001 (Pred. dist.)& 21.7 & 0.033 & 0.599 \\
TC93 (189 dist.)    & 41.0 & 0.284 & 0.856 \\
TC93 (Pred. dist.)  & 14.8 & 0.303 & 0.887 \\
\enddata
\end{deluxetable}

\subsection{Comparison of YMW16 with earlier Galactic $n_e$ models}\label{sec:model-comp}
The most commonly used methods of estimating pulsar distances from
their DM are based on the Galactic electron density models NE2001 and
TC93. It is therefore of interest to compare the Galactic model and
predictions of YMW16 with these earlier models and their
predictions. All three models have the same basic structure for the
Galactic $n_e$ with an extended thick disk, a thin disk largely
confined to the inner Galaxy, spiral arms and a greater or lesser
number of local features.

\begin{figure}[ht]
\includegraphics[angle=270,width=85mm]{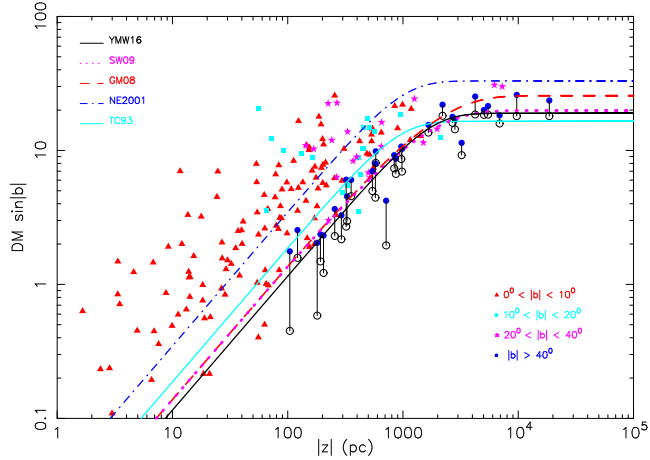}
\caption{DM$_\perp = {\rm DM}\sin|b|$ versus $|z| = D_i\sin|b|$ for
  Galactic pulsars. The lines show the variation of DM$_\perp$ for the
  thick-disk component of several different models of the Galactic
  $n_e$ distribution. For SW09 \citep{sw09} and GM08 \citep{gmcm08}
  the line is based on fits of just the thick disk to DMs and
  distances for high-latitude pulsars, but for TC93, NE2001 and YMW16,
  the line represents the thick-disk component resulting from a global
  fit to the distance data. The thick-disk DM components for the YMW16
  model are shown by open circles for the high-latitude
  pulsars.\label{fg:dm-z}}
\end{figure}

The model DM for high-latitude pulsars is primarily determined by the
thick disk. Figure~\ref{fg:dm-z} shows the variation of DM$_\perp =
{\rm DM}\sin|b|$, with $|z| = D_i\sin|b|$, where $b$ is the Galactic
latitude, for the 189 pulsars with independent distances. Curves
giving the DM$_\perp$ contribution of the thick-disk component for
other Galactic $n_e$ models are also shown. For pulsars at low
latitudes, there is large DM$_\perp$ contribution from other
components of the Galactic $n_e$ distribution, especially the spiral
arms and the thin disk. However, even for high-latitude pulsars, there
is some contribution from these other components. Figure~\ref{fg:dm-z}
shows the effect of subtracting these other components for the YMW16
model, to give the thick-disk component alone. 

Recent models, including YMW16, have largely converged on a larger
scale height $\sim 1600$~pc and a lower DM$_\perp \sim
20$~cm$^{-3}$~pc for the thick disk. For NE2001, the thick disk
clearly includes a much larger contribution from low-$z$ components
that are assigned to spiral arms or the thin disk in other models. As
discussed in \S\ref{sec:intro}, it also substantially over-estimates
the high-$z$ DMs or, equivalently, under-estimates the distances of
high $z$ pulsars which are mostly at high $|b|$. It is notable that
this is not the case for the earlier TC93 model.

The thin disk is very similar in YMW16, NE2001 and TC93, an annulus with
radius about 4~kpc, although for YMW16 the disk scale height is about
65~pc compared to 150~pc for TC93 and 140~pc for NE2001.

For the spiral arms, YMW16 adopts a simple logarithmic-spiral pattern
based on HII-region distances \citep{hh14} in contrast to the somewhat
more complex modified spiral pattern used by TC93 and NE2001. Part of
the modified pattern used by these earlier models was to reduce $n_e$
in the nominal Sagittarius tangential region. In YMW16, this is
accomplished by defining an under-dense zone in the Sagittarius
arm. An over-dense zone in the Carina arm is also included in YMW16.  

TC93 has just one local feature, the Gum Nebula, whereas NE2001 has 82
local clumps, including the Gum Nebula and a clump surrounding the
Vela pulsar, 17 voids and four other local features including the
Local Bubble and Loop I. As described in \S\ref{sec:model} and
\S\ref{sec:results} above, YMW16 takes a more conservative approach
with just seven local features chosen on the basis that each affected
the model distance for a group of pulsars either within or behind the
feature. The parameters of these features are based on relatively
recent studies unavailable to both TC93 and NE2001.

Finally, YMW16 includes components representing the Magellanic Clouds
and the intergalactic medium. These components were not included in
any previous model for pulsar distance estimation. 

Comparisons of the YMW16 model predictions with those of TC93 and
NE2001 are difficult since these earlier models were based on a
smaller sample of independent distances (74 pulsars for TC93 and 112
for NE2001) than that used for our model. However, for the moment, we
will ignore this issue and directly compare the distance predictions
of the YMW16 model with those of earlier models for the currently
known 189 pulsars with independent distance data.

In Table~\ref{tb:errors} we summarise the distribution of distance
errors $D_{err}$ for the YMW16, NE2001 and TC93 models. It is clear
that the YMW16 model has benefited greatly from recent parallax
measurements for relatively high-latitude pulsars from VLBI
\citep[e.g.,][]{cbv+09} and pulsar timing array projects
\citep{rhc+16,mnf+16}. These have allowed a much better definition of
the thick disk and consequently smaller distance errors for
high-latitude pulsars with 19 of the 29 high-latitude pulsars having
$D_{err}$$<$20\%. Overall, the YMW16 model has significantly smaller
distance errors than NE2001 and much smaller distance errors than
TC93.

\begin{deluxetable}{lccccccccc}
\tablecaption{Comparison of distance errors with previous models\label{tb:errors}}
\tablehead{
\colhead{} & \multicolumn{4}{c}{29 pulsars with $|b|>40\degr$} & \colhead{} &
\multicolumn{4}{c}{189 pulsars} \\
\cline{2-5} \cline{7-10} \\
\colhead{Model} & \colhead{0\%} & \colhead{0\%-20\%} & \colhead{20\%-40\%} & \colhead{$>$40\%} &
\colhead{} & \colhead{0\%} & \colhead{0\%-20\%} & \colhead{20\%-40\%} & \colhead{$>$40\%}}
\startdata
YMW16 & 14 & 5 & 4 & 6 &  & 86 & 38 & 25 & 40\\
NE2001 & 6 & 5 & 3 & 15 &   & 77 & 32 & 19 & 61\\
TC93 & 6 & 5 & 4 & 14 &   & 45 & 36 & 29 & 79\\
\enddata
\end{deluxetable}

\begin{figure}[ht]
\includegraphics[angle=270,width=85mm]{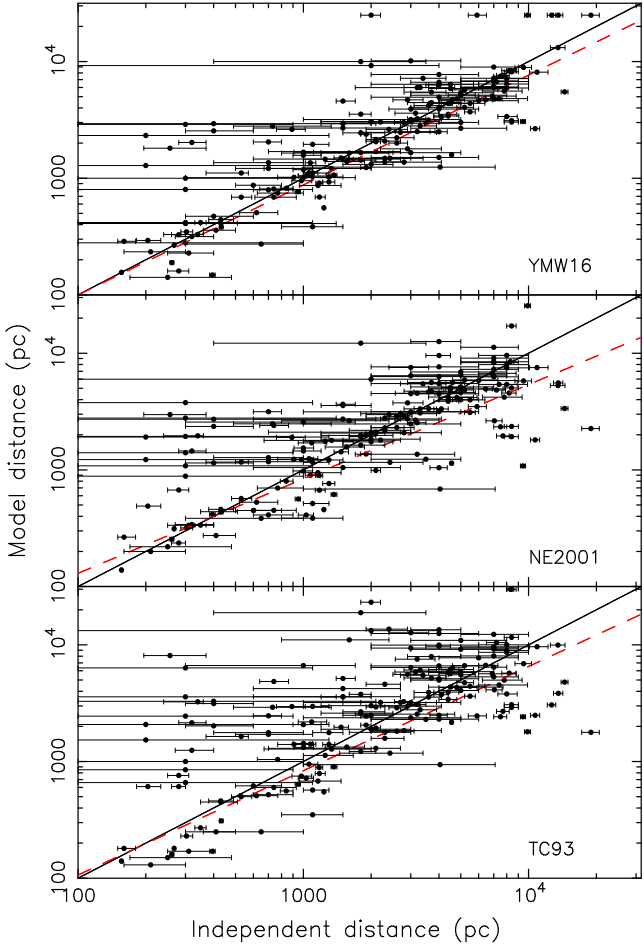}
\caption{Distances from the YMW16, NE2001 and TC93 Galactic $n_e$
  models versus independently measured distances and uncertainties for the 189 pulsars
  where these are known. The sloping black lines correspond to $D_m=D_i$
  and the red dashed lines show the result of a weighted least-squares
  fit of $\log D_m$ as a function of $\log D_i$.\label{fg:dmdi}}
\end{figure}

In Figure~\ref{fg:dmdi} we directly compare the model and independent
distances for the YMW16, NE2001 and TC93 models. We also show the results
of weighted least-square fits of a linear relationship
\begin{equation}\label{eq:DD}
\log D_m=a\log D_i+b
\end{equation}
with weights equal to $1/(\log D_u - \log D_i)$. Table~\ref{tb:dmdi}
gives the results of the fits and also the correlation coefficients
computed using the same weights. Not surprisingly, the YMW16 model is
a better fit to the independent distance data and has a higher
correlation coefficient with it. Compared to the earlier models, many
fewer distances are over-estimated. However, these over-estimated
distances generally have large distance undertainties. In fact, more
of the high-weight distances are under-estimated by the earlier
models, resulting in fitted lines of smaller slope located largely
below the line of equality. Somewhat surprisingly though, the TC93
model has both a slope closer to 1.0 and a higher correlation
coefficient compared to the NE2001 model. 

\begin{deluxetable}{lccc}
\tablecaption{Distance model fits\label{tb:dmdi}}
\tablehead{
\colhead{} & \multicolumn{2}{c}{Least-square fits} & \colhead{Correlation} \\
\colhead{Model} & \colhead{$a$} & \colhead{$b$} & \colhead{coefficient}}
\startdata
YMW16 & 0.946 & 0.104 & 0.932 \\
NE2001 & 0.809 & 0.495 & 0.877 \\
TC93 & 0.894 & 0.239 & 0.892 \\
\enddata
\end{deluxetable}

In \S\ref{sec:results} above, we estimated the reliability of the
YMW16 model by repeatedly fitting the independent distance data to 184
pulsars randomly chosen from the 189 with independent distances
currently available and then comparing model and independent distances
for the omitted five pulsars. For NE2001 and TC93, the process is
simpler, since many independent distances have been published since
these models were created. Unfortunately, however, neither
\citet{tc93} nor \citet{cl02} or \citet{cl03} published the list of
independent distances used in constructing their model. We have
therefore selected the set of 71 independent distance measurements
published since 2008 in order to test the predictive abilities of
these models.\footnote{Despite the TC93 and NE2001 models using 74 and
  112 independent distances respectively, the ATNF Pulsar Catalogue
  and \citet{vwc+12} show that only 41 distances were formally
  published before 1993 and 90 before 2002. We have therefore selected
  a cutoff date that gives somewhat less than $189-112=77$
  distances. It is likely that some unpublished distances were used in
  constructing these models and also that updated distance estimates
  have been published for some pulsars.} Distributions of $\epsilon_m$
and $\epsilon_{mp}$ are shown in Figure~\ref{fg:dhist} and the results
of the robust Gaussian fitting are given in
Table~\ref{tb:dfit}. Unsurprisingly, the two earlier models are
significantly inferior to YMW16 when tested against the full sample of
189 independent distances. However, this inferior performance also
carries over to the ``predictive'' data sets. For NE2001 and TC93, the
rms deviations $\epsilon_{mp}$ are about 60\% and 89\% respectively,
corresponding to 95\% confidence limits of approximately 120\% and
180\% for the two models.

\begin{figure}[ht]
\includegraphics[angle=270,width=170mm]{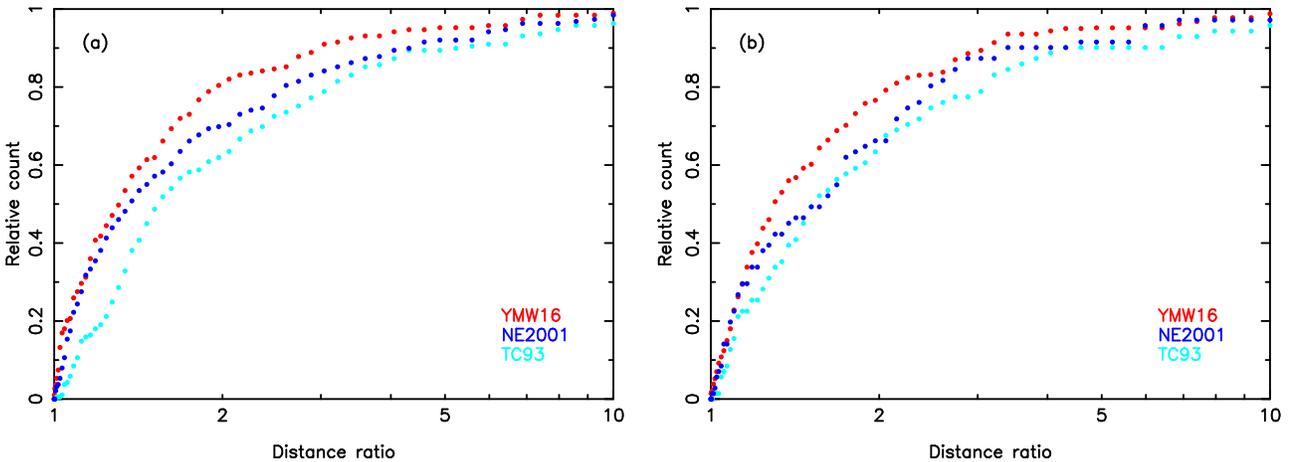}
\caption{Cumulative histograms of distance ratios (a) $\rho_m$ and (b)
  $\rho_{mp}$ for the full sample of 189 independently determined
  distances and the predictive data sets, respectively, for the
  three Galactic $n_e$ models: YMW16, NE2001 and TC93. (See text for
  details.) \label{fg:chist}}
\end{figure}

Differences in the performance of the three models are also clearly
illustrated by cumulative histograms. Following \citet{sch12}, we have
used distance ratios, $\rho_m$, where $\rho_m=D_m/D_i$ for $D_m/D_i\ge
1.0$ and $\rho_m=D_i/D_m$ for $D_m/D_i < 1.0$, rather than the
relative distance errors used for the histograms in
Figure~\ref{fg:dhist}, as this allows use of a logarithmic $x$ axis for
the cumulative histograms, and similarly for
$\rho_{mp}$. Figure~\ref{fg:chist} shows these cumulative histograms
for the three models. For both the current and predictive data sets,
the superior performance of the YMW16 model is evident. Both
Figure~\ref{fg:dhist} and Figure~\ref{fg:chist} show that TC93
signficantly over-estimates many distances. Figure~\ref{fg:chist}b
shows that, for low values of $\rho_{mp}$ i.e. accurately predicted
distances, despite the much smaller number of free parameters in the
YMW16 model, but YMW16 does much better for larger $\rho_{mp}$.

\subsection{Interstellar scattering}
\begin{figure}[ht]
\includegraphics[angle=270,width=85mm]{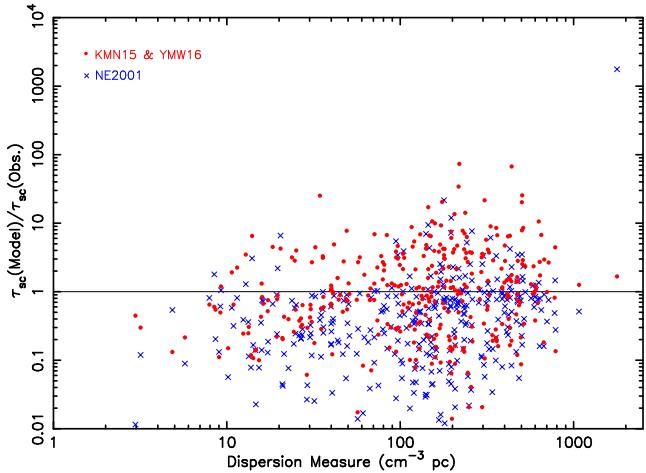}
\caption{Ratio of model scattering times to observed scattering times,
  normalized to 1 GHz, for 355 Galactic pulsars as a function of
  dispersion measure. The red dots are for the YMW16 model which is
  based on the results of \citet{kmn+15} and the blue crosses are for the
  predictions of the NE2001 model.\label{fg:gal_sc}}
\end{figure}

Predicted interstellar scattering times at 1.0 GHz, based on the
\citet{kmn+15} model (Equation~\ref{eq:tscat}) for Galactic pulsars
with independent distances are given in Tables~\ref{tb:a1_px} --
\ref{tb:a1_stars}. The \citet{kmn+15} model was based on a fit to a
total of 358 measurements of scattering times for Galactic pulsars. In
Figure~\ref{fg:gal_sc} we plot the ratios of model predictions to
observed scattering times, scaled to 1.0 GHz assuming $\tau_{\rm sc}
\propto \nu^{-4.0}$, for YMW16 and NE2001. In addition to 354
$\tau_{\rm sc}$ values and corresponding observation frequencies
provided to us by M. Krishnakumar (private communication), we include
the recently measured scattering timescale for the Galactic Center
pulsar, PSR J1745$-$2900, which has a DM = 1778 cm$^{-3}$~pc
\citep{efk+13} in this figure. \citet{ppe+15} measured $\tau_{\rm sc}
= 0.1330\pm 0.0005$~s at 2.0 GHz and a frequency index $\alpha =
-3.71\pm 0.02$ for the scattering timescale of PSR
J1745$-$2900. Scaling this to 1 GHz assuming an index of $-4.0$ gives
$\tau_{\rm sc} = 2.13$~s, whereas an index of $-3.71$ gives
1.74~s. Both of these are close to the YMW16 prediction of 3.55~s
(cf. Table~\ref{tb:a1_neb}). For consistency, we have chosen to plot
the former value. As discussed by \citet{ppe+15}, there is some
indication of time variability in the scattering timescale for this
pulsar, with \citet{sle+14} measuring $\tau_{\rm sc}=1.3\pm 0.2$~s at
1~GHz.

As expected, Figure~\ref{fg:gal_sc} shows the KMN15/YMW16 points
essentially uniformly distributed about $R_{\rm sc}=\tau_{\rm
  sc}$(Model)/$\tau_{\rm sc}$(Obs.) = 1 line. However, with the
exception of the PSR J1745$-$2900 point, the NE2001 values are
systematically biased low. For KMN15/YMW16, the mean and rms deviation
of $\log R_{\rm sc}$ (excluding PSR J1745$-$2900) are $-0.080\pm
0.039$ and 0.735, respectively, whereas for NE2001 the corresponding
values are $-0.573\pm 0.046$ and 0.863. For NE2001, the mean offset in
the log corresponds to $\tau_{\rm sc}$ being low on average by a
factor of $\sim 3.7$. Different assumptions about scaling from
observations at other frequencies may influence these results, but it
is very unlikely that they can account for the general
under-estimation of $\tau_{\rm sc}$ by NE2001.

The NE2001 model predicts a very large scattering timescale for
pulsars within or behind the Galactic Center disk and, in particular
for PSR J1745$-$2900 predicts $\tau_{\rm sc} = 3750$~s, about a factor
of 1800 greater than the observed value. \citet{bdd+14} argue that the
principal scattering screen for PSR J1745$-$2900 is located about
5.9~kpc from the Galactic Center, in the Scutum spiral arm, rather than
in the Galactic Center region as assumed by \citet{cl02}. The close
agreement of the observed $\tau_{\rm sc}$ for this pulsar with the
prediction of the \citet{kmn+15} model for Galactic scattering shows
that the scattering observed for this pulsar is not exceptional given
its large DM and hence supports the conclusions of \citet{bdd+14}.

\subsection{The Magellanic Clouds}\label{sec:mc_results}
The location, size and shape of the Large and Small Magellanic Clouds
are based primarily on the distribution of young stars and
star-forming regions in the Clouds. The models are relatively simple
as the number of pulsars known to be associated with the Clouds is
small (Table~\ref{tb:a1_mc}) and, in most cases, we have no prior
information on the location of individual pulsars within the
Clouds. For the LMC, the derived central electron density is
0.066~cm$^{-3}$, comparable to the electron density averaged over the
Galactic disk. The derived central electron density of the
separately-modelled giant HII region 30 Doradus is about
0.3~cm$^{-3}$, but this is quite uncertain as it is primarly
determined by one pulsar (PSR J0537$-$69) and we don't know the
line-of-sight position of this pulsar relative to the nebula.

Of the five pulsars associated with the SMC (Table~\ref{tb:a1_mc}),
four have relatively small DMs ($<100$~cm$^{-3}$~pc). The fifth,
PSR~J0131$-$7310, has a DM of 175~cm$^{-3}$~pc despite being located
right at the edge of the SMC. The diffuse H$\alpha$ survey of
\citet{bhb13} shows a tongue of emission covering the position of
PSR~J0131$-$7310 and the HII region N90 \citep{lgb+10} is nearby. The
SMC images in the Barbara A. Milulski Archive for Space
Telescopes\footnote{\url{
 https://archive.stsci.edu/prepds/fuse\_mc/overviewimages.html}} show
an H$\alpha$ shell surrounding the pulsar. There is no significant
radio continuum feature at the pulsar position in the 4.8~GHz image of
\citet{lgb+10}. It is possible that this pulsar just happens to lie
behind a relatively high-density clump of ionised gas, possibly part
of the surrounding H$\alpha$ shell. However, we choose not to
explicity model this and omit this pulsar from the SMC fit in order
to give a more representative model for the SMC $n_e$
distribution. The derived central density for the SMC is
0.045~cm$^{-3}$ (Table~\ref{tb:params}).

Model distances for the Magellanic Cloud pulsars are given in
Table~\ref{tb:a1_mc} along with the DM contributions of the Galaxy and
Magellanic Clouds. Because of the relatively high Galactic latitudes
of the Magellanic Clouds, the Galactic contributions to the DM are
usually small compared to the Magellanic Cloud contributions. However,
there are exceptions, in particular, for PSR J0451$-$67. The discovery
paper for this pulsar \citep{mfl+06} recognised its uncertain
association with the LMC, and we confirm that here. For NE2001, the
Galactic model DM in this direction is $\sim 2$~cm$^{-3}$~pc less than
the observed DM ($45\pm 1$~cm$^{-3}$~pc) putting the pulsar in the
LMC, but for YMW16, the Galactic model DM (48.3~cm$^{-3}$~pc) is about
3~cm$^{-3}$~pc more than the observed DM, giving the pulsar an
unrealistic model distance of $\sim 4.2$~kpc
(Table~\ref{tb:a1_mc}). Future revisions of the distance model and/or
better determination of the pulsar parameters may resolve this
uncertainty about the association. For several other nominally LMC pulsars,
DM$_{\rm Gal}$ dominates the observed DM, but the association is
likely to be correct in these cases. The total model DM in the
direction of the SMC pulsar PSR~J0131$-$7310, discussed above, is
about 85~cm$^{-3}$~pc, much less than the pulsar DM of
205~cm$^{-3}$~pc, and so the model distance is just a nominal upper limit.

Predicted scattering timescales for the Magellanic Cloud pulsars are
listed in Table~\ref{tb:a1_mc}. For 20 of the 27 pulsars listed, the
overall scattering time is dominated by the scattering occuring in the
Magellanic Clouds. The largest predicted scattering timescale is
0.27~ms for PSR J0537$-$69. This pulsar has a relatively short period
of 112~ms. Along with the rest of the Magellanic Cloud pulsars, PSR
J0537$-$69 is relatively weak and its best profile \citep{rcl+13} has
insufficient sensitivity to detect a scattering tail. In fact, no
Magellanic Cloud pulsars have measured scattering timescales.

\subsection{Fast Radio Bursts}
Table~\ref{tb:a1_frb} gives the model redshift and co-moving distance
estimates for the 17 currently known FRBs based on the model of
\citet{kat16} which assumes a present-day baryon density $n_{\rm IGM}
= 0.16$~m$^{-3}$. Model Galactic, Magellanic Cloud and IGM
contributions to the DM are listed in Table~\ref{tb:a1_frb}. As
discussed in \S\ref{sec:IGM}, by default, we adopt a value of
100~cm$^{-3}$~pc for $\rm DM_{Host}$, the observed ($z=0$) DM
contribution of the FRB host galaxy. Derived redshifts range between
0.2 and 2.1 and model distances from 900~Mpc to 5~Gpc. Several aspects
of these results are worth further comment.

For the first time, we estimate the contribution of the SMC to the DM
of the original Lorimer Burst \citep{lbm+07}, FRB010724, obtaining
$\rm DM_{MC} \sim 61$~cm$^{-3}$~pc. \citet{lbm+07} assumed $\rm
DM_{Host} = 200$~cm$^{-3}$~pc and a distance of 500~Mpc and hence a
redshift $z\sim 0.12$. The YMW16 model gives a redshift of 0.254 and a
co-moving distance of 1~Gpc. However, as discussed in
\S\ref{sec:mc_results}, the SMC model is based on just four pulsars
and omits one high-DM pulsar, PSR J0131$-$7310, so it is quite
possible that our estimate of $\rm DM_{MC}$ in this direction is
under-estimated. In any case, it is clear that the SMC contribution to
the DM of FRB010724 is significant.

For FRB150418, the only FRB with a (possible) identification
\citep{kjb+16}, with no fine tuning whatsoever, the YMW16 model
redshift is 0.492 (Table~\ref{tb:a1_frb}), exactly the
redshift of the suggested galaxy identification. While it is
interesting that the model redshift is close to the galaxy redshift,
the exact match is no more than a remarkable coincidence. This is
illustrated by the derived redshifts of 0.562 and 0.351 for DM$_{\rm
  Host}$ equal to 50~cm$^{-3}$~pc and 200~cm$^{-3}$~pc, respectively.

 In the discovery paper, \citet{sch+14} discuss the origin of
  FRB121102 and conclude that it is most likely to be
  extra-galactic. This choice was primarily based on the NE2001
  estimate of the Galactic contribution, 188~cm$^{-3}$~pc, being a
  small fraction of the total DM, 557~cm$^{-3}$~pc. However, as
  Table~\ref{tb:a1_frb} shows, the YMW16 model gives $\rm DM_{Gal}
  \sim 287$~cm$^{-3}$~pc, more than 50\% of the total DM. While the
  residual DM is still relatively large, this new result does somewhat
  weaken the argument that the source is extragalactic. Furthermore,
  FRB121102 is the only FRB known to emit multiple pulses
  \citep{ssh+16}. This also marks it as unusual since no repeating
  pulses have been found from other FRBs \citep{pjk+15}.

Both of these points lead one to consider the relationship of FRBs to
RRATs. RRATs are radio sources that emit detectable pulses only
sporadically \citep{mll+06}. In most cases, their DMs suggest that
they are Galactic sources and careful analyses of pulse arrival times
have revealed periodicties within the range of pulsar periods
\citep[e.g.,][]{mll+06,kkl+15}. Consequently, they are generally
considered to be a class of Galactic pulsars with unusual emission
properties. However, the
``RRATalog''\footnote{\url{http://astro.phys.wvu.edu/rratalog/}}, a
list of currently known RRATs, lists 12 sources where only a single
pulse has been detected. Several of these have DMs that are close to
the estimated total Galactic DM in their direction, notably
J1059$-$01, J1354+24 and J1610$-$17, which have DM/DM$_{\rm
  Gal,total}$ ratios of 0.711, 0.975 and 0.815, respectively. It is
possible that these sources and maybe a few others could in fact be
extra-galactic FRBs \cite[cf.,][]{kea16}. Conversely, the properties
of FRB121102 suggest that it could be a Galactic RRAT, a possibilty
that was indeed considered by \citet{sch+14}. Several other FRBs have
low Galactic latitudes and large estimated $\rm DM_{Gal}$ values
(Table~\ref{tb:a1_frb}) leading to doubt about their extra-galactic
origin \citep[e.g.,][]{bm14}. Given the uncertainties inherent in
Galactic $n_e$ models, the identification of burst sources with DMs
near the total Galactic DM in their direction as Galactic or
extra-galactic will remain problematic. Consideration of other
properties such as pulse shape and whether or not there are repeated
pulses may necessary to discriminate between Galactic and
extra-galactic origins for observed burst sources.

Predicted scattering timescales for FRBs are listed in the final two
columns of Table~\ref{tb:a1_frb}. All but three of these are dominated
by the IGM contribution. For the exceptions, FRB010621,
FRB121102 and FRB150418, the observed scattering times are upper limits and the
predicted Galactic scattering time is less than the upper
limit. Consequently, these cases do not significantly affect our FRB
scattering model (Equation~\ref{eq:tsc_igm}). FRB130729 stands out
with the observed scattering timescale \citep{cpk+16} much larger than
is predicted by the model. The observed profile for this FRB has a
rather low signal-to-noise ratio and it is possible that it has an
intrinsic double-peaked profile similar to FRB121002
\citep{cpk+16}. In this case the estimate of the scattering timescale
would be much reduced.

\section{Summary and Conclusions}\label{sec:summary}
In this paper we have constructed a new model, YMW16, for the
distribution of thermal free electrons in the Galaxy with the main aim
of providing more reliable distance estimates for real or simulated
pulsars based on their dispersion measure. The model also includes
contributions from the Magellanic Clouds and the intergalactic medium
(IGM), enabling distance estimates for extra-galactic pulsars and
FRBs.  The Galactic model is based on independently determined
distances for 189 pulsars. As tabulated in Appendix~A, these
model-indepependent distances are derived from measurements of annual
parallax, HI absorption together with a kinematic model for Galactic
rotation, associations with globular clusters, supernova remnants or
(in one case) an HII region or distance estimates for binary companion
stars. While we estimate the amount of pulse broadening due to
interstellar scattering for Galactic and Magellanic Cloud pulsars and
FRBs, we do not use scattering data as an input to the model.

The main components of the model are as follows: a plane parallel
thick disk with a scale height of about 1600~pc, a thin disk of scale
height about 65~pc mainly representing the ``molecular ring'' of the
Galaxy around 4~kpc Galactocentric radius, spiral arms following a
logarithmic spiral pattern based on the Galactic distribution of HII
regions \citep{hh14}, a Galactic Center disk of radius 160~pc and
scale height 35~pc, an ellipsoidal shell with major axis in the $z$
direction representing the Gum Nebula, the Local Bubble, a region of
reduced $n_e$ surrounding the Sun, together with regions of enhanced
density on the periphery of the Bubble, and a hemispherical shell of
enhanced density associated with the nearby Galactic Loop I. The
Carina-Sagittarius spiral arm has a region of enhanced $n_e$ on the
Carina side and a region of depressed $n_e$ on the Sagittarius
side. All Galactic disk components follow a sech$^2(z/H)$
$z$-dependence, where $H$ is the scale height, and have a sech$^2(R)$
cutoff at $R = 15$~kpc with a scale length of 2.5~kpc. Unlike the
NE2001 model, the YMW16 model does not have clumps or voids toward
particular pulsars or highly-scattered galactic or extra-galactic
sources. The $n_e$ distribution in the Magellanic Clouds is based on
the distribution of star-formation regions and young stars. The SMC is
modelled as a spherical distribution and the LMC as an inclined disk
with an additional component representing the giant HII region 30
Doradus. The IGM model is based on the relations given by
\citet{kat16} with a mean present-day intergalactic $n_e$ of
0.16~m$^{-3}$.

In total, the YMW16 model has 82 fixed parameters and 35 fitted
parameters. A non-linear least-squares procedure was used to fit the
32 free parameters of the Galactic model to the 189 independently
measured pulsar distances. The three free parameters of the Magellanic
Cloud model were separately fitted to the 27 Magellanic Cloud pulsars
with known DMs. Values of the fixed and fitted parameters and their
uncertainties are given in Section~\ref{sec:model} and
Table~\ref{tb:params} and the overall form of the model is shown in
Figure~\ref{fg:ne_model}. Of the 189 Galactic pulsars, the model
distance is within the uncertainties of the independent distance for
86 pulsars and within 20\% of the nearest limit for a further 38
pulsars. By repeatedly selecting a random sample containing 184 of the
189 pulsars, refitting the model to these 184 pulsars and checking the
model prediction for the remaining five pulsars, we estimate that 95\%
of model predictions will have a relative error of less than 90\%. 

Comparison of the YMW16 model distances with those of the TC93 and
NE2001 models for the full sample of 189 pulsars showed that the new
model performs substantially better, especially for high-latitude
pulsars. However, this is not a fair comparison as we have the
advantage of fitting the full 189-pulsar sample whereas TC93 and
NE2001 did not. For a fairer comparison we compared the model
predictions from the repeated fitting the randomly chosen 184-pulsar
sample with predictions of the TC93 and NE2001 models for a sample of
71 pulsars with recently determined independent distances. This showed
that YMW16 is expected to perform about 50\% better than NE2001 and
100\% better than TC93 in estimating pulsar distances based on their DM.

While the YMW16 model performs significantly better than earlier
models in predicting the distances of pulsars based on their DM, there
is certainly room for improvement. At present less than 10\% of known
pulsars have independently determined distances and only about 8\%
have sufficiently accurate distance estimates to be useful. More than
75\% of these are located within 5~kpc of the Sun. New relatively
precise independent distance estimates, especially for more distant
pulsars, will enable us both to better test the model and to improve
it. Currently the model contains 15 components describing the $n_e$
distribution in the Galaxy, including five spiral arms and seven
relatively local features such as the Carina excess and the Gum
Nebula. The highly anomalous prediction for PSR J0248+6021 highlights
the effect of HII regions on the predicted distances. YMW16 includes
the several relatively local over-dense regions, but does not attempt
to model more distant HII regions such as W5. A future iteration of
the model would benefit from inclusion of at least the larger and
denser HII regions within a few kpc of the Sun. It is however
impossible to model the detailed structure of the ISM throughout the
Galaxy or even within the relatively local region, and so it is
inevitable that there will be errors and uncertainties in the
predicted distances. 

The YMW16 model is the first to include models (albeit relatively
simple models) for the $n_e$ distribution in the Large and Small
Magellanic Clouds and for the intergalactic medium. These provide
useful information on the expected DMs and distances for
extra-galactic pulsars and redshift and distance estimates for known
and hypothetical FRBs. Notable among the results is the significant
$\rm DM_{MC}$ for the Lorimer Burst.

\acknowledgements

We thank Bill Coles for help with the definition of the
goodness-of-fit statistic, for suggesting the adopted method of
estimating parameter uncertainties and for useful discussions on the
effects of interstellar scattering, and M. Krishnakumar for providing
us with tables of scattering data for Galactic pulsars. We also thank
Matthew Bailes, Simon Johnston, JP Marquart and Ryan Shannon for
useful discussions and George Hobbs and Hailong Zhang for help with
the YMW16 web interface.  Finally, we thank the referee for helpful
comments. This work was supported by National Basic
Research Program of China (973 Program 2015CB857100 and 2012CB821801),
the Strategic Priority Research Programme (B) of the Chinese Academy
of Science (No. XDB09000000), the National Natural Science Foundation
of China (No. 11373006) and West Light Foundation of CAS
(No. ZD201302). We acknowledge extensive use of the SAO/NASA
Astrophysics Data System (\url{http://www.adsabs.harvard.edu/}) in the
preparation of the work presented here.


\clearpage
\appendix

\section{DM-independent distances and YMW16 model distances}
In this Appendix we list by category the DM-independent distances used
to derive the YMW16 Galactic electron density model. Each table gives
the pulsar J2000 name, the Galactic coordinates ($l$,$b$), the DM, the
distance lower limit $D_l$, the estimate distance $D_{i}$ and the
distance upper limit $D_u$. Detailed references for the source of
these data can be obtained from the ATNF Pulsar Catalogue (V1.54). In
each table, we also give $D_m$, the distance estimate based on the
YMW16 Galactic electron density model, and the fractional distance
error $D_{err}$ (expressed as a percentage) defined by
Equation~\ref{eq:derr}. Table~\ref{tb:a1_mc} lists the 27 pulsars with
known DM that are believed to be associated with the Magellanic Clouds
along with the DM contributions from the Galaxy and the Magellanic
Clouds and the model distance. Similarly, Table~\ref{tb:a1_frb} lists
details of the 17 currently known FRBs, the Galactic, Magellanic Cloud
and IGM contributions to the DM and the derived FRB source redshifts
and co-moving distances. The dispersion measure contribution of the
FRB host galaxy, DM$_{\rm Host}$, is assumed to be 100~cm$^{-3}$~pc. All
tables list an estimate of the scattering timescale at 1~GHz for each
source.

\begin{deluxetable}{lrrDrrrrrr}
\tabletypesize{\scriptsize}
\tablecaption{Independent distances from measurements of annual parallax\label{tb:a1_px}}
\tablehead{
\colhead{J2000} & \colhead{$l$} & \colhead{$b$} & \multicolumn2c{DM} & \colhead{$D_l$}
     & \colhead{$D_i$} & \colhead{$D_u$} & \colhead{$D_m$} & \colhead{$D_{err}$}& \colhead{log $\tau_{sc}$}\\
\colhead{Name} & \colhead{($\degr$)} & \colhead{($\degr$)} & \multicolumn2c{(cm$^{-3}$~pc)} & \colhead{(pc)}
   & \colhead{(pc)} & \colhead{(pc)} & \colhead{(pc)} & \colhead{(\%)}& \colhead{(s)}
}
\decimals
\startdata
J0023+0923   &   111.383 &   $-$52.849 &    14.30 &     1000 &  4050 &  7100 & 1244 & 0 & $-$7.700  \\
J0030+0451   &   113.141 &   $-$57.611 &     4.33 &      286 &   303 &   323 & 345  & 7 & $-$8.971  \\
J0034$-$0721 &   110.420 &   $-$69.815 &    11.38 &      950 &  1030 &  1110 & 1052 & 0 & $-$7.966  \\
J0108$-$1431 &   140.930 &   $-$76.815 &     2.38 &      160 &   210 &   300 & 232  & 0 & $-$9.554  \\
J0139+5814   &   129.216 &    $-$4.044 &    73.78 &     2400 &  2600 &  2900 & 2006 & 20  & $-$5.215  \\
J0218+4232   &   139.508 &   $-$17.527 &    61.25 &     2550 &  3150 &  4000 & 2928 & 0 & $-$5.538  \\
J0332+5434   &   144.995 &    $-$1.221 &    26.76 &      900 &  1000 &  1100 & 1183 & 8 & $-$6.868  \\
J0358+5413   &   148.190 &       0.811 &    57.14 &      900 &  1000 &  1200 & 1594 & 33  & $-$5.657  \\
J0437$-$4715\tablenotemark{a} & 253.394 & $-$41.963 & 2.64 & 155 & 156 & 157 & 156  & 0 & $-$9.454  \\
J0452$-$1759 &   217.078 &   $-$34.087 &    39.90 &      300 &   400 &   600 & 2710 & 352 & $-$6.254  \\
J0454+5543   &   152.617 &       7.547 &    14.49 &     1130 &  1180 &  1250 & 631  & 79  & $-$7.684  \\
J0538+2817   &   179.719 &    $-$1.686 &    39.57 &     1100 &  1300 &  1500 & 946  & 16  & $-$6.267  \\
J0613$-$0200 &   210.413 &    $-$9.305 &    38.78 &      950 &  1090 &  1270 & 1024 & 0 & $-$6.299  \\
J0630$-$2834 &   236.952 &   $-$16.758 &    34.47 &      280 &   320 &   370 & 2072 & 460 & $-$6.486  \\
J0633+1746   &   195.134 &       4.266 &     2.89 &      170 &   250 &   480 & 138  & 23  & $-$9.366  \\
J0636+5129   &   163.909 &      18.643 &    11.10 &      182 &   204 &   233 & 209  & 0 & $-$7.994  \\
J0645+5158   &   163.963 &      20.251 &    18.20 &      625 &   769 &  1000 & 669  & 0 & $-$7.400  \\
J0659+1414   &   201.108 &       8.258 &    13.98 &      250 &   280 &   310 & 159  & 57  & $-$7.727  \\
J0737$-$3039A&   245.236 &    $-$4.505 &    48.92 &     1000 &  1100 &  1300 & 1105 & 0 & $-$5.919  \\
J0751+1807   &   202.730 &      21.086 &    30.25 &      300 &   400 &   600 & 428  & 0 & $-$6.686  \\
J0814+7429   &   139.998 &      31.618 &     5.73 &      425 &   432 &   440 & 366  & 16  & $-$8.692  \\
J0820$-$1350 &   235.890 &      12.595 &    40.94 &     1800 &  1900 &  2000 & 2321 & 16  & $-$6.212  \\
J0826+2637   &   196.963 &      31.742 &    19.45 &      270 &   320 &   400 & 313  & 0 & $-$7.313  \\
J0835$-$4510 &   263.552 &    $-$2.787 &    67.99 &      260 &   280 &   300 & 328  & 9 & $-$5.357  \\
J0922+0638   &   225.420 &      36.392 &    27.27 &     1000 &  1100 &  1300 & 1908 & 47  & $-$6.841  \\
J0953+0755   &   228.908 &      43.697 &     2.96 &      256 &   261 &   266 & 186  & 37  & $-$9.343  \\
J1012+5307   &   160.347 &      50.858 &     9.02 &      600 &   700 &   900 & 804  & 0 & $-$8.222  \\
J1017$-$7156 &   291.558 &   $-$12.553 &    94.22 &      196 &   256 &   370 & 1807 & 388 & $-$4.784  \\
J1022+1001   &   231.795 &      51.101 &    10.25 &      610 &   740 &   930 & 691  & 0 & $-$8.083  \\
J1023+0038   &   243.490 &      45.782 &    14.32 &     1328 &  1367 &  1410 & 1057 & 26  & $-$7.699  \\
J1024$-$0719 &   251.702 &      40.516 &     6.49 &      800 &  1100 &  1500 & 381  & 110 & $-$8.566  \\
J1045$-$4509 &   280.851 &      12.254 &    58.17 &      240 &   340 &   360 & 338  & 0 & $-$5.626  \\
J1136+1551   &   241.902 &      69.195 &     4.85 &      330 &   350 &   370 & 415  & 12  & $-$8.859  \\
J1239+2453   &   252.450 &      86.541 &     9.24 &      790 &   840 &   900 & 826  & 0 & $-$8.196  \\
J1300+1240   &   311.310 &      75.414 &    10.17 &      500 &   600 &   800 & 878  & 10  & $-$8.092  \\
J1456$-$6843 &   313.869 &    $-$8.543 &     8.60 &      380 &   430 &   490 & 436  & 0 & $-$8.273  \\
J1509+5531   &    91.325 &      52.287 &    19.61 &     2000 &  2100 &  2200 & 2072 & 0 & $-$7.302  \\
J1537+1155\tablenotemark{a} & 19.848 & 48.341 & 11.61 & 961  &  1162 &  1471 & 876  & 10  & $-$7.944  \\
J1543+0929   &    17.811 &      45.775 &    35.24 &     5400 &  5900 &  6500 & 25000  & 285 & $-$6.451  \\
J1559$-$4438 &   334.540 &       6.367 &    56.10 &     2000 &  2300 &  2800 & 1480 & 35  & $-$5.688  \\
J1600$-$3053 &   344.090 &      16.451 &    52.33 &     1500 &  1800 &  2300 & 2535 & 10  & $-$5.806  \\
J1603$-$7202 &   316.630 &   $-$14.496 &    38.05 &      370 &   530 &   570 & 1129 & 98  & $-$6.330  \\
J1614$-$2230 &   352.636 &      20.192 &    34.50 &      400 &   700 &  1000 & 1395 & 40  & $-$6.484  \\
J1640+2224   &    41.051 &      38.271 &    18.42 &     1450 &  2425 &  3400 & 1502 & 0 & $-$7.384  \\
J1643$-$1224 &     5.669 &      21.218 &    62.41 &      640 &   740 &   860 & 791  & 0 & $-$5.505  \\
J1713+0747   &    28.751 &      25.223 &    15.99 &     1136 &  1176 &  1220 & 919  & 24  & $-$7.564  \\
J1730$-$2304 &     3.137 &       6.023 &     9.62 &      520 &   620 &   770 & 512  & 2 & $-$8.153  \\
J1738+0333   &    27.721 &      17.742 &    33.77 &     1370 &  1470 &  1587 & 1505 & 0 & $-$6.518  \\
J1741+1351   &    37.885 &      21.641 &    24.21 &     1020 &  1075 &  1136 & 1363 & 20  & $-$7.013  \\
J1744$-$1134 &    14.794 &       9.180 &     3.14 &      384 &   395 &   406 & 148  & 159 & $-$9.286  \\
J1756$-$2251 &     6.499 &       0.948 &   121.18 &      490 &   730 &  1330 & 2806 & 111 & $-$4.334  \\
J1900$-$2600 &    10.342 &   $-$13.451 &    37.99 &      500 &   700 &  1100 & 1237 & 12  & $-$6.332  \\
J1909$-$3744 &   359.731 &   $-$19.596 &    10.39 &     1230 &  1234 &  1239 & 564  & 118 & $-$8.068  \\
J1918$-$0642 &    30.027 &    $-$9.123 &    26.55 &      769 &   909 &  1111 & 1026 & 0 & $-$6.880  \\
J1932+1059   &    47.382 &    $-$3.884 &     3.18 &      250 &   310 &   400 & 229  & 9 & $-$9.273  \\
J1935+1616   &    52.436 &    $-$2.093 &   158.52 &     2900 &  3700 &  5000 & 4314 & 0 & $-$3.850  \\
J1939+2134   &    57.509 &    $-$0.290 &    71.04 &     1200 &  1500 &  2000 & 2897 & 45  & $-$5.281  \\
J1944+0907   &    47.160 &    $-$7.357 &    24.34 &      670 &  1335 &  2000 & 1218 & 0 & $-$7.005  \\
J2017+0603   &    48.621 &   $-$16.026 &    23.92 &     1250 &  1560 &  1870 & 1398 & 0 & $-$7.030  \\
J2018+2839   &    68.099 &    $-$3.983 &    14.17 &      890 &   980 &  1090 & 957  & 0 & $-$7.711  \\
J2022+2854   &    68.863 &    $-$4.671 &    24.64 &      700 &  2100 &  2700 & 1707 & 0 & $-$6.988  \\
J2022+5154   &    87.862 &       8.380 &    22.65 &     1600 &  1800 &  2100 & 1424 & 12  & $-$7.106  \\
J2043+1711   &    61.919 &   $-$15.313 &    20.71 &     1000 &  1250 &  1667 & 1473 & 0 & $-$7.229  \\
J2048$-$1616 &    30.514 &   $-$33.077 &    11.46 &      920 &   950 &   970 & 775  & 19  & $-$7.958  \\
J2055+3630   &    79.133 &    $-$5.589 &    97.31 &     4400 &  5000 &  5800 & 5240 & 0 & $-$4.726  \\
J2124$-$3358 &    10.925 &   $-$45.438 &     4.60 &      360 &   410 &   500 & 360  & 0 & $-$8.912  \\
J2129$-$5721\tablenotemark{a} & 338.005 & $-$43.570 & 31.85 & 1700 &3200&4700& 6170 & 31  & $-$6.608  \\
J2144$-$3933 &     2.794 &   $-$49.466 &     3.35 &      150 &   160 &   180 & 289  & 61  & $-$9.223  \\
J2145$-$0750 &    47.777 &   $-$42.084 &     9.00 &      480 &   530 &   590 & 693  & 18  & $-$8.224  \\
J2157+4017   &    90.488 &   $-$11.341 &    70.86 &     2500 &  2900 &  3400 & 4750 & 40  & $-$5.285  \\
J2214+3000   &    86.855 &   $-$21.665 &    22.56 &      909 &  1000 &  1111 & 1674 & 51  & $-$7.112  \\
J2222$-$0137 &    62.018 &   $-$46.075 &     3.28 &      266 &   267 &   268 & 267  & 0 & $-$9.243  \\
J2313+4253   &   104.410 &   $-$16.422 &    17.28 &     1000 &  1060 &  1160 &  1109  & 0 & $-$7.466  \\   
\enddata
\tablenotetext{a}{Distance estimate based on measured $\dot P_b$}
\end{deluxetable}

\begin{deluxetable}{lrrDrrrrrr}
\tabletypesize{\scriptsize}
\tablecaption{Kinematic distances from HI absorption measurements\label{tb:a1_kin}}
\tablehead{
\colhead{J2000} & \colhead{$l$} & \colhead{$b$} & \multicolumn2c{DM} & \colhead{$D_l$}
     & \colhead{$D_i$} & \colhead{$D_u$} & \colhead{$D_m$} & \colhead{$D_{err}$}& \colhead{log $\tau_{sc}$}\\
\colhead{Name} & \colhead{($\degr$)} & \colhead{($\degr$)} & \multicolumn2c{(cm$^{-3}$~pc)} & \colhead{(pc)}
   & \colhead{(pc)} & \colhead{(pc)} & \colhead{(pc)} & \colhead{(\%)}& \colhead{(s)}
}
\decimals
\startdata
J0141+6009   & 129.147 &  $-$2.105  &    34.80 &   1600 &  2300 &  3000 &  1495 & 7 & $-$6.471  \\
J0738$-$4042 & 254.194 &  $-$9.192  &   160.80 &    800 &  1600 &  2400 &  1563 & 0 & $-$3.824  \\
J0742$-$2822 & 243.773 &  $-$2.444  &    73.78 &   1200 &  2000 &  3000 &  3115 & 4 & $-$5.215  \\
J0837$-$4135 & 260.904 &  $-$0.336  &   147.29 &    600 &  1500 &  2700 &  1426 & 0 & $-$3.983  \\
J0908$-$4913 & 270.266 &  $-$1.019  &   180.37 &    300 &  1000 &  1700 &  1023 & 0 & $-$3.617  \\
J0942$-$5552 & 278.571 &  $-$2.230  &   180.20 &    100 &   300 &  1100 & 415 & 0 & $-$3.618  \\
J1001$-$5507 & 280.226 &     0.085  &   130.32 &      0 &   300 &  1400 & 408 & 0 & $-$4.203  \\
J1048$-$5832 & 287.425 &     0.577  &   129.10 &   2200 & 2900 &   4100 & 1793  & 23  & $-$4.220  \\
J1056$-$6258 & 290.292 &  $-$2.966  &   320.30 &   1900 &  2400 &  2900 & 2613  & 0 & $-$2.574  \\
J1141$-$6545 & 295.791 &  $-$3.863  &   116.08 &   1000 &  3000 &  5000 & 1676  & 0 & $-$4.411  \\
J1157$-$6224 & 296.705 &  $-$0.199  &   325.20 &   2000 &  4000 &  6000 & 4758  & 0 & $-$2.546  \\
J1224$-$6407 & 299.984 &  $-$1.415  &    97.47 &   2000 &  4000 &  6000 & 1534  & 30  & $-$4.723  \\
J1243$-$6423 & 302.051 &  $-$1.532  &   297.25 &      0 &  2000 &  4000 & 9411  & 135 & $-$2.710  \\
J1326$-$5859 & 307.504 &     3.565  &   287.30 &   2000 &  3000 &  5000 &  10688  & 114 & $-$2.772  \\
J1327$-$6222 & 307.074 &     0.204  &   318.80 &   2000 &  4000 &  6000 & 6514  & 9 & $-$2.582  \\
J1359$-$6038 & 311.239 &     1.126  &   293.71 &   3000 &  5000 &  7000 & 5472  & 0 & $-$2.732  \\
J1401$-$6357 & 310.568 &  $-$2.140  &    98.00 &   1200 &  1800 &  2500 & 1762  & 0 & $-$4.714  \\
J1453$-$6413 & 315.733 &  $-$4.427  &    71.07 &   2000 &  2800 &  4100 & 1432  & 40  & $-$5.280  \\
J1600$-$5044 & 330.690 &     1.631  &   260.56 &   6000 &  6900 &  8800 & 5110  & 17  & $-$2.949  \\
J1602$-$5100 & 330.688 &     1.286  &   170.93 &   7300 &  8000 &  8900 & 3407  & 114 & $-$3.714  \\
J1644$-$4559 & 339.193 &  $-$0.195  &   478.80 &   4100 &  4500 &  4900 & 4437  & 0 & $-$1.842  \\
J1651$-$4246 & 342.457 &     0.923  &   482.00 &   4600 &  5200 &  7300 & 5753  & 0 & $-$1.830  \\
J1707$-$4053 & 345.718 &  $-$0.198  &   360.00 &   3000 &  4000 &  6000 & 3912  & 0 & $-$2.361  \\
J1709$-$4429 & 343.098 &  $-$2.686  &    75.69 &   2000 &  2600 &  3100 & 2275  & 0 & $-$5.170  \\
J1721$-$3532 & 351.687 &     0.670  &   496.00 &   4000 &  4600 &  5200 & 4705  & 0 & $-$1.777  \\
J1740$-$3015 & 358.294 &     0.238  &   152.15 &    100 &   400 &  2100 & 2945  & 40  & $-$3.924  \\
J1741$-$2054\tablenotemark{a} & 6.425 & 4.909 & 4.70 & 300 & 650 & 1000 &273  & 10  & $-$8.890  \\
J1745$-$3040 & 358.553 &   $-$0.963  &   88.37 &     0 &   200 &  1300  & 2343  & 80  & $-$4.897  \\
J1752$-$2806 &   1.540 &   $-$0.961  &   50.37 &   100 &   200 &  1300  & 1335  & 3 & $-$5.870  \\
J1801$-$2304 &   6.837 &   $-$0.066  & 1073.90 &  3000 &  4000 &  5000  & 6522  & 30  & $-$0.369  \\
J1807$-$0847 &  20.061 &      5.587  &  112.38 &   600 &  1500 &  2700  &2700 & 0 & $-$4.469  \\
J1809$-$1943 &  10.727 &   $-$0.158  &  178.00 &  3100 &  3600 &  4100  &3153 & 0 & $-$3.641  \\
J1820$-$0427 &  25.456 &      4.733  &   84.44 &   100 &   300 &   900  & 2918  & 224 & $-$4.978  \\
J1823+0550   &  34.987 &      8.859  &   66.78 &  1200 &  2000 &  3300  &3089 & 0 & $-$5.388  \\
J1824$-$1945 &  12.279 &   $-$3.106  &  224.65 &  2800 &  3700 &  4300  & 5612  & 31  & $-$3.219  \\
J1825$-$0935 &  21.449 &      1.324  &   19.38 &   100 &   300 &  1000  &261  & 0 & $-$7.317  \\
J1832$-$0827 &  23.272 &      0.298  &  300.87 &  4800 &  5200 &  5700  & 4046  & 19  & $-$2.688  \\
J1833$-$0827 &  23.386 &      0.063  &  411.00 &  4000 &  4500 &  5000  &4381 & 0 & $-$2.120  \\
J1848$-$0123 &  31.339 &      0.039  &  159.53 &  4000 &  4400 &  4800  & 3532  & 13  & $-$3.839  \\
J1852+0031   &  33.523 &      0.017  &  787.00 &  6000 &  8000 & 10000  & 6400  & 0 & $-$0.936  \\
J1857+0212   &  35.617 &   $-$0.390  &  506.77 &  6000 &  8000 & 10000  & 5932  & 1 & $-$1.738  \\
J1901+0331   &  37.213 &   $-$0.637  &  402.08 &  5000 &  7000 &  9000  & 6053  & 0 & $-$2.160  \\
J1901+0716   &  40.569 &      1.056  &  252.81 &  2700 &  3400 &  4300  & 7237  & 68  & $-$3.004  \\
J1902+0556   &  39.501 &      0.210  &  177.49 &  3100 &  3600 &  4200  & 4198  & 0 & $-$3.646  \\
J1902+0615   &  39.814 &      0.336  &  502.90 &  5000 &  7000 & 10000  & 7174  & 0 & $-$1.752  \\
J1903+0135   &  35.727 &   $-$1.955  &  245.17 &  2800 &  3300 &  3900  & 6000  & 54  & $-$3.060  \\
J1906+0641   &  40.604 &   $-$0.304  &  472.80 &  5000 &  7000 &  9000  & 6816  & 0 & $-$1.865  \\
J1906+0746   &  41.598 &      0.147  &  217.75 &  6000 &  7400 &  9900  & 4814  & 25  & $-$3.275  \\
J1909+0254   &  37.605 &   $-$2.713  &  171.73 &  3600 &  4500 &  7700  & 5883  & 0 & $-$3.706  \\
J1909+1102   &  44.832 &      0.992  &  149.98 &  4000 &  4800 &  5900  & 4788  & 0 & $-$3.950  \\
J1915+1009   &  44.707 &   $-$0.651  &  241.69 &  5000 &  7000 &  9000  & 5990  & 0 & $-$3.086  \\
J1916+1312   &  47.576 &      0.451  &  237.01 &  3600 &  4500 &  5700  & 6351  & 11  & $-$3.121  \\
J1917+1353   &  48.260 &      0.624  &   94.54 &  4000 &  5000 &  6000  & 2940  & 36  & $-$4.777  \\
J1921+2153   &  55.777 &      3.501  &   12.44 &   100 &   300 &  1100  &809  & 0 & $-$7.865  \\
J1922+2110   &  55.278 &      2.935  &  217.09 &  2000 &  4000 &  6000  & 7041  & 17  & $-$3.281  \\
J1926+1648   &  51.859 &      0.063  &  176.88 &  4000 &  6000 &  9000  & 4495  & 0 & $-$3.652  \\
J1932+2020   &  55.575 &      0.639  &  211.15 &  3000 &  5000 &  8000  & 4964  & 0 & $-$3.331  \\
J1932+2220   &  57.356 &      1.554  &  219.20 & 10100 & 10900 & 12200  & 7999  & 26  & $-$3.263  \\
J1946+1805   &  55.326 &   $-$3.500  &   16.22 &   100 &   300 &   900  & 1010  & 12  & $-$7.546  \\
J2004+3137   &  69.011 &      0.021  &  234.82 &  7000 &  8000 & 10000  & 7264  & 0 & $-$3.138  \\
J2021+3651\tablenotemark{a} & 75.222 & 0.111 & 367.50 &400 & 1800 &3500 & 10512 & 200 & $-$2.324  \\
J2113+4644   &  89.003 &   $-$1.266   &   141.26 &   3000 &  4000 &5000 & 4123  & 0 & $-$4.058  \\
J2257+5909   & 108.831 &   $-$0.575   &   151.08 &   2000 &  3000 &4000 & 3151  & 0 & $-$3.937  \\
J2321+6024   & 112.095 &   $-$0.566   &    94.59 &   1800 &  2700 &3900 & 2513  & 0 & $-$4.777  \\
\enddata
\tablenotetext{a}{Distance estimate based on X$-$ray absorption}
\end{deluxetable}

\begin{deluxetable}{llrrDrrrrrr}
\tabletypesize{\scriptsize}
\tablecaption{Independent distances from globular cluster associations\label{tb:a1_gc}}
\tablehead{
\colhead{J2000} & Globular & \colhead{$l$} & \colhead{$b$} & \multicolumn2c{DM} & \colhead{$D_l$}
     & \colhead{$D_i$} & \colhead{$D_u$} & \colhead{$D_m$} & \colhead{$D_{err}$}& \colhead{log $\tau_{sc}$}\\
\colhead{Name} & Cluster & \colhead{($\degr$)} & \colhead{($\degr$)} & \multicolumn2c{(cm$^{-3}$~pc)} & \colhead{(pc)}
   & \colhead{(pc)} & \colhead{(pc)} & \colhead{(pc)} & \colhead{(\%)}& \colhead{(s)}
}
\decimals
\startdata
J0024$-$7204C & 47Tuc    &  305.923 &  $-$44.892 &    24.40 &  3650 &  4000 &  4350 & 2547  & 43  & $-$7.002  \\
J0514$-$4002A & NGC1851  &  244.514 &  $-$35.036 &    52.15 & 12100 & 12650 & 13200 & 25000 & 89  & $-$5.812  \\
J1312+1810    & M53      &  332.954 &     79.763 &    24.00 & 17200 & 18900 & 20600 & 25000 & 21  & $-$7.025  \\
J1342+2822A   & M3       &   42.209 &     78.709 &    26.50 &  9600 &  9900 & 10200 & 25000 & 145 & $-$6.883  \\
J1518+0204A    & M5       &    3.870 &     46.802 &    29.40 &  7000 &  7500 & 8000 &7533 & 0 & $-$6.729  \\
J1623$-$2631  & M4       &  350.976 &     15.960 &    62.86 &  1600 &  1800 &  2000 & 3651  & 83  & $-$5.493  \\
J1641+3627A    & M13      &   59.000 &     40.914 &    30.60 &  7200 &  7700 & 8200 &7760 & 0 & $-$6.669  \\
J1701$-$3006A & M62      &  353.578 &      7.322 &   114.97 &  6470 &  7050 &  7630 & 4881  & 33  & $-$4.428  \\
J1721$-$1936  & NGC6342  &    4.857 &      9.738 &    75.70 &  7800 &  8400 &  9000 & 3070  & 154 & $-$5.170  \\
J1740$-$5340A  & NGC6397  &  338.165 &  $-$11.967 &    71.80 &  1500 &  2200 & 2700 & 3140  & 16  & $-$5.262  \\
J1748$-$2021A & NGC6440  &    7.728 &      3.801 &   219.40 &  7600 &  8200 &  8800 & 8551  & 0 & $-$3.262  \\
J1748$-$2446A  & Ter5     &    3.836 &      1.696 &   239.00 &  4600 &  5500 & 6400 &4410 & 4 & $-$3.106  \\
J1750$-$3703A & NGC6441  &  353.532 &   $-$5.009 &   233.82 & 12500 & 13500 & 14500 & 13527 & 0 & $-$3.146  \\
J1801$-$0857A & NGC6517  &   19.225 &      6.762 &   182.56 &  6500 &  7200 &  7900 &6679 & 0 & $-$3.595  \\
J1803$-$30    & NGC6522  &    1.025 &   $-$3.926 &   192.00 &  6240 &  7800 &  9360 &5749 & 9 & $-$3.504  \\
J1804$-$0735  & NGC6539  &   20.792 &      6.773 &   186.32 &  7800 &  8400 &  9000 &8156 & 0 & $-$3.558  \\
J1807$-$2459A & NGC6544  &    5.837 &   $-$2.203 &   134.00 &  2540 &  2790 &  3040 &3015 & 0 & $-$4.153  \\
J1823$-$3021A & NGC6624  &    2.788 &   $-$7.913 &    86.88 &  7800 &  8400 &  9000 & 3145  & 148 & $-$4.927  \\
J1824$-$2452A & M28      &    7.797 &   $-$5.578 &   120.50 &  5200 &  5500 &  5800 & 3737  & 39  & $-$4.344  \\
J1835$-$3259A & NGC6652  &    1.532 &  $-$11.371 &    63.35 & 10200 & 10700 & 11200 & 2711  & 276 & $-$5.480  \\
J1836$-$2354A & M22      &    9.886 &   $-$7.561 &    89.11 &  2900 &  3200 &  3500 &3269 & 0 & $-$4.882  \\
J1905+0154A   & NGC6749  &   36.208 &   $-$2.201 &   193.69 & 13950 & 14450 & 14950 & 5509  & 153 & $-$3.488  \\
J1910$-$5959A  & NGC6752  &  336.525 &  $-$25.730 &    33.28 &  4490 &  4550 & 4610 & 1642  & 173 & $-$6.540  \\
J1911+0101A   & NGC6760  &   36.111 &   $-$3.918 &   199.00 &  8700 &  9500 & 10300 &8869 & 0 & $-$3.439  \\
J1953+1846A    & M71      &   56.744 &   $-$4.563 &   117.00 &  6000 &  6450 & 6900 & 4505  & 33  & $-$4.397  \\
J2129+1210A    & M15      &   65.012 &  $-$27.312 &    67.00 & 12900 & 13550 &14200 & 25000 & 76  & $-$5.383  \\
J2140$-$2310A & M30      &   27.179 &  $-$46.837 &    25.06 &  9200 &  9450 &  9700 & 3112  & 196 & $-$6.963  \\
\enddata
\end{deluxetable}

\begin{deluxetable}{llrrDrrrrrr}
\tabletypesize{\scriptsize}
\tablecaption{Independent distances from nebular associations\label{tb:a1_neb}}
\tablehead{
\colhead{J2000} & Nebula & \colhead{$l$} & \colhead{$b$} & \multicolumn2c{DM} & \colhead{$D_l$}
     & \colhead{$D_i$} & \colhead{$D_u$} & \colhead{$D_m$} & \colhead{$D_{err}$}&\colhead{log $\tau_{sc}$} \\
\colhead{Name} & & \colhead{($\degr$)} & \colhead{($\degr$)} & \multicolumn2c{(cm$^{-3}$~pc)} & \colhead{(pc)}
   & \colhead{(pc)} & \colhead{(pc)} & \colhead{(pc)} & \colhead{(\%)}& \colhead{(s)}
}
\decimals
\startdata
J0205+6449   & SNR:3C58         &   130.719 &      3.084 &   140.70 &  2560 &  3200 &  3840 & 2784  & 0 & $-$4.065  \\
J0248+6021   & HII:W5           &   136.903 &      0.697 &   370.00 &  1800 &  2000 &  2200 &  25000  & 1036  & $-$2.311  \\
J0534+2200   & PWN:Crab         &   184.558 &   $-$5.784 &    56.79 &  1500 &  2000 &  2500 & 1311  & 14  & $-$5.667  \\
J1119$-$6127 & SNR:G292.2$-$0.5 &   292.151 &   $-$0.537 &   707.40 &  8000 &  8400 &  8800 & 6414  & 25  & $-$1.130  \\
J1124$-$5916 & SNR:G292.0+1.8   &   292.038 &      1.752 &   330.00 &  3000 &  5000 &  8000 & 2678  & 12  & $-$2.520  \\
J1400$-$6325 & SNR:G310.6$-$1.6 &   310.592 &   $-$1.593 &   563.00 &  5000 &  7000 &  9000 &  9169 & 2 & $-$1.547  \\
J1513$-$5908 & SNR:G320.4$-$1.2 &   320.321 &   $-$1.162 &   252.50 &  3600 &  4400 &  5700 &4450 & 0 & $-$3.006  \\
J1550$-$5418 & SNR:G327.24$-$0.1 &  327.237 &   $-$0.132 &   830.00 &  3500 &  4000 &  4500 & 6291  & 40  & $-$0.839  \\
J1745$-$2900 & Galactic Center  &   359.944 &   $-$0.047 &  1778.00 &  8000 &  8300 &  8600 & 8289  & 0 & 0.550 \\
J1803$-$2137 & SNR:G8.7$-$0.1   &     8.395 &      0.146 &   233.99 &  3800 &  4400 &  4900 & 3422  & 11  & $-$3.145  \\
J1833$-$1034 & SNR:G21.5$-$0.9  &    21.501 &   $-$0.885 &   169.50 &  3800 &  4100 &  4400 & 3355  & 13  & $-$3.729  \\
J1856+0113   & SNR:W44          &    34.560 &   $-$0.497 &    96.74 &  2700 &  3300 &  3900 & 2820  & 0 & $-$4.737  \\
J1930+1852   & SNR:G54.1+0.3    &    54.096 &      0.265 &   308.00 &  5000 &  7000 & 10000 & 6191  & 0 & $-$2.645  \\
J1952+3252   & SNR:CTB80        &    68.765 &      2.823 &    45.01 &  1000 &  3000 &  5000 & 3221  & 0 & $-$6.057  \\
J2229+6114   & SNR:G106.6+2.9   &   106.647 &      2.949 &   204.97 &  2400 &  3000 &  3600 & 5037  & 40  & $-$3.385  \\
J2337+6151   & SNR:G114.3+0.3   &   114.284 &      0.233 &    58.41 & 600 &   700 &   800   & 2079  & 160 & $-$5.619  \\
\enddata
\end{deluxetable}

\begin{deluxetable}{llrrDrrrrrr}
\tabletypesize{\scriptsize}
\tablecaption{Independent distances from stellar companions\label{tb:a1_stars}}
\tablehead{
\colhead{J2000} & Star & \colhead{$l$} & \colhead{$b$} & \multicolumn2c{DM} & \colhead{$D_l$}
     & \colhead{$D_i$} & \colhead{$D_u$} & \colhead{$D_m$} & \colhead{$D_{err}$}& \colhead{log $\tau_{sc}$}\\
\colhead{Name} & Type & \colhead{($\degr$)} & \colhead{($\degr$)} & \multicolumn2c{(cm$^{-3}$~pc)} & \colhead{(pc)}
   & \colhead{(pc)} & \colhead{(pc)} & \colhead{(pc)} & \colhead{(\%)}& \colhead{(s)}
}
\decimals
\startdata
J0337+1715   & WD   &   169.990 &  $-$30.039 &    21.32 & 1220 & 1300 & 1380 &  817 & 49  & $-$7.189  \\
J0348+0432   & WD   &   183.337 &  $-$36.774 &    40.46 & 1900 & 2100 & 2300 &   2255 & 0 & $-$6.231  \\
J0614$-$3329 & WD   &   240.501 &  $-$21.827 &    37.05 &  760 &  890 & 1020 &   2691 & 164 & $-$6.372  \\
J1227$-$4853 & MS-G &   298.965 &     13.796 &    43.42 & 1800 & 1900 & 2000 &   1244 & 45  & $-$6.116  \\
J1231$-$1411 & WD   &   295.531 &     48.385 &     8.09 &  350 &  430 &  510 & 420  & 0 & $-$8.338  \\
J1302$-$6350 & Be   &   304.184 &   $-$0.992 &   146.72 & 1900 & 2300 & 2700 &  2213  & 0 & $-$3.990  \\
J1544+4937   & UL   &    79.172 &     50.166 &    23.22 & 2000 & 3500 & 5000 &  2988  & 0 & $-$7.071  \\
J1903+0327   & MS   &    37.336 &   $-$1.014 &   297.52 & 6000 & 7000 & 8000 & 6122 & 0 & $-$2.708  \\
J2032+4127   & Be   &    80.224 &      1.028 &   114.65 & 1400 & 1500 & 1700 &  4623  & 172 & $-$4.433  \\
\enddata
\end{deluxetable}

\newpage
\begin{deluxetable}{lcrrDrrrDDrr}
\tabletypesize{\scriptsize}
\tablecaption{Pulsars in the Magellanic Clouds\label{tb:a1_mc}}
\tablehead{
\colhead{J2000} & Cloud & \colhead{$l$} & \colhead{$b$} & \multicolumn2c{DM} & \colhead{$D_l$}
     & \colhead{$D_i$} & \colhead{$D_u$} & \multicolumn2c{DM$_{\rm
    Gal}$} &\multicolumn2c{DM$_{\rm MC}$} & \colhead{$D_m$} &\colhead{log $\tau_{sc}$} \\
\colhead{Name} & & \colhead{($\degr$)} & \colhead{($\degr$)} &
\multicolumn2c{(cm$^{-3}$~pc)}
& \colhead{(pc)} & \colhead{(pc)} & \colhead{(pc)} & \multicolumn2c{(cm$^{-3}$~pc)}
& \multicolumn2c{(cm$^{-3}$~pc)} & \colhead{(pc)}&\colhead{(s)} 
}
\decimals
\startdata
J0045$-$7042 & SMC & 303.652 &  $-$46.418 &  70    & 54000 & 59700 & 66000 & 28.87 &  41.13 &  58520   &  $-$6.505   \\
J0045$-$7319 & SMC & 303.514 &  $-$43.804 & 105.4  & 54000 & 59700 & 66000 & 30.70  &  74.70  &  58725   &  $-$5.494   \\
J0111$-$7131 & SMC & 300.669 &  $-$45.510 &  76    & 54000 & 59700 & 66000 & 29.31 &  46.69 &  59164   &  $-$6.298   \\
J0113$-$7220 & SMC & 300.615 &  $-$44.688 & 125.49 & 54000 & 59700 & 66000 & 30.02 &  95.47 &  60998   &  $-$5.061   \\
J0131$-$7310 & SMC & 298.944 &  $-$43.648 & 205.2  & 54000 & 59700 & 66000 & 31.00 &  54.47 &  100000  &  $-$6.039   \\
J0449$-$7031 & LMC & 282.286 &  $-$35.512 &  65.83 & 44000 & 49700 & 56000 & 49.99 &  15.84 &  50463   &  $-$6.184   \\
J0451$-$67   & LMC & 278.410 &  $-$36.290 &  45    & 44000 & 49700 & 56000 &45.00 &  0.00  &  4184    &  $-$6.057   \\
J0455$-$6951 & LMC & 281.290 &  $-$35.187 &  94.89 & 44000 & 49700 & 56000 & 51.13 &  43.76 &  50591   &  $-$6.146   \\
J0456$-$69   & LMC & 280.457 &  $-$35.334 & 103    & 44000 & 49700 & 56000 & 50.87 &  52.13 &  50514   &  $-$6.113   \\
J0456$-$7031 & LMC & 282.049 &  $-$34.966 & 100.3  & 44000 & 49700 & 56000 & 51.53 &  48.77 &  51132   &  $-$6.133   \\
J0457$-$69   & LMC & 281.143 &  $-$35.113 &  91    & 44000 & 49700 & 56000 & 51.36 &  39.64 &  50356   &  $-$6.138   \\
J0458$-$67   & LMC & 278.670 &  $-$35.518 &  97    & 44000 & 49700 & 56000 & 50.35 &  46.65 &  50185   &  $-$6.172   \\
J0502$-$6617 & LMC & 276.869 &  $-$35.504 &  68.9  & 44000 & 49700 & 56000 & 49.98 &  18.92 &  49066   &  $-$6.184   \\
J0519$-$6932 & LMC & 280.287 &  $-$33.254 & 119.4  & 44000 & 49700 & 56000 & 56.88 &  62.52 &  49741   &  $-$5.803   \\
J0521$-$68   & LMC & 279.124 &  $-$33.262 & 136    & 44000 & 49700 & 56000 & 56.86 &  79.14 &  49777   &  $-$5.393   \\
J0522$-$6847 & LMC & 279.348 &  $-$33.168 & 126.45 & 44000 & 49700 & 56000 & 57.17 &  69.28 &  49581   &  $-$5.626   \\
J0529$-$6652 & LMC & 276.974 &  $-$32.763 & 103.2  & 44000 & 49700 & 56000 & 57.84 &  45.36 &  49581   &  $-$5.937   \\
J0532$-$6639 & LMC & 276.675 &  $-$32.481 &  69.3  & 44000 & 49700 & 56000 & 58.59 &  10.71 &  47853   &  $-$5.915   \\
J0532$-$69   & LMC & 280.333 &  $-$32.163 & 124    & 44000 & 49700 & 56000 & 60.40  &  63.60  &  49602   &  $-$5.774   \\
J0534$-$6703 & LMC & 277.129 &  $-$32.279 &  94.7  & 44000 & 49700 & 56000 &59.45 &  35.25 &  48820   &  $-$5.890   \\
J0535$-$66   & LMC & 276.884 &  $-$32.197 &  75    & 44000 & 49700 & 56000 & 59.60  &  15.40  &  48067   &  $-$5.886   \\
J0535$-$6935 & LMC & 280.076 &  $-$31.936 &  93.7  & 44000 & 49700 & 56000 & 61.18 &  32.52 &  48857   &  $-$5.841   \\
J0537$-$69   & LMC & 279.768 &  $-$31.730 & 273    & 44000 & 49700 & 56000 & 61.89 &  211.11   &  49901   &  $-$3.633   \\
J0540$-$6919 & LMC & 279.717 &  $-$31.516 & 146.5  & 44000 & 49700 & 56000 & 62.62 &  83.88 &  48907   &  $-$5.290   \\
J0542$-$68   & LMC & 278.449 &  $-$31.406 & 114    & 44000 & 49700 & 56000 &62.85 &  51.15 &  48893   &  $-$5.794   \\
J0543$-$6851 & LMC & 279.125 &  $-$31.235 & 131    & 44000 & 49700 & 56000 & 63.57 &  67.43 &  48805   &  $-$5.673   \\
J0555$-$7056 & LMC & 281.460 &  $-$30.119 &  73.4  & 44000 & 49700 & 56000 &67.41 &  5.99  &  48786   &  $-$5.673   \\
\enddata
\end{deluxetable}

\newpage
\begin{deluxetable}{lrrcclDDDrrrr}
\tabletypesize{\scriptsize}
\tablecaption{Fast Radio Bursts\label{tb:a1_frb}}
\tablehead{
  \colhead{Name} & \colhead{$l$} & \colhead{$b$} & \colhead{DM}
   & \colhead{Obs.~$\tau_{\rm sc}$} & \colhead{Ref.} & \multicolumn2c{DM$_{\rm Gal}$} 
   & \multicolumn2c{DM$_{\rm MC}$} & \multicolumn2c{DM$_{\rm IGM}$} & \colhead{$z$}
   & \colhead{$D_m$}&\colhead{log $\tau_{sc}$} & \colhead{$\tau_{sc}$} \\
   & \colhead{($\degr$)} & \colhead{($\degr$)} & \colhead{(cm$^{-3}$~pc)} &\colhead{(ms)} 
  & & \multicolumn2c{(cm$^{-3}$~pc)} & \multicolumn2c{(cm$^{-3}$~pc)} & \multicolumn2c{(cm$^{-3}$~pc)}
  &  & \colhead{(Mpc)}&\colhead{(s)}&\colhead{(ms)}
}
\decimals
\startdata
FRB010125 & 356.641 & $-$20.020 &   790(3) & $<1.4$  & bb14   & 75.91 & 0.00  & 614.09  & 0.861 & 2769 & $-$2.775  & 1.68  \\
FRB010621 &  25.433 &  $-$4.003 &  745(10) & $<2.1$  & kkl+11 &321.56  & 0.00  & 323.44  & 0.453 & 1667 & $-$2.868  & 1.36  \\
FRB010724 & 300.653 & $-$41.805 &   375    &   1.2   & lbm+07 & 32.65 & 61.38 & 180.97  & 0.254 & 1008 & $-$3.465  & 0.34  \\
FRB090625 & 226.443 & $-$60.030 & 899.55(1)& 3.7(7)  & cpk+16 & 25.48 & 0.00  & 774.07  & 1.085 & 3276 & $-$2.645  & 2.26  \\
FRB110220 &  50.828 & $-$54.766 & 944.38(5)& 1.9(1)  & tsb+13 & 24.12 & 0.00  & 820.26  & 1.150 & 3412 & $-$2.612  & 2.44  \\
FRB110523 &  56.119 & $-$37.819 & 623.30(6)& 4.1(4)  & mls+15 & 33.00 & 0.00  & 490.30  & 0.687 & 2332 & $-$2.902  & 1.25  \\
FRB110627 & 355.861 & $-$41.752 & 723.0(3) & $<0.5$  & tsb+13 & 33.57 & 0.00  & 589.43  & 0.826 & 2685 & $-$2.798  & 1.59  \\
FRB110703 &  80.997 & $-$59.019 & 1103.6(7)& $<1.5$  & tsb+13 & 23.08 & 0.00  & 980.52  & 1.375 & 3855 & $-$2.511  & 3.08  \\
FRB120127 &  49.287 & $-$66.203 &  553.3(3)& $<0.4$  & tsb+13 & 20.63 & 0.00  & 432.67  & 0.607 & 2114 & $-$2.973  & 1.06  \\
FRB121002 & 308.219 & $-$26.264 &1629.18(2)& 6.7(7)  & cpk+16 &  60.50 & 0.00  & 1468.68  & 2.059 & 4984 & $-$2.283  & 5.21  \\
FRB121102 & 174.950 &  $-$0.225 &   557(2) & $<1.5$  & sch+14 &287.12  & 0.00  & 169.88  & 0.238 & 952 & $-$3.074  & 0.84  \\
FRB130626 &   7.450 &   +27.420 & 952.4(1) & 2.9(7)  & cpk+16 & 65.09 & 0.00  & 787.31  & 1.104 & 3316 & $-$2.635  & 2.32  \\
FRB130628 & 225.955 &   +30.655 & 469.88(1)& 1.24(7) & cpk+16 & 46.99 & 0.00  & 322.89  & 0.453 & 1665 & $-$3.138  & 0.73  \\
FRB130729 & 324.787 &   +54.744 &   861(2) &  23(2)  & cpk+16 & 25.42 & 0.00  & 735.58  & 1.031 & 3159 & $-$2.673  & 2.12  \\
FRB131104 & 260.549 & $-$21.925 &   779(1) & 0.3(1)  & rsj15  &220.20  & 0.00  & 458.80  & 0.643 & 2214 & $-$2.940  & 1.15  \\
FRB140514 &  50.841 & $-$54.611 &  562.7(6)& 5.4(1)  & pbb+15 & 24.17 & 0.00  & 438.53  & 0.615 & 2136 & $-$2.965  & 1.08  \\
FRB150418 & 232.665 &  $-$3.234 &  776.2(5)& $<0.25$ & kjb+16 &325.54  & 0.00  & 350.66  & 0.492 & 1782 & $-$2.845  & 1.43  \\
\enddata   
\tablerefs{bb14: \citet{bb14}; cpk+16: \citet{cpk+16}; kjb+16: \citet{kjb+16};
  kkl+11: \citet{kkl+11}; lbm+07: \citet{lbm+07}; mls+15: \citet{mls+15};
  pbb+15: \citet{pbb+15}; rsj15: \citet{rsj15}; sch+14: \citet{sch+14};
  tsb+13: \citet{tsb+13}
}
\end{deluxetable}

\clearpage
\setcounter{figure}{0}
\renewcommand{\thefigure}{B.\arabic{figure}}

\section{Perpendicular distances}
In this section we give the equations that are used to compute the
perpendicular distances from spiral arms and the Gum Nebula
shell. These distances are required to estimate the electron density
near these features using the equations given in \S\ref{sec:spiral}
and \S\ref{sec:gn} respectively. Other features such as the nebula at
the Galactic Center (\S\ref{sec:galctr}) and Loop I
(\S\ref{sec:loop1}) have circular or spherical symmetry and so the
calculations of perpendicular distances are relatively simple as given
by the equations in the relevant sub-section of
\S\ref{sec:model}. These equations only approximate the true
perpendicular distances from a point ($x,y$), but their implementations
are faster than the more precise (iterative) solutions and are
adequate for present purposes.

\subsection{Spiral arms}
\begin{figure}[h!]
\includegraphics[angle=270,width=85mm]{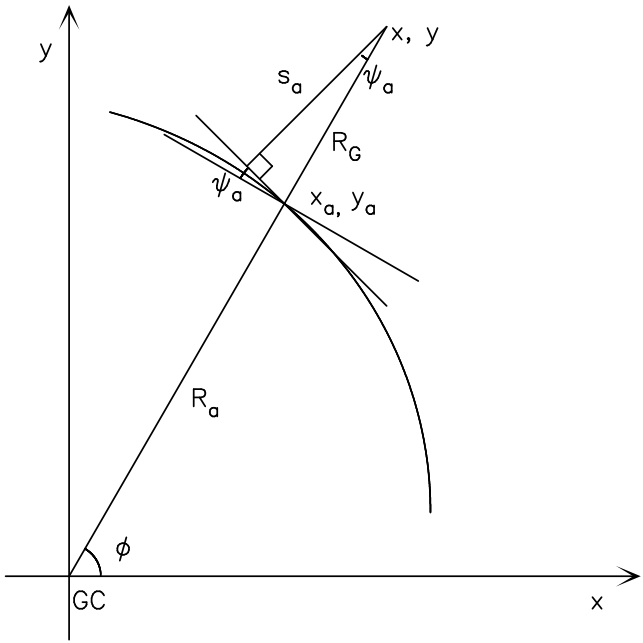}
\caption{Geometry for estimating the perpendicular distance from a
  point ($x,y$) in the plane of the Galaxy to the central axis of a
  spiral arm.}\label{fg:spiral_s}
\end{figure}

To compute the $n_e$ profile of spiral arms
(Equation~\ref{eq:arm_ne}), we require the perpendicular distance
$s_a$ from the central axis of the spiral arm in the Galactic ($x,y$)
plane. The geometry is illustrated in Figure~\ref{fg:spiral_s}. The
polar coordinates of ($x,y$) are ($R,\phi$) where the origin is at
the Galactic Center. From Equation~\ref{eq:spiral}, $R_a$, the radial
distance of the point ($x_a,y_a$) on the arm axis at azimuth $\phi$,
is given by
\begin{equation}
R_a = R_{a_i} \exp[(\phi-\phi_{a_i})\tan\psi_a]
\end{equation}
where $\psi_a$ is the pitch angle of the spiral arm and $R_{a_i}$ and
$\phi_{a_i}$ give the start location of the spiral
(see Table~\ref{tb:spiral}). The
approximate perpendicular distance from the arm axis is then given by:
\begin{equation}
s_a=(R-R_a)\cos\psi_a
\end{equation}

\clearpage

\subsection{Gum Nebula}
\begin{figure}[ht]
\includegraphics[angle=270,width=85mm]{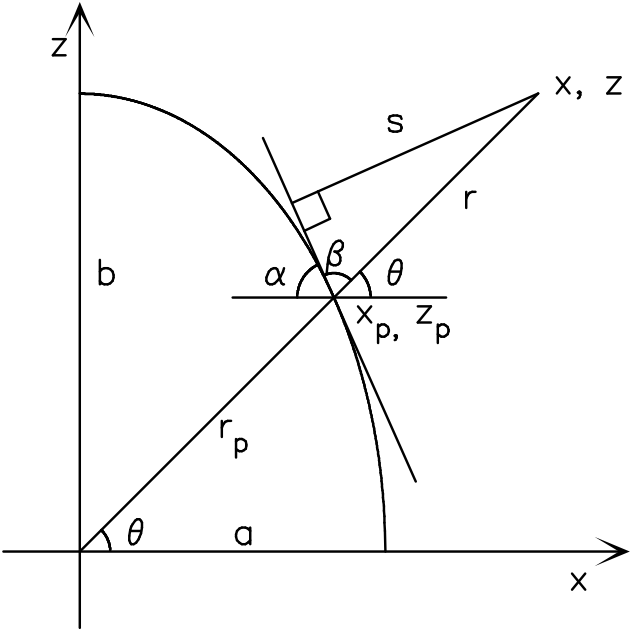}
\caption{Geometry of the Gum Nebula model. The origin of this local coordinate system is
  the center of the Nebula with the $z$ axis is parallel to the
  Galactic $z$ axis. }\label{fg:gns}
\end{figure}

We model the Gum Nebula as an ellipsoidal shell with major axis in the
$z$ direction where the local ($x,y,z$) system
is centered on the Gum Nebula and these axes are parallel to the
corresponding Galactic axes.  We require the perpendicular
distance to the shell mid-line from either inside or outside the shell
to compute the local electron density (Equation~\ref{eq:gn}). The shell
is circular in the $x-y$ plane and hence, for the purposes of this
calculation, we simplify the problem to two dimensions, with the $x-z$
plane passing through the point of interest. Figure~\ref{fg:gns} shows
the geometry in the $x-z$ plane. For compactness we use $a$ and $b$
for the semi-minor axis and semi-major axis, respectively. 

A point ($x_p,z_p$) on the shell is then given by:
\begin{equation}
\frac{{x_p}^2}{a^2}+\frac{{z_p}^2}{b^2}=1
\end{equation}
and we require the perpendicular distance $s$ from a point ($x,z$) to
the shell mid-line. The polar angle $\theta$ is given by:
\begin{equation}
\tan{\theta}=\frac{z}{x}=\frac{z_p}{x_p}.
\end{equation}
Given ($x,z$), we can then obtain $x_p$ and $z_p$ as follows:
\begin{equation}
x_p=\frac{ab}{(b^2+a^2\tan^2\theta)^{1/2}}
\end{equation}
\begin{equation}
z_p=\frac{ab\tan\theta}{(b^2+a^2\tan^2\theta)^{1/2}}
\end{equation}
From Figure~\ref{fg:gns}, $\tan(-\alpha)$ is the gradient of the tangent at ($x_p$, $z_p$):
\begin{equation}
\tan(-\alpha)=\frac{-bx_p}{a(a^2-x_p^2)^{1/2}},
\end{equation}
the angle $\beta$ is:
\begin{equation}
\beta=180\degr-\alpha-\theta
\end{equation}
and
\begin{equation}
r-r_p=[(x-x_p)^2+(z-z_p)^2]^{1/2}.
\end{equation}
We can then approximate the perpendicular distance from ($x,z$) to the
shell mid-line by:
\begin{equation}
s=(r-r_p)\sin{\beta}.
\end{equation}

\clearpage
\setcounter{figure}{0}
\renewcommand{\thefigure}{C.\arabic{figure}}

\section{Coordinate systems for the Large Magellanic Cloud}
\begin{figure}[ht]
\includegraphics[angle=270,width=85mm]{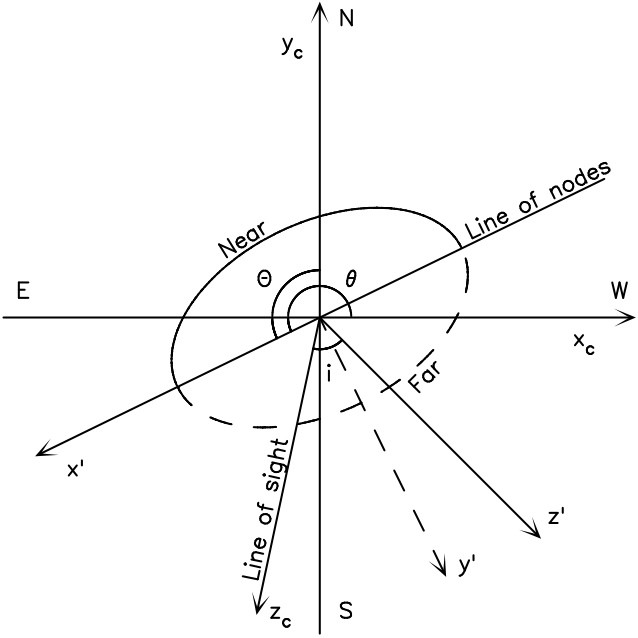}
\caption{Coordinate systems for the Large Magellanic Cloud. The
  ($x_c,y_c,z_c$) system is centered on the LMC with the $+x_c$ axis
  in the direction of decreasing right ascension (W), the $+y_c$ axis
  toward north and the $+z_c$ axis toward the observer. The
  ($x',y',z'$) system is also centered on the LMC with the galaxy disk
  in the ($x',y'$) plane and $+x'$ along the line of nodes at
  position angle $\theta$ measured counter-clockwise from the $x_c$
  axis. The $+y'$ axis lies behind the plane of the sky and the disk normal $+z'$ is
  oriented at inclination angle $i$ to the $z_c$ axis in front of the
  plane of the sky. }\label{fg:lmc}
\end{figure}

To compute electron densities in the Large Magellanic Cloud we need to
transform from the Galactic ($l,b,D$) system to the
($x',y',z'$) system centered on the LMC with the
$x'$ axis along the line of nodes as described in
\S\ref{sec:mc}. Figure~\ref{fg:lmc} illustrates the relevant axes and
angles.

Since the parameters of the LMC are described in
celestial (J2000) coordinates, we first convert to the
($\alpha,\delta,D$) system using:
LMC ($\alpha_{\rm LMC}, \delta_{\rm LMC}$) and $\phi$ is a position
angle measured counter-clockwise from the direction of decreasing
right ascension:
\begin{align}\label{eq:c-angle}
  \cos\rho &= \cos\delta\cos\delta_{\rm LMC}\cos(\alpha
  -\alpha_{\rm  LMC})+\sin\delta\sin\delta_{\rm LMC} \\ \nonumber
\sin\rho\cos\phi &= -\cos\delta\sin(\alpha-\alpha_{\rm LMC})\\ \nonumber
\sin\rho\sin\phi &= \sin\delta\cos\delta_{\rm LMC}
-\cos\delta\sin\delta_{\rm LMC}\cos(\alpha-\alpha_{\rm LMC}) 
\end{align}
The ($x_c,y_c,z_c$) system is then defined by:
\begin{align}\label{eq:c-xyz}
x_c &= D\sin{\rho}\cos{\phi} \\ \nonumber
y_c &= D\sin{\rho}\sin{\phi} \\ \nonumber
z_c &= D_{\rm LMC} -D\cos{\rho} 
\end{align}
and the ($x',y',z'$) system by:
\begin{align}\label{eq:lmc1}
x' &= x_c\cos\theta+y_c\sin\theta \\ \nonumber
y' &= -x_c\sin\theta\cos i +y_c\cos\theta\cos i -z_c\sin i \\ \nonumber
z' &= -x_c\sin\theta\sin i +y_c\cos\theta\sin i +z_c\cos i. 
\end{align}
where $i$ is the LMC plane inclination angle, $\theta =
\Psi+90\degr$ and $\Psi$ is the astronomical position angle of the
line of nodes. Using Equations~\ref{eq:c-xyz} and \ref{eq:lmc1} we
obtain:
\begin{align}\label{eq:lmc2}
x' &= D\sin\rho\cos(\phi-\theta) \\ \nonumber
y' &= D[\sin\rho\cos i\sin(\phi-\theta)+\cos\rho\sin i]-
      D_{\rm  LMC}\sin i \\ \nonumber
z' &= D[\sin\rho\sin i\sin(\phi-\theta)
       -\cos\rho\cos i]+D_{\rm LMC}\cos i 
\end{align}

\section{The YMW16 Distance -- DM Program}
The program {\sc ymw16} computes distances for Galactic pulsars,
Magellanic Cloud pulsars and Fast Radio Bursts (FRBs) from their
Galactic coordinates and DMs using the YMW16 model parameters. It also
does the reverse calculation, computing DMs that correspond to given
Galactic coordinates and distances. An estimate of the scattering
timescale $\tau_{\rm sc}$ is output for Galactic and Magellanic Cloud
pulsars and FRBs.

The program is written in C and is
publically available at the following websites:
\begin{itemize}
\item http://www.xao.ac.cn/ymw16/
\item http://www.atnf.csiro.au/research/pulsar/ymw16/
\item https://bitbucket.org/psrsoft/ymw16/
\end{itemize}
The first two websites also have an interactive facility enabling
on-line execution of the {\sc ymw16} program and provide for download
of the latest version of the program. The third website includes the
full development history of the program, a download facility and an
``issues'' reporting system.

Following the definitions in section \ref{sec:model}, {\sc ymw16}
includes eight functions for Galactic components, three for Magellanic
Cloud components and one for FRBs. The corresponding C files are {\tt
  thick.c}, {\tt thin.c}, {\tt spiral.c}, {\tt galcen.c}, {\tt gum.c},
{\tt localbubble.c}, {\tt nps.c}, {\tt fermibubble.c}, {\tt lmc.c},
{\tt dora.c}, {\tt smc.c} and {\tt frb\_d.c}. The two remaining C
files are the main program {\tt ywm16.c} and {\tt ymw16par.c} which
reads in the model parameters (Table~\ref{tb:params}) from {\tt
  ymw16par.txt}. The spiral parameters (Table~\ref{tb:spiral}) are
contained in {\tt spiral.txt} which is read in by {\tt spiral.c}.  The
Sun is located at $z_\odot = +6.0$~pc above the Galactic plane and a
warp in the outer Galactic disk is included in the model.

To compute the distance $D_m$ corresponding to a given DM, the local
$n_e$ is evaluated at steps of 5~pc along the path, or steps of
$D_t/200$ if the nominal number of steps along the path is less than
200, where $D_t={\rm DM}/0.013$~pc is a nominal distance. The model DM
is accumulated at each step until the input DM is reached (or the
distance limit is reached). For the inverse process, the DM is
accumulated out to the input distance. In order to improve the
  computational efficiency of the program, we cease calculation of all model
  components when the distance from the central point or axis of a
  given component is greater than six times the scale length of that
  component. For a sech$^2$ dependence, the value of $n_e$ at this
  point is less than 10$^{-5}$ times the central value and hence
  negligible.

To compile and execute the {\sc ymw16} program:
\begin{enumerate}
\item Download the source code from one of the above websites and
    unpack.
\item  Run {\tt make\_ymw16} to compile the code and
create the executable {\sc ymw16}.
\item  To run {\sc ymw16}: {\tt ymw16 [-h] [-t $\langle$text$\rangle$] [-d $\langle$dirname$\rangle$] [-v] [-V] $\langle$mode$\rangle$ $\langle$gl$\rangle$
  $\langle$gb$\rangle$ $\langle$DM/Dist$\rangle$
  [$\langle$DM\_{Host}$\rangle$] $\langle$ndir$\rangle$}, where
  optional inputs are enclosed in square brackets and\\
{\tt -h}: prints the help page \\
{\tt -t $\langle$text$\rangle$}: where $\langle${\tt text}$\rangle$ (no
space, maximum 64 characters) is appended to the output line\\
{\tt -d $\langle$dirname$\rangle$}: where $\langle${\tt dirname}$\rangle$ is a directory containing YMW16 data files\\
{\tt -v}: prints diagnostics \\
{\tt -V}: prints more diagnostics \\
{\tt mode} is one of:  {\tt Gal}, {\tt MC}, or {\tt IGM} \\
{\tt gl}: Galactic longitude (deg.)\\
{\tt gb}: Galactic latitude (deg.)\\
{\tt DM/Dist}: One of DM (cm$^{-3}$~pc) or distance, depending on
{\tt ndir}. Distance has units of pc for modes {\tt Gal} and {\tt MC}
and Mpc for mode {\tt IGM} \\ 
{\tt DM\_Host}: Dispersion measure of the FRB host galaxy in the
observer frame (default
100~cm$^{-3}$~pc). (Note: if present, {\tt DM\_Host} is ignored for {\tt Gal} and
{\tt MC} modes.) \\
{\tt ndir}: ndir$=$1 converts from DM to distance and ndir$=$2 converts
from distance to DM.
The output $\tau_{\rm sc}$ has units of seconds. 
\end{enumerate}

Output formats are as follows. \\ 
For ndir = 1: \\
{\tt Gal: gl= $\langle$val$\rangle$  gb= $\langle$val$\rangle$ DM=
  $\langle$val$\rangle$ DM\_Gal: $\langle$val$\rangle$ Dist:
  $\langle$val$\rangle$ log(tau\_sc): $\langle$val$\rangle$} $\langle${\tt text}$\rangle$ \\
{\tt MC: gl= $\langle$val$\rangle$  gb= $\langle$val$\rangle$ DM=
  $\langle$val$\rangle$ DM\_Gal: $\langle$val$\rangle$ DM\_MC:
  $\langle$val$\rangle$ Dist: $\langle$val$\rangle$ log(tau\_sc):
  $\langle$val$\rangle$} $\langle${\tt text}$\rangle$ \\
{\tt IGM: gl= $\langle$val$\rangle$  gb= $\langle$val$\rangle$ DM=
  $\langle$val$\rangle$ DM\_Gal: $\langle$val$\rangle$ DM\_MC:
  $\langle$val$\rangle$ DM\_IGM: $\langle$val$\rangle$ DM\_Host:
  $\langle$val$\rangle$ z: $\langle$val$\rangle$ 
Dist: $\langle$val$\rangle$ log(tau\_sc): $\langle$val$\rangle$}
$\langle${\tt text}$\rangle$. \\ \\
For ndir = 2: \\
{\tt Gal: gl= $\langle$val$\rangle$  gb= $\langle$val$\rangle$ D= $\langle$val$\rangle$ DM: $\langle$val$\rangle$ log(tau\_sc): $\langle$val$\rangle$}  $\langle${\tt text}$\rangle$\\
{\tt MC: gl= $\langle$val$\rangle$  gb= $\langle$val$\rangle$ D= $\langle$val$\rangle$ DM\_Gal: $\langle$val$\rangle$  DM\_MC: $\langle$val$\rangle$  DM: $\langle$val$\rangle$ log(tau\_sc): $\langle$val$\rangle$} $\langle${\tt text}$\rangle$ \\
{\tt IGM: gl= $\langle$val$\rangle$  gb= $\langle$val$\rangle$ D= $\langle$val$\rangle$ DM\_Gal: $\langle$val$\rangle$ DM\_MC: $\langle$val$\rangle$ DM\_IGM: $\langle$val$\rangle$ DM\_Host: $\langle$val$\rangle$ z: $\langle$val$\rangle$ DM: $\langle$val$\rangle$ log(tau\_sc): $\langle$val$\rangle$} $\langle${\tt text}$\rangle$. \\

For ndir = 1 and mode {\tt Gal}, if the input DM exceeds the
range of the Galactic model, the distance is set set to 25000~pc. For
ndir = 1 and mode {\tt MC}, the upper limit is 100000~pc. There
is no upper limit for {\tt IGM} mode.

\end{document}